\documentclass[prd,aps,superscriptaddress,nofootinbib,showpacs,twocolumn]{revtex4}

\usepackage[intlimits]{amsmath}
\usepackage{amsfonts}
\usepackage{amssymb,amscd}
\usepackage{color}
\usepackage[dvips]{epsfig}                   

\newcommand{\cH}{\mathcal{H}}

\newcommand{\beq}{\begin{equation}}
\newcommand{\eeq}{\end{equation}}
\newcommand{\be}{\begin{equation}}
\newcommand{\ee}{\end{equation}}

\newcommand*{\pd}{\partial}
\newcommand*{\pref}[1]{(\ref{#1})}
\newcommand*{\mn}{{\mu\nu}}
\newcommand*{\nn}{\nonumber}
\newcommand*{\no}{\noindent}

\def\eq#1{(\ref{#1})}

\def\Tr{{\rm Tr}}
\def\s0#1#2{\mbox{\small{$ \frac{#1}{#2} $}}}
\def\0#1#2{\frac{#1}{#2}}

\begin{document}

\title{On the infrared behavior of Landau gauge Yang-Mills theory}

\author{Christian~S.~Fischer}
\affiliation{Institut f\"ur Kernphysik, 
  Technische Universit\"at Darmstadt,
  Schlossgartenstra{\ss}e 9,\\ 
  D-64289 Darmstadt, Germany}
\affiliation{GSI Helmholtzzentrum f\"ur Schwerionenforschung GmbH, 
  Planckstr. 1  D-64291 Darmstadt, Germany.}
\author{Axel Maas}
\affiliation{Institut f\"ur Physik, Karl-Franzens Universit\"at Graz, 
Universit\"atsplatz 5, A-8010 Graz, Austria.}
\author{Jan M.~Pawlowski}
\affiliation{Institut f\"ur Theoretische Physik, University of
  Heidelberg, Philosophenweg 16, D-62910 Heidelberg, Germany.}

\date{\today}
\begin{abstract}
We discuss the properties of ghost and gluon propagators 
in the deep infrared momentum region of Landau gauge Yang-Mills theory. 
Within the framework of Dyson-Schwinger equations and the functional 
renormalization group we demonstrate that it is only a matter of infrared 
boundary conditions whether infrared scaling or decoupling occurs. We argue 
that the second possibility is at odds with global BRST symmetry in the 
confining phase. For this purpose we improve upon existing truncation
schemes in particular with respect to transversality and renormalization.
\end{abstract}

\maketitle

\section{Introduction}

In the past decade much progress has been made in the understanding of
the infrared sector of QCD described in terms of Green's functions.
Various methods have been employed for this purpose, in particular
Dyson-Schwinger equations (DSEs) \cite{von
  Smekal:1997vx,Alkofer:2000wg,Watson:2001yv,
  Lerche:2002ep,Fischer:2002eq,Alkofer:2004it,Schleifenbaum:2004id,
  Fischer:2006ub,Schleifenbaum:2006bq,Fischer:2006vf,Alkofer:2008jy,%
  Kellermann:2008iw, Alkofer:2008dt, Cornwall:1981zr,
  Binosi:2007pi,Aguilar:2007nf,Aguilar:2008xm,
  Boucaud:2006if,Boucaud:2008ji,Boucaud:2008ky, Bloch:2002eq},
functional Renormalization group equations (FRGs)
\cite{Fischer:2006vf,Wetterich:1992yh,Litim:1998nf,Gies:2006wv,%
  Ellwanger:1994iz,Bonini:1994kp,Pawlowski:2005xe,%
  Litim:2000ci,Ellwanger:1995qf,Pawlowski:2003hq,%
  Pawlowski:2004ip,Kato:2004ry,Fischer:2004uk,Braun:2007bx}, stochastic
quantization methods \cite{Zwanziger:2003cf,gzwanziger}, effective
theory approaches
\cite{Dudal:2005na,Dudal:2007cw,Capri:2007ix,Dudal:2008sp}, and
lattice gauge theory \cite{Oliveira:2007dy,4d,4d2,sternbeck06,
  Cucchieri:2006xi,cucchieril7,Cucchieri:2007rg,Bogolubsky,
  limghost,gggvertex,Ilgenfritz:2006he,Cucchieri:2008qm}.  However, as
the full Green's functions are non-perturbative, their determination
is notoriously complicated, and each and every method faces individual
technical problems.

Already in pure Yang-Mills theory the determination of these  
Green's functions poses a considerable challenge. It 
is complicated by the fact that these are in general 
gauge-dependent quantities. However, this fact can also be 
used to simplify matters by an appropriate choice of gauge.
Most commonly Landau gauge has been used in this respect.

Over the past ten years substantial progress has been made 
using functional methods in 
this gauge. Based on the pioneering works  
ref.~\cite{von Smekal:1997vx,Lerche:2002ep,gzwanziger} powerful
tools have been developed to analyze the whole tower of 
Dyson-Schwinger and functional renormalization-group equations
in the deep infrared \cite{Alkofer:2004it,Fischer:2006vf}. 
As a result a self-consistent scenario has been identified which 
consists of a dressed ghost propagator more divergent in the 
infrared than its tree-level counterpart and an infrared 
suppressed gluon propagator with a finite or even vanishing 
dressing function at zero momentum. This solution agrees well and
in fact supports the Kugo-Ojima confinement
scenario developed in ref.~\cite{Kugo}. As we shall argue below 
this solution is the only one that has this property.

In lattice gauge theory substantial progress has been made in the past
years to overcome intrinsic problems related to discretization and
volume artefacts as well as gauge fixing ambiguities. Recent results
on very large volumes \cite{cucchieril7,Cucchieri:2007rg,limghost}
produce an infrared finite gluon propagator and, more important, an
infrared finite ghost in disagreement with the continuum results
discussed above. However, the situation on the lattice is still not
settled yet. These results are affected from discretization artefacts
\cite{lvs} and effects due to Gribov copies 
\cite{Bogolubsky,4d2,Bakeev:2003rr,gc}, to a
significant extent. We will discuss these issues below in section
\ref{sec:lattice}.

Nevertheless, the lattice results have inspired a number of recent works in
the continuum theory aiming at solutions of Dyson-Schwinger equations
with a finite ghost in the infrared
\cite{Aguilar:2008xm,Boucaud:2008ky,Dudal:2007cw}. In this work we
confirm that these solutions are present if the functional equations
are taken on their own correctly. However, as we will argue in section
\ref{sec:ggc}, these solutions cannot represent the infrared behavior
of Yang-Mills theory in the confined phase while at the same time
maintaining global color symmetry and BRST symmetry.

This work is organized as follows.  We will discuss the basic setting,
the non-perturbative regime of Yang-Mills theory, and one set of
methods to determine these Green's functions, functional equations
including boundary conditions, in section \ref{sec:funceq}. In
particular, in section \ref{sec:ggc} we will discuss the roles of
global symmetries and identify boundary conditions necessary to obtain
a BRST-symmetric, confining solution. The derivation of the functional
equations requires a discussion of certain formal aspects,
renormalization and gauge invariance done in section \ref{sec:stiren}.
In section \ref{sec:sec2} we then discuss the impact of the boundary
conditions on the infrared behavior of the ghost and gluon
propagators, the simplest Green's functions. As an example we
reconsider a truncation scheme for the DSEs which has been introduced
in \cite{Fischer:2002eq}. For the ghost DSE this truncation is
identical to those chosen by Boucaud et al.\ \cite{Boucaud:2008ky}
and Aguilar et al.\ \cite{Aguilar:2008xm}.  We also introduce a
modified scheme which maintains transversality in the gluon DSE and is
free of quadratic divergencies.  We solve the coupled system of ghost
and gluon DSEs with different choices of boundary conditions and
verify the appearance of two different types of solutions in the
infrared. We compare our approach with others and comment on various
technical issues.  In section \ref{sec:FRG} we repeat this exercise in
the framework of the functional renormalization group. We compare and
comment on recent lattice results in section \ref{sec:lattice}.  We
also discuss in section \ref{sec:funceq} and \ref{sec:lattice} the
role of different non-perturbative extensions of the Landau gauge. In
addition, we show in section \ref{sec:selec} also that neither type of
solution can be associated with a massive gluon characterized by a
gauge-independent pole mass.  Finally, in section \ref{sec:sec6} we
summarize and discuss our results further, the main result being an
understanding of how and why both types of solutions exist, and, in
the framework of local quantum field theory, which could be the
preferred one.

\section{Non-perturbative Yang-Mills theory}\label{sec:funceq}

\subsection{Gauge-fixing}

The starting point for the derivation of functional 
equations is the gauge-fixed path-integral. Here we shall 
use Landau gauge, 
\begin{equation} 
\pd_\mu A_\mu^a=0\,.\label{eq:landau}
\end{equation}
\no Within the standard Faddeev-Popov gauge fixing procedure 
this leads to the gauge fixed action in Euclidean space-time
\beq 
S=\int d^dx\left(F_\mn^a F_\mn^a+\bar c^a\pd_\mu D_\mu^{ab} c^b 
\right)\,,\label{eq:fixedaction} 
\eeq 
\no with the ghost fields $\bar c^a$ and $c^a$ and  
\begin{eqnarray}
  F^a_\mn&=&\pd_\mu A^a_\nu-\pd_\nu A_\mu^a+gf^{abc} A_\mu^b 
  A_\nu^c\nn\\
  D_\mu^{ab}&=&\delta^{ab}\pd_\mu+g f^{acb} A_\mu^c\,.\label{eq:defFD}
\end{eqnarray} 
\no This procedure is only well-defined 
within perturbation theory. In general there are several gauge 
equivalent confingurations satisfying the gauge condition (\ref{eq:landau}).
These are called Gribov copies \cite{Gribov,Singer:dk}. Indeed, the related 
path integral vanishes identically, as contributions from different 
gauge copies cancel out\cite{Neuberger:1986xz}. 

The latter problem can be resolved by restricting the  
configuration space to the first Gribov region 
\cite{Gribov}. This region is characterized by the condition that 
the Faddeev-Popov operator $M^{ab}=-\pd_\mu D_\mu^{ab}$ is 
positive semi-definite. This gauge prescription is called 
minimal Landau gauge. However, it is still not sufficient 
to single out a unique gauge copy \cite{vanBaal:1997gu}. 
Furthermore it breaks global BRST-invariance, since global BRST
transformations relate different Gribov copies \cite{Neuberger:1986xz}. 

A unique selection of gauge copies can be achieved by restricting the
configuration space to the fundamental modular region. This region
is characterized by the global minimum of the gauge fixing 
potential \cite{vanBaal:1997gu}, 
\be
{\cal F(A)} =c\int d^dx A_\mu^a A_\mu^a,\label{eq:abslancon}
\ee
\no with $c$ a positive constant. The corresponding gauge is also 
called the absolute Landau gauge \cite{gc}. Based on the work
\cite{Zwanziger:2003cf} one may hope that absolute Landau gauge 
is compatible with global BRST symmetry in the thermodynamic limit. 
This gauge is discussed further in section \ref{sec:lattice}.

\subsection{Global symmetries and confinement}\label{sec:ggc}

The appearance of Gribov copies is related to the local (i.e.
perturbative) nature of the standard Faddeev-Popov procedure without
the supplemental condition of minimising \pref{eq:abslancon}. Consequently, to
obtain a well-defined non-perturbative functional integral, it is
necessary to improve upon this procedure.

A promising approach in this respect is to upgrade the gauge fixing 
procedure to a globally defined (i.e.\ non-perturbative) BRST-symmetry 
that admits global BRST-charges. Then, the path integral is non-vanishing
and well-defined. This is also necessary for a successful implementation 
of BRST symmetry on the lattice \cite{Neuberger:1986xz}. Contrary to 
the perturbative setting such a procedure implies a weighting of 
Gribov copies such that the contributions from different gauge copies 
do not cancel out. Very recently within the framework of topological 
field theories a globally valid BRST quantisation has been put
forward in Ref.~\cite{vonSmekal:2007ns}.

Functional methods such as Dyson-Schwinger or functional
Renormalization Group equations have the advantage that 
such an explicit construction of a gauge fixing in terms of a
topological field theory can be avoided. 
The reason is that global constraints from
the gauge fixing procedure do not alter the form of the functional
equations but constrain the boundary conditions of Green's functions
at vanishing momenta in the infrared. The related infrared constraint
has been deduced by consistency arguments for a local covariant
quantum field theory, see \cite{Lerche:2002ep}. This constraint
is directly related to the global BRST-charge: the Kugo-Ojima confinement
scenario is based on a formulation with well-defined global
BRST-charges \cite{Kugo}. These enable the definition of the physical
part $\cH_{phys}$ of the state space of Yang-Mills theory in terms of
the cohomology of the BRST charge operator and is the basis of the
well-known BRST-quartet mechanism.  Given furthermore that one is
interested in a confining solution of infrared Yang-Mills theory a
well-defined global color charge is then necessary to ensure that
$\cH_{phys}$ contains only colorless states.  In Landau gauge this
enforces an infrared enhancement of the ghost propagator \cite{Kugo}; 
the ghost dressing function is then infrared divergent.

In functional methods this enhancement can be implemented as an
infrared renormalization condition. This condition leads to a unique
\cite{Fischer:2006vf} (scaling) solution of the whole tower of
functional equations for the one-particle irreducible Green's
functions of Yang-Mills theory. Hence it implicitly defines the unique
gauge fixing with well-defined global BRST-charges.  

In turn, given
confinement, an infrared solution with finite ghost at zero momentum
(termed 'decoupling' below) implies broken global gauge and BRST
symmetries. Indeed, all known BRST-quantizations that are compatible
with an infrared finite ghost even break off-shell BRST
\cite{Dudal:2007cw,Dudal:2008sp,Tissier:2008nw}.
The only possibility for the decoupling solution to coexist with a globally
well-defined BRST charge is in a Higgs phase, where the breaking of
global color symmetry implies the existence of super-selection
sectors. 

Hence, in the continuum we have a unique scenario with global
BRST-charges and an infrared enhanced ghost. These global constraints
supplementing Landau gauge are most easily implemented within
functional methods with consistent renormalization conditions fixed in
the infrared, as will be demonstrated in detail in section
\ref{sec:sec2}.

\subsection{Functional relations}\label{sfunrel}

We proceed by briefly discussing the functional relations for the
Green's functions we use in the present work following
\cite{Fischer:2006vf}.

A quantum field theory or statistical theory can be defined uniquely
in terms of its renormalized correlation functions. They are generated
by the effective action $\Gamma$, the generating functional of 1PI
Green's functions. For the functional RG equation we also consider the
effective action in the presence of an additional scale $k$, where the
propagation is modified via $k$-dependent terms
\begin{eqnarray}\label{dSk}
  \Delta S_k=
  \s012 \int \, A^\mu_a\, R_{\mu\nu}^{ab} \, A^\nu_b 
  +\int \, \bar C_a\, R^{ab}\,C_b\,, 
\end{eqnarray}  
where $R_{\mu\nu}^{ab}$ and $ R^{ab}$ are $k$-dependent regulator
functions.  Within the standard choice $k$ is an infrared cut-off
scale, and the functions $R$ cut-off the propagation for momenta
smaller than $k$.  The
regularized effective action $\Gamma_k$ is expanded in gluonic and
ghost vertex functions and reads schematically
\begin{eqnarray}\label{eq:Gk}
  \Gamma_k[\phi]=\sum_{m,n} \0{1}{m! n!^2 } 
  \Gamma_k^{(2n,m)}\, \bar C^n\, C^n\,A^m\,,
\end{eqnarray} 
in an expansion about vanishing fields $\phi=(A,C,\bar C)$. In
\eq{eq:Gk} an integration over momenta and a summation over indices is
understood.  The effective action $\Gamma_k$ satisfies functional
relations such as the quantum equations of motion, the Dyson-Schwinger
equations (DSEs); symmetry relations, the Ward or Slavnov-Taylor
identities (STIs); as well as functional RG or flow equations (FRGs).
All these different equations relate to each other, for a detailed
discussion see \cite{Pawlowski:2005xe}. Indeed, the Slavnov-Taylor
identities are projections of the quantum equations of motion,
whereas flow equations can be read as differential DSEs, or DSEs as
integrated flows. Written as a functional relation for the effective
action $\Gamma_k$ the DSE reads,
{\it e.g.\ }\cite{Pawlowski:2005xe}, 
\begin{eqnarray}
  \0{\delta \Gamma_k}{\delta \phi}[\phi]&=&
  \0{\delta S_{\rm cl}}{\delta \phi}[\phi_{\rm op}]\,,\label{eq:DSE} 
\end{eqnarray} 
where the operators $\phi_{\rm op}$ are defined as 
\begin{eqnarray} 
  \phi_{\rm op}(x) = \int d^4 y\, G_{\phi \phi_i}[\phi](x,y) 
  \0{\delta}{\delta\phi_i(y)}+\phi(x)\,,\nn
\end{eqnarray}
and 
\begin{eqnarray}\label{eq:G}
  G_{\phi_1\phi_2}[\phi]=\left(\0{1}{\Gamma_k^{(2)}[\phi]
      +R_k}\right)_{\phi_1\phi_2}\,
\end{eqnarray} 
is the full field dependent propagator for a propagation from $\phi_1$
to $\phi_2$. The functional derivatives in \eq{eq:DSE} act on
the corresponding fields and generate one loop and two loop diagrams
in full propagators. The functional DSE \eq{eq:DSE} relates 1PI
vertices, the expansion coefficients of $\Gamma_k$, to a set of one
loop and two loop diagrams with full propagators and full vertices,
but one classical vertex coming from the derivatives of $S_{\rm cl}$.
We emphasize that the DSE \eq{eq:DSE} only implicitly depends on the
regularization via the definition of the propagator in \eq{eq:G}. 

%%%%%%%%%%%%%%%%%%%%%%%%%%%%%%%%%%%%%%%%%%%%%%%%%%%%%%%%%%%%%%%%%%%%%%%%
\begin{figure}[h]
\centerline{\epsfig{file=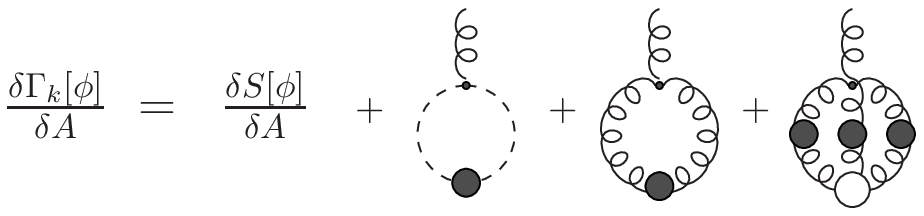,width=8.5cm}}
\ \vspace{.4cm}\ 
\centerline{\epsfig{file=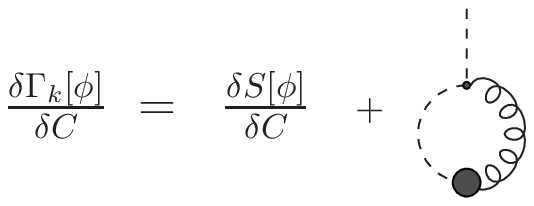,width=5cm}}
\caption{Functional DSE for the effective 
action. Filled circles denote fully dressed field dependent propagators 
\eq{eq:G}. Empty circles denote fully dressed field dependent vertices, 
dots denote field dependent bare vertices.}
\label{fig:funDSE}
\end{figure}
%%%%%%%%%%%%%%%%%%%%%%%%%%%%%%%%%%%%%%%%%%%%%%%%%%%%%%%%%%%%%%%%%%%%%%%%%

For Yang-Mills theory a diagrammatic representation of the structure
of the functional DSE \eq{eq:DSE} is shown in Fig.~\ref{fig:funDSE}.
The rhs is given in powers of the field-dependent fully dressed
propagator $G_{\phi \phi}[\phi]$, and its derivatives, as well as the
field dependent bare vertices. The momentum scaling of Green's
functions is directly related to the scaling of these building blocks.

\begin{figure}[h]
\centerline{\epsfig{file=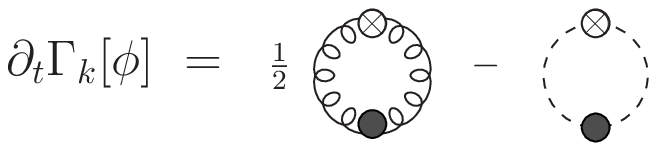,width=5.5cm}}
\caption{Functional flow for the effective 
action. Filled circles denote fully dressed field dependent propagators 
\eq{eq:G}. Crosses denote the regulator insertion $\partial_t R$. }
\label{fig:funflow}
\end{figure}

Wetterich's flow equation \cite{Wetterich:1992yh} for the effective action of
pure Yang-Mills theory reads, e.g. \cite{Ellwanger:1995qf,%
Pawlowski:2003hq,Kato:2004ry,Fischer:2004uk,Pawlowski:2005xe}, 
\begin{eqnarray}
  \partial_t \Gamma_k[\phi]&=&  
  \frac{1}{2} \int \0{d^4 p}{(2\pi)^4} \ G_{ab}^{\mu\nu}[\phi](p,p)
  \ {\partial_t} R_{\mu\nu}^{ba}(p)\nn\\ 
 & &  -
  \int \0{d^4 p}{(2\pi)^4} \ G_{ab}[\phi](p,p)
  \ {\partial_t} R^{ba}(p)\,,\label{eq:flow}
\end{eqnarray} 
where $t=\ln k$. The flow \eq{eq:flow} relates the cut-off scale
derivative of the effective action to one loop diagrams with fully
dressed field-dependent propagators. We can contrast
the diagrammatic representation of the DSE in Fig.~\ref{fig:funDSE}
with that of \eq{eq:flow}, given in Fig.~\ref{fig:funflow}.

 %%%%%%%%%%%%%%%%%%%%%%%%%%%%%%%%%%%%%%%%%%%
%%%%%%%%%%%%%%%%%%%%%%%%%%
%%%%%%%%%%%%%%%%%%%%%%%%%%%%%%%%%%%%%%%%%%%%%%%%%%%%%%%%%%%%%%%%%%%%%%%
 
Fig.~\ref{fig:funflow} shows the structure of the functional flow
\eq{eq:flow}. The rhs is given by the field-dependent fully dressed
propagator $G_{\phi \phi}[\phi]$ and the regulator insertion
$\partial_t R$. The standard use of
\eq{eq:flow} is to take a regulator function $R(p^2)$ which tends
towards a constant in the infrared and decays sufficiently fast in the
ultraviolet, and hence implements an infrared cut-off. 
\begin{figure}[h]
\centerline{\includegraphics[width=8cm]{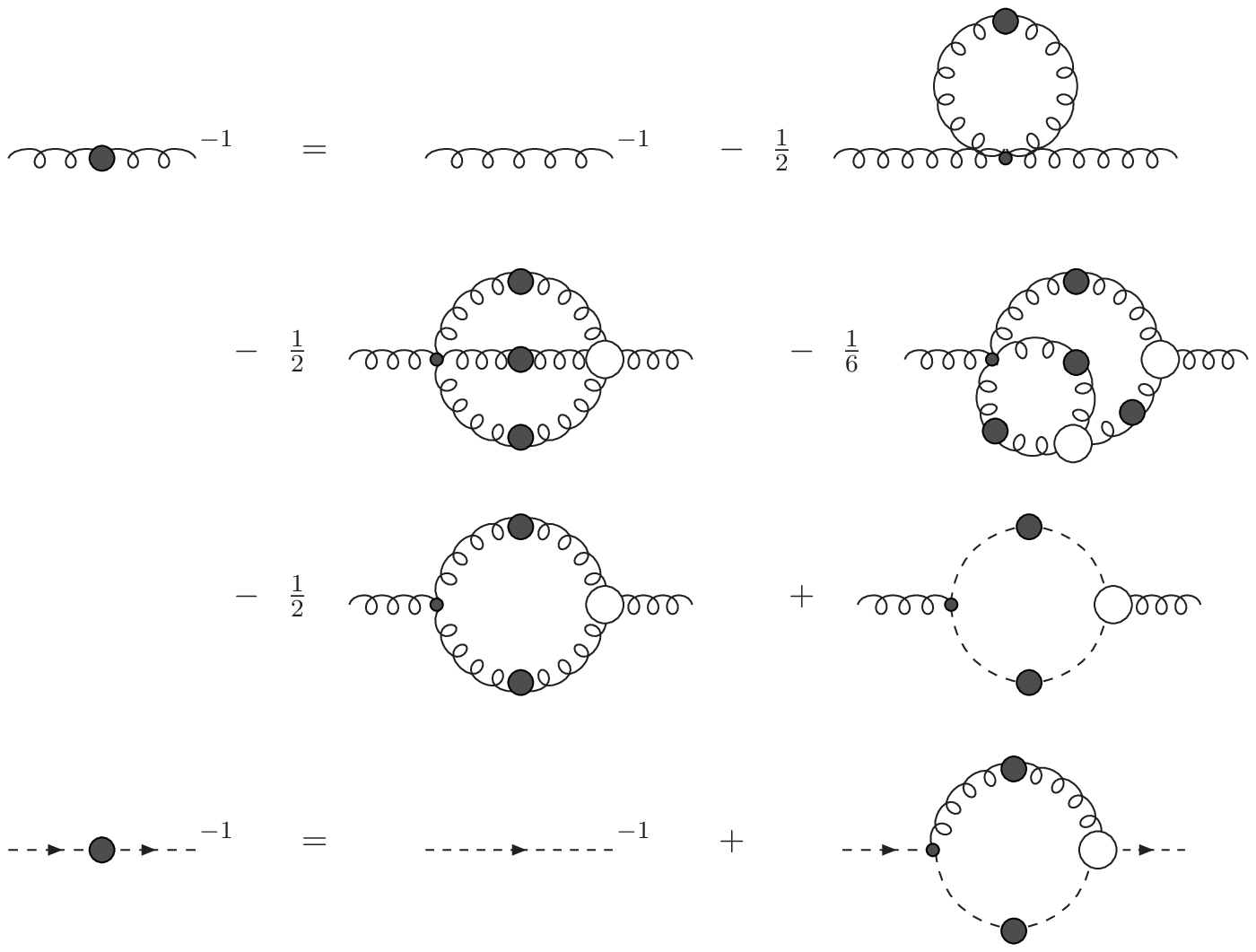}}
\caption{Dyson-Schwinger equations for the gluon and
  ghost propagator.  Filled circles denote dressed propagators and
  empty circles denote dressed vertex functions.}
\label{fig:DSE-gl}
\end{figure}

%%%%%%%%%%%%%%%%%%%%%%%%%%%%%%%%%%%%%%%%%%%%%%%%%%%%%%%%%%%%%%%%%%%%%%%%%%%%%%%%%%%%%%%%%%%%%%%%%%%%
%%%%% 
%%%%%%%%%%%%%%%%%%%%%%%%%%%%%%%%%%%%%%%%%%%%%%%%%%%%%%%%%%%%%%%%%%%%%%%%%%%%%%%%%%%%%%%%%%%%%%%%%%%%
\begin{figure}[h]
\centerline{\includegraphics[width=8cm]{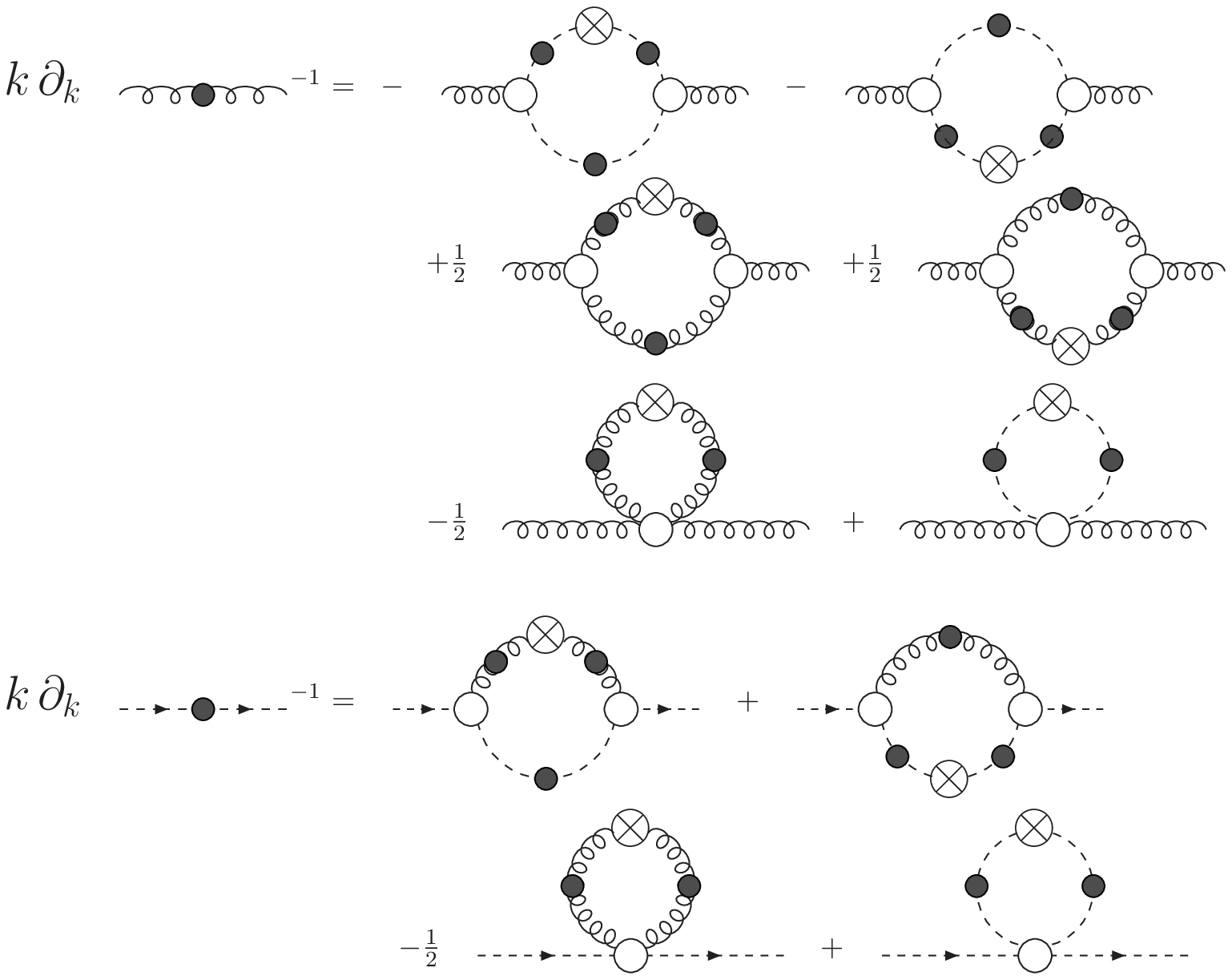}}
\caption{Functional RG equations for the gluon and
  ghost propagator.  Filled circles denote dressed propagators and
  empty circles denote dressed vertex functions. Crosses denote insertions
  of the regulators.}
\label{fig:FRG-gl}
\end{figure}

%%%%%%%%%%%%%%%%%%%%%%%%%%%%%%%%%%%%%%%%%%%%%%%%%%%%%%%%%%%%%%%%%%%%%%%%%%%%%%%%%%%%%%%%%%%%%%%%%%%%
%%%%% 

The diagrammatical expressions for the exact Dyson-Schwinger equations
(DSEs) for the ghost and gluon propagators is given in
Fig.~\ref{fig:DSE-gl}. The related FRG equations for the propagators
are displayed in Fig.~\ref{fig:FRG-gl}.  The vertices again satisfy
DSE/FRG equations which contain vertex functions with a larger number
of external legs and so on. Thus in practice, solving DSEs/FRGs for
general momenta requires a truncation scheme. Such truncation schemes
will be discussed starting with section \ref{sec:sec2}. Many
statements in the remainder of this section apply to the correct
solution, and may be violated by any given truncation. E.\ g., any
given truncation scheme will in general violate constraints, which are
imposed by the symmetries of the theory. Furthermore, any truncation
scheme will neglect part of the physics. The best which can be
achieved by truncation is that for a given topic these violations are
sub-leading effects. Depending on the question to be answered and the
truncation the irrelevance of violations may be established a-priori
or a-posteriori. This will be discussed further in the following
sections.

In Landau gauge, there is another property of the DSE/FRG equations
for general amputated Green's functions. The equations for 
fully transversal Green's functions $\Gamma^{(m,n)}_{\rm transversal}$ 
close on themselves, 
\begin{eqnarray}\label{eq:closedFun} 
\Gamma^{(m,n)}_{\rm transversal}
={\rm FRG/DSE}[\{\Gamma^{(m,n)}_{\rm transversal}\}]\,.
\end{eqnarray} 
This follows directly from the fact that all internal gluon legs are
transversal due to the transversality of the gluon propagator in
Landau gauge, and transversal projections on the external gluon legs
leaves us with purely transversal vertices on the rhs of the DSE/FRG
equations.  It is this subset of equations, \eq{eq:closedFun}, which
carries the full dynamics of the theory. 

The DSE/FRG equations for Green's functions with at least one 
longitudinal gluon leg, $\Gamma_{\rm longitudinal}^{(m,n)}$, depend both 
on the set of longitudinal as well as transversal Green's functions, 
\begin{equation} 
\Gamma_{\rm longitudinal}^{(m,n)}={\rm FRG/DSE}
[\{\Gamma_{\rm longitudinal}^{(m,n)}\}\,,\,
\{\Gamma_{\rm transversal}^{(m,n)}\}]\,, 
\label{eq:longFun}\end{equation} 
and hence is not closed. The $\Gamma_{\rm longitudinal}^{(m,n)}$ also
have to satisfy STIs, that in general imply non-trivial cancellations
between different diagrams in the DSE/FRG equations in
\eq{eq:longFun}. This will be be discussed in detail in section
\ref{sec:stiren}.

Here we discuss standard perturbation theory as a prime example for
these non-trivial cancellations: In leading order
perturbation theory, in which only the one-loop graphs on the
right-hand side in Fig.~\ref{fig:DSE-gl} or the integrated flow of
Fig.~\ref{fig:FRG-gl} contribute, none of the loops is transverse
individually, and only their sum is, up to higher order in the
coupling constant.

\subsection{Analytical solutions and confinement}

There are so far two limits in which an analytical determination of
the leading contribution of Green's functions can be performed without
referring to ad-hoc truncations.

One case is the far ultraviolet, where, due to asymptotic freedom,
it is possible to determine the leading behavior of Green's functions
with analytical means. This leading part can be determined with
great accuracy, and including higher order corrections of the leading
part. Asymptotic freedom also grants the luxury of proving that the
leading part of the Green's functions, as determined by perturbation
theory, is in fact uniquely the correct one.

The other kinematical limit is that of the far infrared, i.e.\ for
external momentum scales $p^2 \ll \Lambda^2_{\tt QCD}$. There, a
general power law behavior of the dressing functions of 
one-particle irreducible Green's
functions with $2n$ external ghost legs and $m$ external gluon legs
has been derived \cite{Alkofer:2004it,Huber:2007kc}:
\begin{equation}
  \Gamma^{(n,m)}(p^2) \sim (p^2)^{(n-m)\kappa + (1-n)(d/2-2)}\,.
  \label{IRsolution}
\end{equation}
Here, $d$ is the space-time dimension. Below we shall restrict
ourselves to the most important case $d=4$. Eq.~(\ref{IRsolution}) is
the unique self-consistent 'scaling' solution of the full, untruncated
tower of DSEs and FRGs \cite{Fischer:2006vf}. 

An important consequence of (\ref{IRsolution}) is the presence of a
nontrivial infrared fixed point in the running couplings related to
the primitively divergent vertex functions of Yang-Mills theory
\cite{Alkofer:2004it}:
\begin{eqnarray}
  \displaystyle \alpha^{gh-gl}(p^2) &=&
  \frac{g^2}{4\pi} \, { G^2(p^2)} \, { Z(p^2)}
  \sim \frac{const_{gh-gl}}{N_c} ,
  \nonumber \\
  \displaystyle \alpha^{3g}(p^2) &=&
  \frac{g^2}{4\pi} \, { [\Gamma^{0,3}(p^2)]^2} \, { Z^3(p^2)}
  \sim \frac{const_{3g}}{N_c} ,
  \nonumber \\
  \displaystyle \alpha^{4g}(p^2) &=&
  \frac{g^2}{4\pi} \, {  \Gamma^{0,4}(p^2)} \, { Z^2(p^2)}
  \sim \frac{const_{4g}}{N_c}\,, \label{coupling}
\end{eqnarray}
for $p^2 \rightarrow 0$. Here $G(p^2)$ is the dressing function of the
ghost propagator and $Z(p^2)$ the corresponding one for the gluon.
The appearance of these infrared fixed points is independent of the
precise value of $\kappa$, although the numerical values of the
pre-factors are influenced indirectly \cite{Alkofer:2004it}.

In terms of gluon and ghost propagators 
\begin{eqnarray}\nonumber 
 D_{\mu \nu}(p) &=& \left(\delta_{\mu \nu} -
    \frac{p_\mu p_\nu}{p^2}\right) D(p^2)\\\nonumber  
&=& \left(\delta_{\mu \nu} -
    \frac{p_\mu p_\nu}{p^2}\right)
  \frac{Z(p^2)}{p^2}\;\\[2ex]
  D_G(p) &=& -\frac{G(p^2)}{p^2}
  \label{props} \end{eqnarray}  
the solutions (\ref{IRsolution}) yield the power
laws \beq Z(p^2) \sim (p^2)^{-\kappa_A}; \hspace*{2cm} G(p^2) \sim
(p^2)^{-\kappa_C} \label{typeI} \eeq with
$\kappa=\kappa_C=-\kappa_A/2$ in four dimensions. For this solution
the anomalous dimension $\kappa$ is known to be positive 
\cite{Watson:2001yv,Lerche:2002ep}, $\kappa > 0$, one therefore finds
an infrared divergent ghost dressing function and an infrared 
vanishing gluon dressing
function. For $\kappa=1/2$ the gluon propagator (\ref{props}) is
finite at zero momentum, $0 < D(0) < \infty$, whereas for $\kappa >
1/2$ even the gluon propagator is vanishing in the infrared.

Note that the conservation of global color charge implies that 
$\kappa_C >0$ in the Kugo-Ojima confinement scenario \cite{Kugo}. 
This already enforces scaling \cite{Fischer:2006vf} with the
relation $\kappa_A=-2\kappa_C$. In fact the Gribov-Zwanziger scenario of
confinement \cite{Zwanziger:2003cf,gzwanziger,Gribov} even implies
$\kappa=\kappa_C=-\kappa_A/2 > 1/2$. In addition, a confinement
criterion for quarks was recently put forward that links
the infrared behavior of ghost and gluon propagators to the order
parameter of quark confinement, the Polyakov loop \cite{Braun:2007bx}.
Here quark confinement  results in the constraint $\kappa > 1/4$ for the
scaling solution \cite{Braun:2007bx}. Numerical and analytical
infrared solutions for propagators obtained in truncation schemes
\cite{von
  Smekal:1997vx,Lerche:2002ep,Fischer:2002eq,Pawlowski:2003hq,%
  Fischer:2004uk,pawlowski} satisfy the set of constraints by the
Kugo-Ojima and the Gribov-Zwanziger scenario as well as the
confinement criterion of \cite{Braun:2007bx}.
 
The absence of scaling implies the decoupling of (some) degrees of
freedom. A solution of this type has been discussed in
\cite{Aguilar:2007nf,Binosi:2007pi,Aguilar:2008xm,Boucaud:2006if,%
Boucaud:2008ji,Boucaud:2008ky,Dudal:2005na,Dudal:2007cw,Capri:2007ix,%
Dudal:2008sp} and is given by: 
\begin{equation} 
Z(p^2)\sim p^2; \hspace*{2cm}
G(p^2) \sim const.  \label{typeII} 
\end{equation} 
We refer to this type of solution as the 'decoupling solution'. It
does not satisfy the set of constraints by the Kugo-Ojima and the
Gribov-Zwanziger scenario, but satisfies the quark confinement
criterion of \cite{Braun:2007bx}: $3 \kappa_A-2 \kappa_C<-2$. Note
that the latter criterion is independent of the scenario and comes
from the demand of a vanishing order parameter for quark confinement.
We also emphasize that the difference between the scaling solution
(\ref{typeI}) and the decoupling solution (\ref{typeII}) is not the
possible appearance of an infrared finite gluon propagator,
$\kappa_A=-1$, but the presence or absence of the scaling relation
$\kappa_A=-2\kappa_C$.  We will come back to this point in detail in
section \ref{sec:sec2}, where we discuss numerical solutions of the
Dyson-Schwinger equations for the ghost and gluon propagators.

\section{Slavnov-Taylor identities and renormalization}\label{sec:stiren}

\subsection{Slavnov-Taylor identities}\label{ssti}

We now discuss the relevance of Slavnov-Taylor identities when it 
comes to identifying acceptable solutions from functional equations.
We present general arguments why in transverse gauges these 
identities cannot serve to generate constraints on the transversal 
Green's functions of the theory. This is then demonstrated explicitly
for the case of the infrared behavior of the ghost dressing function.

Slavnov-Taylor identities (STI) relate various Green's functions. STIs
are most conveniently represented in terms of the Zinn-Justin equation
for the effective action, see e.g.\
\cite{Ellwanger:1994iz,Pawlowski:2005xe},
\begin{equation} 
\int_x\,\left( \0{\delta \Gamma}{\delta
    K^a_\mu} \0{\delta \Gamma}{\delta A^a_\mu}+ \0{\delta
    \Gamma}{\delta L^a} \0{\delta \Gamma}{\delta c^a}
  +\0{\partial_\mu A_\mu^a}{\xi}\0{\delta \Gamma}{\delta \bar
    c^a} \right) = \mbox{cut-off\ terms}\,,
 \label{eq:ZJ} \end{equation} 
together with translation invariance of the anti-ghost,
\begin{eqnarray} \label{eq:antighost} 
\int_x \left(\0{\delta \Gamma}{\delta
    \bar c^a} - \partial_\mu  \0{\delta \Gamma}{\delta
    K^a_\mu}\right)=0\,. \label{eq:ZJ2}
\end{eqnarray}
Here $\xi$ denotes the parameter of linear covariant gauges, which has
to be set to zero for Landau gauge.  The cut-off terms on the right
hand side of eq.~(\ref{eq:ZJ}) depend on the chosen renormalization
scheme and are in particular present when a momentum cut-off is used,
as necessary in all systematic numerical treatments.  We come back to
this point in subsection \ref{stiandcut} below. Within the FRG
approach the rhs of \eq{eq:ZJ} is given in a closed form as a one loop
term similar to the flow equation \eq{eq:flow}, see
\cite{Litim:1998nf,Pawlowski:2005xe,Gies:2006wv,Ellwanger:1994iz,%
  Bonini:1994kp}. In general the identities (\ref{eq:ZJ}) and
(\ref{eq:ZJ2}) are solved in a given truncation to the effective
action $\Gamma$ that lead to truncated FRG/DSE-systems for propagators
and vertices.

In Landau gauge the constraints from the STIs 
decouple from the dynamics of the system, if no further assumptions
are made \cite{pawlowski,nedelkopawlowski}. The reason is that
the STIs relate the longitudinal parts, or more precisely the
BRST-projected parts, of vertex functions and propagators to loops of
the transversal gluon propagator, the ghost propagator and vertices, 
schematically written as 
\begin{eqnarray} 
\Gamma_{\rm longitudinal}^{(m,n)}={\rm STI}
[\{\Gamma_{\rm longitudinal}^{(m,n)}\}\,,\,
\{\Gamma_{\rm transversal}^{(m,n)}\}]\,, 
\label{eq:schemSTI}\end{eqnarray} 
to be compared with the FRG/DSE--equations for the longitudinal
Green's functions, \eq{eq:longFun}. In turn the FRG/DSE--equations for
the transversal propagator and vertices close on themselves,
\eq{eq:closedFun}, as already argued in section \ref{sfunrel}.
Formally, \eq{eq:longFun}, \eq{eq:schemSTI} can be resolved for
$\Gamma_{\rm longitudinal}^{(m,n)}[\Gamma_{\rm transversal}^{(m,n)}]$.
In non-Abelian gauge theories the above hierarchy does not close,
there is no self-contained subset of relations for a subset of
longitudinal vertex functions $\{\Gamma_{\rm longitudinal}^{(m,n)}\}$,
whereas in Abelian theories this is indeed possible
\cite{nedelkopawlowski}.

However, most truncations rely on an expansion in vertex functions and
hence $n,m\leq n_{\rm max},m_{\rm max}$. Then also the hierarchy of
non-Abelian STIs might be closed. In such a case the STIs
\eq{eq:schemSTI} can be resolved and provide integral constraints for
the transversal Green's functions $\Gamma_{\rm transversal}^{(m,n)}$.

We conclude that strictly speaking the STIs can only fix the
longitudinal parts of 1PI correlation functions.  If, however, the
following further regularity assumption would be implemented,
 \begin{eqnarray}\label{eq:regularity} 
 |\partial_{p_i} \Gamma^{(m,n)}| <\infty\,, \\\nonumber 
 \end{eqnarray} 
 transversal and longitudinal correlation functions are linked, and
 the STIs also directly constrain the transversal correlation
 functions, for a detailed discussion see \cite{nedelkopawlowski}.
 Such assumptions have been widely used in Abelian theories, see e.g.
 \cite{nedelkopawlowski,Fischer:2004nq,Ball:1980ax}. In non-Abelian
 theories results obtained from \eq{eq:regularity} have to be
 discussed cautiously as \eq{eq:regularity} certainly fails already at
 the level of the propagators.

Having said this, we proceed by exploring the consequences of
\eq{eq:ZJ}
for the correlation functions.
Instead of using the functional identity \eq{eq:ZJ} that allows
for a systematic discussion of the truncation scheme involved, the
STIs for correlation functions are sometimes also derived
by BRST variations of vacuum expectation values of product of field
operators. 

Similar to functional equations, the STIs relate Green's
functions with a different number of external legs. We  
would like to discuss this fact within a relevant 
example. The BRST variation of the expectation value of $A_\mu^a
A_\nu^b \bar{c}^c$ will produce a Slavnov-Taylor identity involving
the ghost and gluon propagators, the three-gluon vertex and the
ghost-gluon vertex, and the irreducible ghost-gluon scattering kernel.
A simple way to obtain this identity is by using BRST invariance on
the expectation value $<A_\mu^a A_\nu^b\bar{c}^c>$, yielding
\begin{widetext}
\begin{eqnarray} 
0&=&-<A_\nu^b\bar{c}^c\pd_\mu c^a>-<A_\mu^a\bar{c}^c\pd_\nu
c^b>+\frac{1}{\xi}<A_\mu^a A_\nu^b \pd_\rho A_\rho^c>
-g(f^{ade}<A_\mu^e A_\nu^b c^d \bar{c}^c>+f^{bde}<A_\mu^a A_\nu^e c^d
\bar{c}^c>).\label{eq:stig3v} \end{eqnarray} \no 
\end{widetext}
The last term can be dropped in lowest order perturbation theory due
to the explicit factor $g$, but this is not possible in
non-perturbative calculations.

Nonetheless, the standard truncation of the effective action used in
the literature within numerical investigations of FRG- and DSE systems
takes into account full momentum-dependent propagators, and
RG-improved ghost-gluon, three-gluon and four-gluon vertices. All
higher vertices, in particular the ghost-gluon scattering kernel, are
usually left out.

As the DSEs for the propagators depend on the bare and full
ghost-gluon and three- and four gluon vertices, but do not depend
explicitly on the ghost-gluon scattering kernel $\langle A^2 c\bar
c\rangle$, this is apparently a viable truncation. Moreover, the
latter vertex has been also left out in most applications to vertex
DSEs, which labels such a truncation as a minimally self-consistent
DSE-approach in a vertex truncation.

However, the scattering kernel does appear in diagrams in the FRG
equations for both the gluon and ghost propagators. This comes about
as both hierarchies of equations represent different resummation
schemes and using the same vertex truncation at the level of the
effective action implements different truncation schemes. Note that it is
precisely this structure that allowed for the uniqueness proof of the
IR-asymptotics from both sets of equations \cite{Fischer:2006vf}, and
is also the key ingredient for the important consistency checks for
the results of correlation functions.

Let us nonetheless proceed and neglect the four-point function
appearing in \eq{eq:stig3v}.  This has been done previously, either
completely \cite{von
  Smekal:1997vx,Boucaud:2008ji,Ball:1980ax,Boucaud:2008ky} or only its
connected part \cite{BarGadda:1979cz}.

Following \cite{Ball:1980ax},
as has been done in general \cite{von
  Smekal:1997vx,Boucaud:2008ji,Boucaud:2008ky}, the STI takes the form
\begin{widetext}
  \be q_\nu
  \Gamma_{\mu\nu\rho}^{(0,3)}(p,q,k)=\frac{G(q^2)}{Z(p^2)}(\delta_{\mu\nu}
  p^2-p_\mu p_\nu)\Gamma_{\nu\rho}(p,q;k)-\frac{G(q^2)}{Z(k^2)}(
  \delta_{\nu\rho}k^2-k_\nu k_\rho)\Gamma_{\mu\nu}(k,q;p)\label{esti}.
  \ee
\end{widetext}
Herein $\Gamma_\mn$ is implicitly defined by the ghost-gluon vertex
\be \Gamma_\mu^{(2,1)}(p,q;k)=p_\nu\Gamma_{\nu\mu}(p,q;k)\nn \ee \no
where $k$ is the gluon momentum. It is the ghost-gluon scattering
kernel, but with two external legs contracted, and thus not the same
quantity as appearing in \eq{eq:stig3v}, which is not contracted and
has thus one momentum argument more. By virtue of Lorentz symmetry it
can be expanded as \cite{Ball:1980ax} \be
\Gamma_{\nu\mu}(k,q;p)=\delta_\mn a-q_\mu p_\nu b+p_\mu k_\nu c+p_\nu
k_\mu d+k_\mu k_\nu e\nn, \label{a-e} \ee \no with five unknown
functions $a$,...,$e$.

When contracting the left-hand side of \pref{esti} with $p_\mu
k_\rho$, it vanishes \cite{Ball:1980ax}. From this the relation
\begin{widetext} 
  \be \frac{G(q^2)}{Z(k^2)}a(p,q,k)-\frac{G(p^2)}{Z(k^2)}a(q,p,k)
  -pq\left(\frac{G(q^2)}{Z(k^2)}b(p,q,k)-\frac{G(p^2)}{Z(k^2)}
    b(q,p,k)\right) +pk\frac{G(q^2)}{Z(k^2)}d(p,q,k)-qk
  \frac{G(p^2)}{Z(k^2)}d(q,p,k)=0\label{sres}, \ee \end{widetext} and
cyclic permutations thereof, have been obtained \cite{Ball:1980ax}.
However, this condition is actually weaker than possible: By the
assumption of a tree-level color structure, the three-gluon vertex has
to be totally antisymmetric in any pair of momentum and
Lorentz-indices. Hence, it is sufficient to contract \pref{esti} just
with either $p_\mu$ or $k_\nu$ to make the left-hand side vanish.
Since on the right-hand side always one term is transverse w.r.t.\ the
momentum which has been used to contract it, it follows that, e.\ g.\
when contracting with $k_\rho$ that \be
0=\frac{G(q^2)}{Z(p^2)}(\delta_\mn p^2-p_\mu
p_\nu)\Gamma_{\nu\rho}(p,q;k)k_\rho\nn.  \ee \no This leaves the
condition (after relabeling) \be a(p,q,k)-pq \, b(p,q,k)+pk \,
d(p,q,k)=0,\label{cyc} \ee \no and cyclic permutations. This is a
condition implying certain cancellations between the dressing
functions of the ghost-gluon vertex, independent of the values of the
dressing functions of the propagators. In particular, the structure is
very reminiscent of the one which will be obtained in section
\ref{sec:sec2} for the longitudinal part of the ghost-gluon vertex to
ensure transversality of the gluon propagator in the far infrared.

Furthermore the equality \pref{cyc} implies the relation \pref{sres},
independent of the value of the propagator dressing functions. The
dressing functions $a$,...,$e$ in eq.~(\ref{a-e}) are an (arbitrary)
splitting of the two independent dressing functions of the ghost-gluon
vertex \cite{Ball:1980ax}, and therefore can exhibit in principle any
type of behavior in the infrared satisfying \pref{cyc}. However, as
the non-longitudinal component of the ghost-gluon vertex is infrared
finite, while the longitudinal one seems to be vanishing
\cite{Schleifenbaum:2004id}, the behavior of $a$,...,$e$ is not
necessarily regular (as assumed in \cite{Boucaud:2008ky}).

We conclude that these results do not provide any constraints on the
ghost dressing function. Furthermore, the condition (\ref{cyc})
has been derived from the approximation (\ref{esti}) to the 
complete STI \eq{eq:stig3v} and hence cannot be used to exclude
certain types of solutions for the Green's functions as has been
attempted in \cite{Boucaud:2008ky}.

\subsection{Renormalization \label{stiandcut}}

We now come back to the issue of renormalization-dependent terms
in the STIs.
The functional equations displayed in Figs.~\ref{fig:DSE-gl} and 
\ref{fig:FRG-gl} and the corresponding STIs apply to the renormalized
propagators and hence are equations for finite Green's functions.
Whereas this finiteness trivially carries over to general truncation
schemes for the functional RG equation due to the regulator insertion,
it is not evident in general truncations of the DSE. The reason for
the latter intricacy originates in the fact that the finiteness of the
propagator DSEs is achieved by cancellations of divergencies
stemming from different diagrams in Fig.~\ref{fig:DSE-gl}. In gauge
theories these cancellations are partially governed by Slavnov-Taylor
identities.  We conclude that finiteness of the DSEs requires a
careful discussion of the renormalization procedure as well as the
Slavnov-Taylor identities in the presence of a given regularization.

For any numerical implementation such a regularization involves a
momentum cut-off. It is well-known that a momentum cut-off scheme might
necessitate additive renormalization, in particular in gauge theories
where a momentum cut-off deforms the Slavnov-Taylor identities.
Indeed, within the flow equation approach the momentum cut-off is
explicitly introduced at the level of the functional integral and the
corresponding modified Slavnov-Taylor identities are explicitly known
\cite{Pawlowski:2005xe}. Furthermore it has been shown how to deduce a
self-consistent renormalization scheme for DSEs with additive
counter terms in the presence of a momentum cut-off from the
corresponding FRG equations \cite{Pawlowski:2005xe}. In particular the
latter entail a gluonic mass term that tends towards zero if the
momentum cut-off is removed.  For the DSEs this translates into
quadratic divergencies that have to be canceled identically
by mass-type counter-terms \cite{Fischer:2005en}. This
leads to a well-defined unique treatment of truncational divergencies
in the DSEs that renders the equation finite within a given general
truncation. In particular these considerations enable to interpret the
longitudinal part of the gluon propagator DSE: It has to be fully
canceled by the related longitudinal counter-term in order to restore
transversality of the fully renormalized propagator. In turn this
uniquely fixes the transversal part. Previously, this procedure has 
implicitly been applied within numerical applications to
DSEs \cite{Fischer:2002eq,Fischer:2003zc,Fischer:2005en,Maas:2005hs}.

Having said this we emphasize that multiplicative renormalization
should still be possible for the DSEs within specific schemes.
Indeed such a scheme is much wanted for as the modified Slavnov-Taylor
identities for the DSEs in the presence of non-transversal counter
terms are only implicitly known via the relations to the according flow
equations. In a multiplicative scheme the standard Slavnov-Taylor
identities can be used which completely decouple the longitudinal parts
of the Green's functions from the DSEs of the transversal Green
functions and render the latter system fully self-contained.
Consequently such a choice optimizes the physics content of a given
truncation. Moreover, for a multiplicative scheme it is possible to
renormalize the complete system with five renormalization constants in
Landau gauge, the two wave-function renormalization constants of the
ghost and the gluon, $\widetilde{Z}_3$ and $Z_3$, as well as the three
primitively divergent vertices, the ghost-gluon vertex with
$\widetilde{Z}_1$, the three-gluon vertex with $Z_1$, and the
four-gluon vertex with $Z_4$.  The latter three can all be related to
a single renormalization constant, $Z_g$ of the coupling $g_0$.

\section{Ghosts and gluons from Dyson-Schwinger equations \label{sec:sec2}}

In this section we report on explicit solutions for the Dyson-Schwinger 
equations of the ghost and gluon propagators. These equations
involve three fully 
dressed vertices: the ghost-gluon, three-gluon and four-gluon vertices. 
For the purpose of the present work we shall restrict ourselves to the 
well-established method of constructing ansaetze for the vertices taking
into account as much information from the vertex DSEs as possible. We
would like to emphasize, however, that by now we have accumulated enough   
information on the details of these vertex equations 
\cite{Alkofer:2004it,Schleifenbaum:2004id,Alkofer:2008jy,Kellermann:2008iw,
Alkofer:2008dt,sternbeck06,Cucchieri:2008qm,Alkofer:2008tt} 
such that a simultaneous treatment of the DSEs for propagators and 
vertices is within reach in the near future.

Here we investigate the possibility of two different types of
solutions for the ghost and gluon propagators in the infrared. In the
course of this study we need to carefully address technical issues
like boundary conditions on the DSEs, transversality of the gluon
propagator in Landau gauge and the appearance of quadratic
divergencies in the gluon DSE once a hard momentum cutoff is
introduced for the numerical treatment of the equations. To illustrate
our points we use two different truncation schemes for the DSEs. The
first one has been developed in
Refs.~\cite{Fischer:2002eq,Fischer:2003zc} and produced results in
quantitative agreement with corresponding lattice calculations in the
ultraviolet and low-momentum regions. Deviations in the mid-momentum
regime around one GeV are of the order of 20 percent and well
understood from the nature of the truncation scheme. Qualitative
differences in the deep infrared are at the heart of our study here
and will be addressed below. The second scheme will be constructed in
the next subsections. It serves to study the influence of technical
issues related to (the breaking of) gauge invariance on the possible
types of solutions of the DSEs. In turn we will address a connection
between renormalization and boundary conditions in the ghost DSE and
the issues of quadratic divergencies and transversality in the gluon
DSE.

\subsection{A boundary condition in the ghost DSE}

The ghost DSE is given by
\begin{widetext}
\begin{eqnarray}\label{eq:ghostDSE}
  \frac{1}{G(p^2;\mu^2)} = \widetilde{Z}_3 -\widetilde{Z}_1 
  \frac{g^2(\mu^2) N_c}{(2\pi)^4}
  \int d^4q \: \Gamma^{(2,1)(0)}_\mu (p,q;\mu) 
  \left(\delta_{\mu \nu} - \frac{k_\mu k_\nu}{k^2}\right) 
  \frac{Z(k^2;\mu^2) G(q^2;\mu^2)}{k^2 \,q^2} 
  \:\Gamma^{(2,1)}_\nu(q,p;\mu) 
\end{eqnarray}
\end{widetext}
with the bare ghost-gluon vertex $\Gamma^{(2,1)(0)}_\mu$, its dressed
counterpart $\Gamma^{(2,1)}_\mu$ and the momentum routing $k=p+q$. The
ghost renormalization function $\widetilde{Z}_3$ and the ghost-gluon
vertex renormalization function $\widetilde{Z}_1$ both depend on the
renormalization point $\mu^2$.

The equation as it stands is multiplicatively renormalizable as
recalled explicitly in appendix \ref{MR}. It then follows that the
choice of $g^2(\mu^2)$, which implicitly fixes $\mu^2$, cannot have
any impact on the type of solutions one finds in the DSEs.  All
multiplicatively renormalizable solutions must exist between
infinitesimally small $g^2$ (asymptotic freedom; large $\mu^2$) and a
potential maximum value that can be derived from the infrared behavior
of the couplings \pref{coupling}. To discuss this further we rewrite
the ghost DSE schematically as
\begin{equation}
\frac{1}{G(p^2;\mu^2)} = \widetilde{Z}_3 - \widetilde{Z}_1 g^2(\mu^2)
I(p;\mu) \label{unsub}
\end{equation}
and subtract the equation at $p^2=0$
\begin{equation}
\frac{1}{G(p^2;\mu^2)} = \frac{1}{G(0;\mu^2)} - \widetilde{Z}_1
g^2(\mu^2) (I(p;\mu)-I(0;\mu))\,. \label{subsub}
\end{equation} 
This way we got rid of the renormalization factor $\widetilde{Z}_3$ in
exchange for the boundary condition $G(0;\mu^2)$. One can now
explicitly choose between solutions of the scaling type (\ref{typeI})
or the decoupling type (\ref{typeII}) by varying $G(0;\mu^2)$
\cite{Lerche:2002ep,Boucaud:2008ky}.  Clearly, $G(0;\mu^2)^{-1}=0$
corresponds to an infrared diverging ghost dressing function. This
choice corresponds to the 'horizon condition' derived by Zwanziger to
account for the presence of the Gribov-horizon when evaluating
correlation functions \cite{Zwanziger:2003cf,gzwanziger,gzwanziger2}.
The other option, $G(0;\mu^2)= const.$, produces an infrared finite
ghost dressing function by construction. If the un-subtracted equation
(\ref{unsub}) is used these conditions translate into equivalent ones
for the renormalization constant $\widetilde{Z}_3$. From this it is
clear that it is instructive but not necessary to subtract the ghost
equation exactly at zero momentum.  One can find both solutions for an
arbitrary subtraction point $s$, if the boundary conditions on
$G(s;\mu^2)$ are selected appropriately. This is a consequence of
renormalization group invariance. For the scaling solution the
equivalence of boundary conditions for $G(0;\mu^2)$ and $G(s;\mu^2)$
has been explicitly demonstrated in \cite{Cucchieri:2007ta}.

In \cite{Boucaud:2008ky} the two types of solutions have been shown
to coexist in a numerical treatment of the ghost DSE alone. Below we
demonstrate that self-consistent solutions of both types exist when the 
ghost and gluon Dyson-Schwinger equations are solved simultaneously. 

\subsection{The gluon DSE: quadratic divergencies}

Having discussed the ghost DSE on general grounds we now focus on the
gluon DSE. To this end we need to specify an explicit truncation scheme
for the fully dressed vertices involved. As stated above, one such scheme
has been constructed in refs.~\cite{Fischer:2002eq,Fischer:2003zc}. We
first shortly summarize this truncation before we qualitatively improve 
upon it.

In general, the dressed ghost-gluon vertex $\Gamma^{(2,1)}_\mu$ can be
written as
\begin{equation}
  \Gamma^{(2,1)}_\mu (p,q;k) = i q_\mu A(q,k) + 
  i k_\mu B(q,k) \label{ghost-gluon}
\end{equation}
where $p$ and $q$ are the incoming and outgoing ghost momenta
respectively and $k$ is the gluon momentum. $A(q,k)$ and $B(q,k)$
denote the two dressing functions of the vertex with $A(q,k)
\rightarrow 1$ and $B(q,k) \rightarrow 0$ return the tree-level
vertex.  The dressing functions $A(q,k)$ and $B(q,k)$ have been
subject to investigations in the continuum
\cite{Lerche:2002ep,Schleifenbaum:2004id} and $A(q,k)$ also on the
lattice \cite{Ilgenfritz:2006he,Cucchieri:2008qm} and found to only
mildly deviate from the tree-level behaviors.  Furthermore it has been
shown in \cite{Lerche:2002ep} that the infrared behavior of the
ghost-gluon system is only mildly affected by possible vertex
dressings. This justifies the truncation
\begin{equation}
\Gamma^{(2,1)}_\mu (p,q;k) = i q_\mu,
\end{equation}
i.\ e., the replacement of the fully dressed ghost-gluon vertex in
the DSEs by its tree-level counterpart. This truncation has been used
in \cite{Fischer:2002eq,Fischer:2003zc} and \cite{Boucaud:2008ky}. 

The situation is somewhat more complicated for the three-gluon vertex.
Here an ansatz $\Gamma^{(0,3)}_{\mu \nu \rho}(p,q) =
\Gamma^{(0,3)(0)}_{\mu \nu \rho} \Gamma^{3g}(p,q)$ has been chosen in
\cite{Fischer:2002eq} with the tree-level vertex
$\Gamma^{(0,3)(0)}_{\mu \nu \rho}$ and dressing function
$\Gamma^{3g}(p,q)$ that leads to solutions for the ghost and gluon DSE
with the correct one-loop running from resummed perturbation theory.
The ansatz is given by eq.~\ref{3g-vertex} in the appendix.
Furthermore it has been shown in \cite{Fischer:2003zc} that this
choice of the vertex also leads to solutions which respect
multiplicative renormalizability of the gluon DSE.  Contributions
involving the four-gluon interaction have been neglected in
\cite{Fischer:2002eq}. For the scaling solution this affects only
the intermediate momentum regime, as shown there. For the decoupling 
solution the omission of the four-gluon vertex may also have an impact 
at low momenta. This needs to be explored in future work. 

In general a truncation scheme such as the one summarized here faces
the problem of quadratic divergencies appearing in the gluon DSE.
These have been dealt with in the past by explicit subtraction
procedures on the level of the integrands
\cite{Fischer:2002eq,Maas:2005hs,Cucchieri:2007ta}, numerical
subtractions on the level of the right hand side of the gluon DSE
\cite{Fischer:2005en} or subtractions of purely quadratically
divergent term involving only the perturbative form of the ghost and
gluon dressing functions \cite{Aguilar:2008xm}. Although conceptually
different all these procedures lead to similar results if correctly
implemented.

Here we introduce yet another procedure which takes into account the
general considerations of subsection \ref{stiandcut}. Recall that the
introduction of a hard momentum cut-off into the theory modifies the
Slavnov-Taylor identities. This modification has its biggest effects
on momentum scales near the cut-off. An improved truncation scheme
that avoids the related problem of quadratic divergencies therefore
should work with vertex modifications at these scales. Such a scheme
is constructed in the following. We devise ansaetze for the
ghost-gluon and the three-gluon vertices that resolve the issue of
quadratic divergencies.  The resulting ghost and gluon DSEs are then
multiplicatively renormalizable in the first place without the need
for additional subtraction procedures.

The procedure turns out to be straightforward. For the 
ghost-gluon vertex we start with the general ansatz 
\begin{widetext}
\begin{eqnarray}
\Gamma^{(2,1)}_\mu(p,q,k) &=& i q_\mu \left( 1 + A + B \frac{p^2}{k^2} + C
  \frac{p^2}{q^2} + D \frac{q^2}{k^2}+ E \frac{q^2}{p^2}  + F
  \frac{k^2}{p^2} + G \frac{k^2}{q^2} + H \frac{(p.q)^2}{p^2 q^2} + I
  \frac{(p.k)^2}{p^2 k^2} + J \frac{(q.k)^2}{q^2 k^2} \right)\,,\nn\\
\label{new-ghost-vertex}
\end{eqnarray} 
\end{widetext}
which includes all possible modifications of the vertex in terms of
dimensionless combinations up to powers of two in momenta and angles.
Here $p$ is the incoming ghost momentum, $q$ is the one of the
outgoing ghost and $k=p-q$ is the momentum attached to the gluon leg.
The coefficients $A..J$ are assumed to be constant in momentum.  One
can then perform an ultraviolet analysis of the gluon DSE and the
ghost DSE along the lines of ref.~\cite{Fischer:2002eq,Fischer:2003zc}
and restrict the possible values of the coefficients by demanding the
following conditions:
\begin{itemize}
\item No quartic and quadratic divergencies should appear in the ghost
  loop of the gluon DSE and the ghost-gluon loop in the ghost DSE.
\item The logarithmic divergencies of these loops should remain the
  same as for the bare ghost-gluon vertex, i.e. as in perturbation
  theory.
\end{itemize}
This restricts the coefficients of (\ref{new-ghost-vertex}) to a
family of solutions which depend on the explicit values of two of the
constants $A..J$. Setting as many of these constants to zero as
possible one arrives at the surprisingly simple form
\begin{equation}
  \bar{\Gamma}_\mu^{(2,1),UV}(p,q,k) = i q_\mu \left( 1 
    -\frac{q^2}{p^2} \right)\,.
\label{new-ghost-vertex2} 
\end{equation} 
A similar procedure for the three-gluon vertex produces the result 
\begin{widetext}
\begin{equation} 
  \bar{\Gamma}_{\rho \nu \sigma}^{(0,3),UV}(q,p) =
  \Gamma^{(0,3)(0)}_{\rho \nu \sigma}(q,p) \,\Gamma^{3g}(p,q)\, 
  \left(1-\frac{140}{51} - \frac{52}{17}
    \frac{p^2}{k^2} + \frac{89}{51} \frac{p^2}{q^2} + \frac{52}{17}
    \frac{q^2}{k^2} - \frac{26}{17} \frac{k^2}{q^2} + \frac{104}{17}
    \frac{(q.k)^2}{q^2 k^2}\right)\,,
\label{new-gluon-vertex}
\end{equation}
\end{widetext}
where $\Gamma^{(0,3)(0)}_{\rho \nu \sigma}(q,p)$ is the bare
three-gluon vertex and $\Gamma^{3g}(p,q)$, given in appendix
\ref{kernels}, has been introduced in
\cite{Fischer:2002eq,Fischer:2003zc} to reproduce the correct
logarithmic running of the gluon in the ultraviolet. The additional,
power-like vertex dressings in (\ref{new-ghost-vertex2}) and
(\ref{new-gluon-vertex}) serve to avoid quadratic divergencies in the
ultraviolet momentum regime.  As discussed above, the power-like
modifications should affect the vertices only for momenta close to the
ultraviolet regulator.  We therefore multiply
(\ref{new-ghost-vertex2}) and (\ref{new-gluon-vertex}) with
appropriate damping factors that cancel these terms smoothly with
decreasing momenta. The explicit form of the damping factors will be
given in the next subsection.

\subsection{Transversality}

Next we come back to the issue of transversality, continuing the
discussion from section \ref{ssti}. In Landau gauge, the gluon
propagator is transverse and this property is reflected by the
structure of the full, untruncated gluon DSE. For symmetry conserving
regularization schemes this property can be maintained in truncation
schemes that satisfy the Slavnov-Taylor identities.  However, even
truncation schemes satisfying a finite number of STIs are extremely
hard to implement in numerical calculations. There one usually has to
resort to a momentum cut-off. As a consequence spurious longitudinal
terms are generated in the gluon DSE that need to be projected out by
contracting the gluon DSE with a transverse projector. This problem
has been overlooked in \cite{Aguilar:2008xm}. There a truncation has
been used that is formally transverse under dimensional regularisation
while it is not when a hard cutoff is used in the numerical calculations.
Consequently, in contradiction to their claims, the numerical solutions 
presented in \cite{Aguilar:2008xm} are not strictly transversal.

In general, projection onto the transverse part of the right hand side
of the gluon DSE, as done in \cite{Fischer:2002eq} guarantees reliable
results. Nevertheless it is interesting to construct a truncation
scheme that in addition minimizes the spurious longitudinal terms.
This is the purpose of this subsection.

As is apparent from the analysis of ref.~\cite{Fischer:2002eq}
transversality is naturally maintained in the ultraviolet where
perturbation theory is at work. In the infrared, however, the
situation is different and sizeable longitudinal components of the
gluon are present.  To improve this situation we consider the most
general form for the dressed ghost-gluon vertex,
eq.~(\ref{ghost-gluon}), and contract the ghost-loop in the gluon DSE
with a longitudinal projector, thereby projecting onto the spurious
terms.  Apart from trivial factors one obtains the combination
\begin{eqnarray} 
q_\mu \frac{p_\mu p_\nu}{p^2} \Gamma^{(2,1)}_\nu
(p,q;k) &=& q_\mu \frac{p_\mu p_\nu}{p^2} \left( k_\nu A(k,p) + p_\nu
  B(k,p) \right) \nonumber\\
  &\stackrel{!}{=}& 0 
\end{eqnarray} 
where $p$ is the gluon momentum and $q$ and $k=q-p$ are the two ghost
momenta.  This is reminiscent of equation \pref{cyc}. From this
relation we find the condition $B(k,p) = -A(k,p) \frac{k.p}{p^2}$
which eliminates the spurious terms and makes the ghost-loop
transverse. As will be shown later this construction is 
sufficient to render the gluon DSE transversal on the level
of numerical accuracy. Note, however, that a more general 
construction for $B$ is possible such that exact transversality 
of the gluon DSE for all momenta is guaranteed.  

As a result the dressed ghost-gluon vertex in the
infrared can be written as
\begin{equation}
  \bar{\Gamma}^{(2,1),IR}_\mu (p,q;k) = i q_\mu A(q,k) - i
  k_\mu \frac{k.q}{k^2}
  A(q,k) \label{ghost-gluon2} 
\end{equation} 
in the momentum labeling of
eq.~(\ref{ghost-gluon}) where $k$ is the gluon momentum.
It is important to note that the additional term in
eq.~(\ref{ghost-gluon2}) is longitudinal in the gluon momentum, i.e.\
it does not contribute to the ghost-loop in Landau gauge but merely
serves to eliminate non-transverse artefacts. Thus the infrared
analysis of the ghost loop performed in \cite{Lerche:2002ep} and the
value of $\kappa$ remain unchanged.

Longitudinal contributions to the ghost-gluon vertex have been
calculated from the vertex-DSE in \cite{Schleifenbaum:2004id} and
found to vanish for large momenta, in agreement with perturbation
theory.  We therefore need to multiply the longitudinal structure in
(\ref{ghost-gluon2}) by appropriate damping factors to account for
this behavior. Furthermore we choose $A(q,k)=1$ as in the truncation
scheme of the previous section.

Taking into account the ultraviolet modifications
(\ref{new-ghost-vertex2}) and (\ref{new-gluon-vertex}) of the previous
subsection we then find our final ansaetze for the ghost-gluon and
three-gluon vertices
\begin{widetext} 
\begin{eqnarray} 
\hspace{-1.5cm}\bar{\Gamma}_\mu^{(2,1)} (p,q;k) &=& i q_\mu \left( 1 -
  \frac{q^2}{p^2} \,f_{UV}(p,q;k)\right)
- i k_\mu \left(\frac{k.q}{k^2} \,f_{IR}(p,q;k) \right) 
\label{new-ghost-vertex3}\\
\bar{\Gamma}_{\rho \nu \sigma}^{(0,3)}(q,p) &=& 
\Gamma^{(0,3)(0)}_{\rho \nu \sigma}(q,p)\Gamma^{3g}(q,p)
\left(1-\left[\frac{140}{51} - \frac{52}{17} \frac{p^2}{k^2} +
    \frac{89}{51} \frac{p^2}{q^2} + \frac{52}{17} \frac{q^2}{k^2} -
    \frac{26}{17} \frac{k^2}{q^2} + \frac{104}{17} \frac{(q.k)^2}{q^2
      k^2}\right]f_{UV}(p,q;k)\right) \label{new-gluon-vertex3}
\end{eqnarray} 
with appropriate damping factors
\begin{eqnarray}
  f_{UV}(p,q;k) = \left( \frac{p^2 q^2 k^2}{(p^2
      +\Lambda_{UV}^2)(q^2+\Lambda_{UV}^2)(k^2+\Lambda_{UV}^2)}\right)^2 ;\hspace*{3mm}
  %\nonumber\\
  f_{IR}(p,q;k) =
  \frac{\Lambda_{IR}^6}{(p^2+\Lambda_{IR}^2)(q^2+\Lambda_{IR}^2)(k^2
    +\Lambda_{IR}^2)} \,, 
  \end{eqnarray} 
  \end{widetext}
  for the infrared and ultraviolet modifications of the vertices.  The
  dressing $\Gamma^{3g}(p,q)$ is given in appendix \ref{kernels}.  The
  resulting forms of the equations for the ghost and gluon DSEs are
  given by
\begin{widetext}
\begin{eqnarray}\label{eq:ghostgluonDSEs} 
  \frac{1}{Z(p^2)} &=& {Z}_3 + g^2\frac{N_c}{3} 
  \int \frac{d^4q}{(2 \pi)^4} \frac{\bar{M}(p^2,q^2,(p-q)^2)}{p^2q^2} 
  G(q^2) G((p-q)^2) \nonumber\\
  &&\hspace{0.6cm}+ 
  g^2 \frac{N_c}{3} \int \frac{d^4q}{(2 \pi)^4} 
  \frac{\bar{Q}(p^2,q^2,(p-q)^2)}{p^2q^2} Z(q^2) Z((p-q)^2) \,
  \Gamma^{3g}(q,p) \; ,
\label{gluon-dse2} \\
\frac{1}{G(p^2)} &=& \tilde{Z}_3 - g^2N_c \int \frac{d^4q}{(2 \pi)^4}
\frac{\bar{K}(p^2,q^2,(p-q)^2)}{p^2q^2}
G(y) Z((p-q)^2) \; . \label{ghost-dse2} 
\end{eqnarray}
\end{widetext}
These forms are similar to the ones of the previously used truncation
scheme of \cite{Fischer:2002eq,Fischer:2003zc}. The effect of the
vertex modifications (\ref{new-ghost-vertex3}) and
(\ref{new-gluon-vertex3}) is hidden in the kernels $\bar{K}, \bar{M},
\bar{Q}$ which are given in appendix \ref{kernels} together with the
original kernels ${K}, {M}, {Q}$ of ref.~\cite{Fischer:2002eq}.

The two free scale parameters $\Lambda_{IR}$ and $\Lambda_{UV}$ are
fixed by the following procedure: Once a numerical solution of the
DSEs has been obtained we project the right hand side of the gluon DSE
onto the longitudinal component of the gluon propagator. This
component is then minimized by changing the two scales. We find that
this procedure is insensitive to $\Lambda_{UV}$ in a window of roughly
$\Lambda/2 < \Lambda_{UV} < \Lambda$, where $\Lambda$ is the
ultraviolet cutoff in our numerical procedure. In practice we choose
the lower bound of this interval. The infrared modification of the
ghost-gluon vertex is important in the regime where the infrared
asymptotics sets in. Consequently we find an optimized value of
$\Lambda_{IR} \approx 300$ MeV.  This choice leads to spurious
longitudinal components of the gluon propagator in the mid-momentum regime
of the order of less than one permille, i.e. of the order of the 
numerical error of our calculations. In the infrared and ultraviolet
momentum regime these terms are even smaller
by orders of magnitude.

Finally we wish to point out that the technical improvements of this
section are also extremely useful when it comes to calculations at
finite temperature, where quadratic divergencies represent a much more
severe problem than at zero temperatures. Improvements along the lines
suggested here have been anticipated in \cite{Cucchieri:2007ta}.

\subsection{Renormalization conditions}

Having specified the truncation scheme it remains to detail the
renormalization conditions imposed on our system of equations. To
discuss these we need to recall the definition of the running coupling
from the ghost-gluon vertex \cite{von Smekal:1997vx}
\begin{equation}
  \alpha(p^2) = \alpha(\mu^2) \frac{G(p^2;\mu^2)^2 
    Z(p^2;\mu^2)}{G(\mu^2;\mu^2)^2 Z(\mu^2;\mu^2)} \label{coupl}
\end{equation}
where the dependence of the dressing functions on the renormalization
point $\mu^2$ is given explicitly. Note that the right hand side of
the equations is independent of $\mu^2$, i.e., it is a
renormalization group invariant. This definition follows
straightforwardly from the Slavnov-Taylor identity
\begin{equation}
\widetilde{Z}_1 = Z_g Z_3 \widetilde{Z}_3^2 \label{STI}
\end{equation}
where the renormalization function $\widetilde{Z}_1$ of the
ghost-gluon vertex is always finite in Landau gauge. For a bare
ghost-gluon vertex, as chosen here, $\widetilde{Z}_1=1$.  Then, in the
momentum subtraction scheme ($\widetilde{MOM}$) defined in \cite{von
  Smekal:1997vx} the renormalization condition $G(\mu^2;\mu^2)^2
Z(\mu^2;\mu^2)=1$ is imposed, which we also adopt here. Then
$G(\mu^2;\mu^2)$ and $Z(\mu^2;\mu^2)$ are not independent of each
other.  In practice one chooses a value for $\alpha(\mu^2)$, which in
turn determines the renormalization point $\mu^2$. In
\cite{Boucaud:2008ky} the special choice
$G(\mu^2;\mu^2)=Z(\mu^2;\mu^2)=1$ has been used, which in general
leads to $\widetilde{Z}_1 = const \not= 1$ by virtue of the STI
(\ref{STI}).

The definition (\ref{coupling}) of the running coupling
represents a mass independent renormalization scheme. This, however, 
may not be adequate for the decoupling solution with $0< D(0) < \infty$.  
In this case one should separate the 'massive' part of the gluon 
propagator from the part proportional to $p^2$. Therefore we introduce
\begin{equation}
  D^{-1}(p^2) = \frac{p^2}{Z(p^2)} = \frac{1}{\bar{Z}(p^2)}\left( 
  p^2 + m^2 \right) \,, \label{masseq}
\end{equation}
with the renormalization group invariant mass parameter $m$. Note that
this splitting is not unique and one obtains an additional free
normalization parameter $\bar{Z}(0)$. The running of $\bar{Z}$ then
enters the running coupling by
\begin{equation}
\alpha(p^2) = \alpha(\mu^2) \frac{G(p^2;\mu^2)^2 \bar{Z}(p^2;\mu^2)}
{G(\mu^2;\mu^2)^2 \bar{Z}(\mu^2;\mu^2)}
\label{coupl2}
\end{equation}   
and the renormalization condition is given by $G(\mu^2;\mu^2)^2
\bar{Z}(\mu^2;\mu^2)=1$. Independently of this renormalization
condition one has the freedom to vary $\bar{Z}(0)$, which together
with the boundary condition $G(0)$ determines the value of the
infrared fixed point of this coupling. This is further discussed in
the next subsection.

For all our numerical results except the ones presented in the 
appendix \ref{MR}, where we show the independence on this choice,
we have used $\alpha(\mu^2)=1$.

\subsection{Numerical results}

%%%%%%%%%%%%%%%%%%%%%%%%%%%%%%%%%%%%%%%%%%%%%%%%%%%%%%%%%%%%%%%%%%%%%
\begin{figure}[th!]
  \includegraphics[width=0.95\columnwidth]{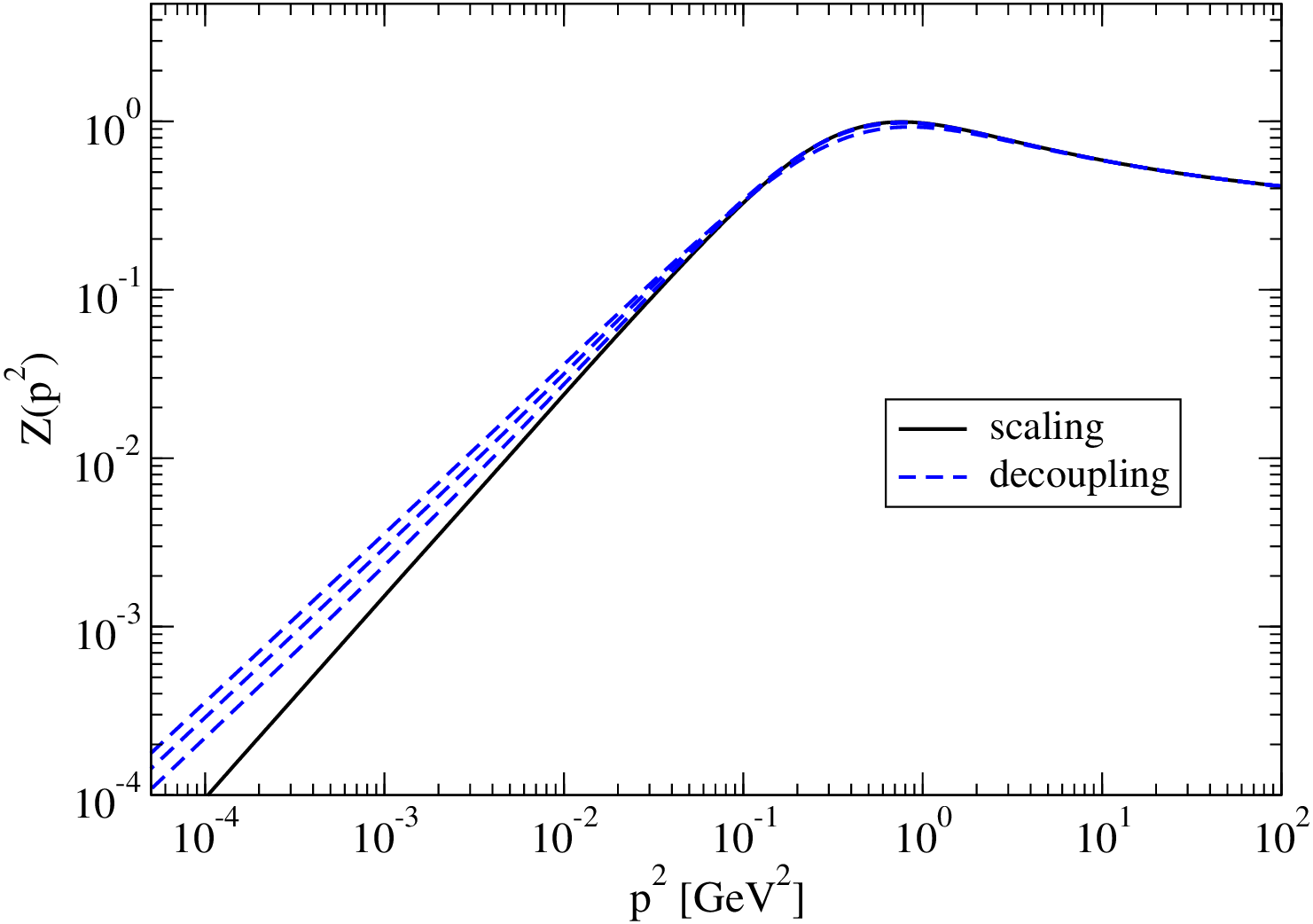}
  \includegraphics[width=0.95\columnwidth]{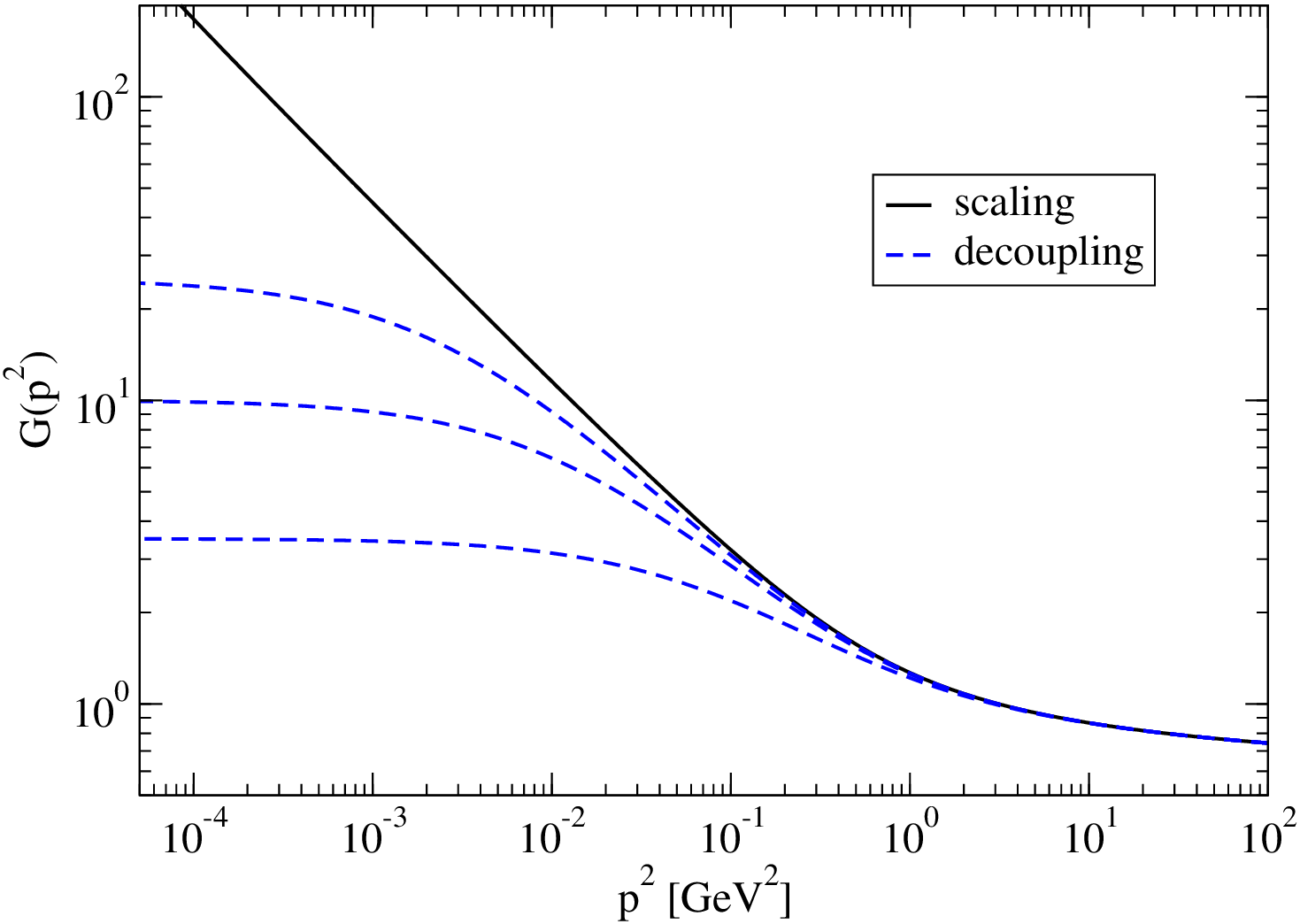}
  \caption{Numerical solutions for the ghost and gluon
    dressing function with different boundary conditions $G(0)$.
    The (artificial) longitudinal components of the gluon 
    propagator are not displayed, since they are of the order of less 
    than one permille, i.e. of the order of the numerical error of 
    our calculations. All results shown here are obtained from 
    our novel truncation scheme. Differences to the scheme defined
    in \cite{Fischer:2002eq,Fischer:2003zc} are, however, only very 
    small and would not be visible in the plots.}
\label{res:sec3}
\end{figure}
\begin{figure}[th!]
  \includegraphics[width=0.95\columnwidth]{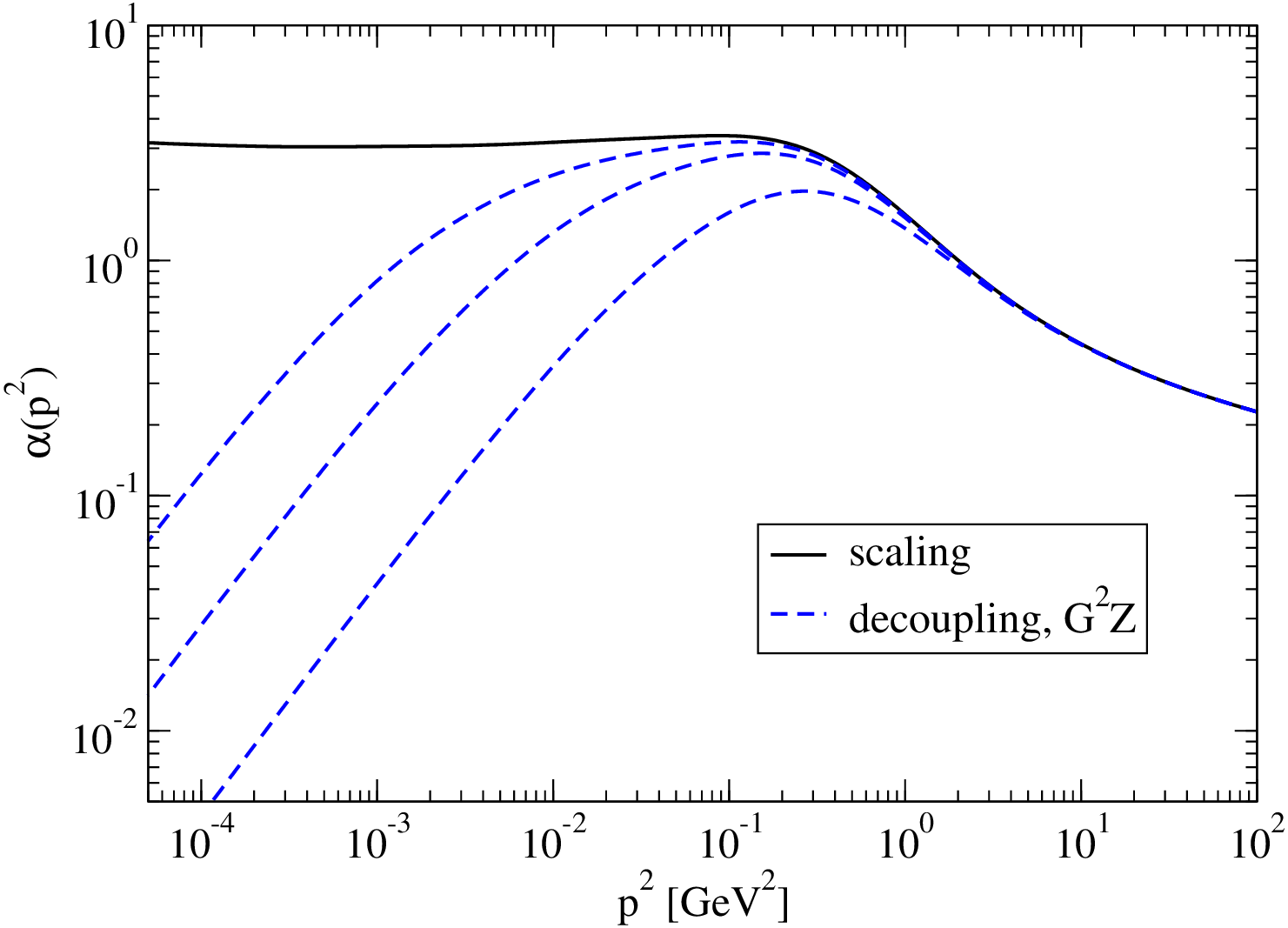}
  \includegraphics[width=0.95\columnwidth]{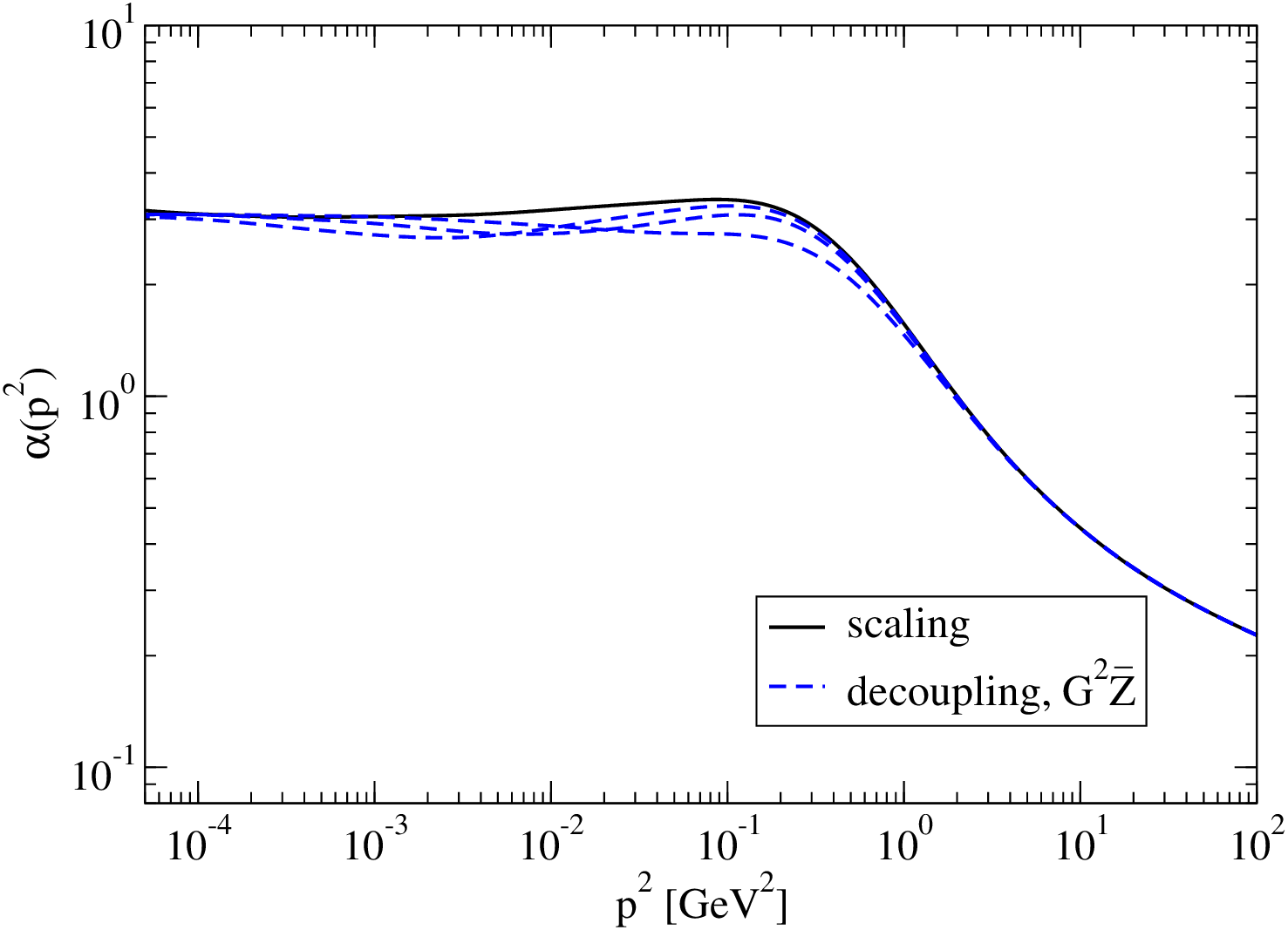}
  \caption{The running coupling for the two types of solutions, defined as
    $\alpha(p^2)=\alpha(\mu^2)G(p^2)^2 Z(p^2)$ (top diagram) and
    $\alpha(p^2)=\alpha(\mu^2)G(p^2)^2 \bar{Z}(p^2)$ (bottom diagram).} 
 \label{res:sec3:coupling}
\end{figure}
%%%%%%%%%%%%%%%%%%%%%%%%%%%%%%%%%%%%%%%%%%%%%%%%%%%%%%%%%%%%%%%%%%%%%%

Our numerical solutions for the ghost and gluon dressing functions are
shown in Fig.~\ref{res:sec3}. The corresponding momentum scale has been 
fixed by best-possible matching of the gluon dressing function
to the corresponding one on the lattice, cf. \section{sec:lattice}. 
Thus we inherit the lattice scale. All results displayed are obtained 
from our novel truncation scheme. Differences to the scheme defined
in \cite{Fischer:2002eq,Fischer:2003zc} are, however, only very small 
and would not be visible in the plots. This provides additional 
justification that the old scheme already represented a reliable result.
Transversality is manifest in our new truncation scheme; the longitudinal 
components of the propagator (not shown in the figure) are smaller than 
one permille and therefore of the same size as the numerical error of 
our calculation. Quadratic divergencies do not appear and the numerical 
solutions respect multiplicative renormalizability, see appendix \ref{MR}.

By varying the boundary condition $G(0,\mu^2)$ we are able to generate
both, a solution of the scaling type I, eq.\ (\ref{typeI})), with
$G^{-1}(0,\mu^2)=0$ and a continuous set of decoupling solutions
characterized by a finite value $G(0,\mu^2)$. Shown are results for
three different values of $G(0,\mu^2)$. The corresponding gluon
propagator is either massive in the sense that $D(0)=\lim_{p^2
  \rightarrow 0} Z(p^2)/p^2 = const.$ for decoupling, or has the power
like behavior (\ref{typeI}) with $\kappa=\kappa_C=(93-\sqrt{1201})/98
\approx 0.595353$ \cite{Lerche:2002ep} in the case of scaling. In the
ultraviolet momentum region, both types of solutions are almost
identical, as expected. The running couplings are shown in
Fig.~\ref{res:sec3:coupling}. For the scaling solution one observes
the infrared fixed point known from previous studies, whereas the
coupling for the decoupling solution falls with $p^2$ in the infrared
when the mass-independent definition (\ref{coupl}) is used (top
diagram of Fig.~\ref{res:sec3:coupling}). If, however, we employ the
mass-dependent definition (\ref{coupl2}) (bottom diagram of
Fig.~\ref{res:sec3:coupling}) the resulting coupling develops a fixed
point in the infrared. The value of this fixed point is an additional
free parameter and has been fixed at $\alpha(0) \approx 3$, i.e. at
approximately the value of the fixed point of the scaling solution.
Note that these couplings still show variations with changes of $G(0)$
at intermediate momenta from $0.1-1$ GeV.

In the light of these findings we wish to make the following comments:
\begin{itemize}
\item In principle one has to face the problem that an inadequate
  ultraviolet renormalization may result in an infrared mass for the
  gluon reflecting the breaking of BRST-invariance by the momentum
  cutoff.  We explicity checked that this is not the case for both our
  scaling and the decoupling solution; the solutions are independent
  of the cutoff $\Lambda$ imposed in the loop integrals. In particular
  this is true for $D(0)$.
\item We also assessed the behavior of the dressing functions under a
  change of the renormalization point $\mu^2 \rightarrow \nu^2$, i.e.
  we chose $\alpha(\nu^2)=0.5$ and $G(\nu^2;\nu^2)^2
  Z(\nu^2;\nu^2)=1$.  Again, by varying $G(0,\nu^2)$ we find both
  types of solutions and the resulting ghost and gluon dressing
  functions are related to the ones at $\alpha(\mu^2)=1$ by
  renormalization factors, as they should. The explicit results can be
  found in appendix \ref{MR}.  We therefore conclude that there is no
  notion of a 'critical' coupling distinguishing the two types of
  solutions as claimed in \cite{Boucaud:2008ky}. Instead it is solely
  the boundary condition $G(0,\mu^2)$ that matters.
\item We wish to point out that the decoupling solution with an
  infrared finite gluon propagator necessarily entails an infrared
  finite ghost dressing function (up to logarithms) as long as the
  ghost-gluon vertex does not contain any nontrivial power-laws
  \cite{Fischer:2006vf,Boucaud:2008ky}.  This situation is also
  realized in \cite{Aguilar:2008xm}, where henceforth $\kappa_C=0$
  must be realized.  This is compatible with part of the fits they
  deduce from their numerical solutions.
\item The issue of transversality is separate from the question of
  which types of solutions are realized in a particular truncation
  scheme. In non-transverse truncation schemes longitudinal components
  of the gluon propagator only constitute a problem if they are
  back-fed into the equation by the use of a non-transverse projection
  method of the gluon DSE. This has carefully been avoided in the
  truncation scheme of \cite{Fischer:2002eq} which consequently
  delivered very similar results as the new scheme employed in this
  section. 
\item We wish to emphasize again that either (i) global BRST symmetry
  is unbroken, then the decoupling solution would imply the breaking
  of global color symmetry as indicative for a Higgs phase of the
  theory. Or, (ii), in a confining phase the decoupling solution
  implies the breaking of global BRST symmetry and therefore does not
  agree with the Kugo-Ojima confinement scenario. Indeed, all known
  BRST-quantizations indicating towards an infrared finite ghost even
  break off-shell BRST \cite{Dudal:2007cw,Dudal:2008sp}.  Therefore it
  is not clear how to construct a physical state space in the
  decoupling case. In terms of Green's functions this implies that it
  is not known how to construct physical observables in general. In
  our opinion this is a serious problem. In addition the decoupling
  solution is in contradiction with the Gribov-Zwanziger confinement
  scenario due to its finite ghost dressing function
  \cite{Zwanziger:1997}.  All in all, the status of the decoupling
  solutions is therefore clearly different from the scaling solution,
  which agrees with both, the Kugo-Ojima and the Gribov-Zwanziger
  scenarios.
\end{itemize}

It is furthermore interesting to compare our results to corresponding
ones from DSEs on a torus. In \cite{Fischer:2002eq,Fischer:2005ui}
solutions from the ghost and gluon DSE on a torus have been discussed
which did not connect to a scaling type of solution even if very large
volumes were taken.  In a later work \cite{Fischer:2007pf} the
renormalization conditions for the torus solutions have been
reconsidered and adapted such that the scaling solution in the
infinite volume/continuum limit has been reproduced.  The status of
the solutions found in \cite{Fischer:2002eq,Fischer:2005ui}, however,
remained somewhat unclear. From the results of this work we are now in
a position to clarify this issue. The different renormalization
conditions on the torus employed in
\cite{Fischer:2002eq,Fischer:2005ui} and \cite{Fischer:2007pf} are in
one-to-one correspondence to the boundary condition $G^{-1}(0,\mu^2)$
for the continuum DSEs investigated in this work. The infinite
volume/continuum limit of the solutions found in
\cite{Fischer:2002eq,Fischer:2005ui} is therefore given by one of the
decoupling type of solutions reported in this work.

Finally we wish to point out that the parameter $G^{-1}(0,\mu^2)$
corresponds to a related one in the Zwanziger-Lagrangian approach 
\cite{Zwanziger:1992qr} to infrared Yang-Mills theory. There, a 
mass parameter can be introduced which produces decoupling solutions 
for non-vanishing mass \cite{Dudal:2008sp} and scaling solutions 
for vanishing mass, see e.g. \cite{Gracey:2005cx} and refs. therein.
The latter is in agreement with the original Gribov-Zwanziger scenario.

\section{Ghosts and gluons from the functional renormalization group 
\label{sec:FRG}}

The DSE-analysis of the previous sections can be repeated within the
functional RG. In order to facilitate the access to the FRG-literature
on Landau gauge Yang-Mills, e.g.\ \cite{Litim:1998nf,Pawlowski:2005xe,
  Gies:2006wv,Ellwanger:1994iz,Bonini:1994kp,Ellwanger:1995qf,%
Pawlowski:2003hq,Pawlowski:2004ip,%
Kato:2004ry,Fischer:2004uk,Braun:2007bx},
we present the results in standard FRG-notation and provide the
dictionary to the DSE-computations. The results shown here are
preliminary results taken from \cite{pawlowski}. We shall be brief on
the details of this analysis and only discuss the relation between the
renormalization procedure in the DSE of section~\ref{sec:sec2} and
that in the FRG.  Furthermore we comment on the truncation and discuss
its relation to that used in the DSE. The renormalization conditions
are contained in appropriately chosen initial conditions for the
effective action $\Gamma_\Lambda$ at the ultraviolet scale $\Lambda$.
Indeed it has been shown in \cite{Pawlowski:2005xe} that the
integrated flow equations define a set of DSEs within a consistent
BPHZ-type renormalization scheme. This system reads for the effective
action
\begin{eqnarray}\label{eq:FRG}
 \Gamma= \Gamma_\Lambda+\012 \Tr \ln \Gamma^{(2)} 
  -\012 \int_\Lambda^0 \0{dk'}{k'} \Tr\, \dot \Gamma^{(2)}_{k'} 
\0{1}{\Gamma_{k'}^{(2)}+R_{k'}} \,.
\end{eqnarray} 
The first term on the rhs of \eq{eq:FRG} comprises the renormalization
conditions as an initial condition as can be seen via the comparison
of the second field derivatives of \eq{eq:FRG} and
\eq{eq:ghostgluonDSEs}. The second term on the rhs of \eq{eq:FRG}
contains one loop contributions to Green's functions $\Gamma^{(2n,m)}$
with full propagators and vertices, the third integral term contains
RG-improvement terms such as the two-loop diagrams in the DSEs. In the
present truncation we need the (inverse) propagators $ \Gamma_A^{(2)}=
\Gamma^{(0,2)}$, $ \Gamma_C^{(2)}= \Gamma^{(2,0)}$,
\begin{eqnarray} \nonumber 
{\Gamma_A^{(2)}}^{ab}_{\mu\nu}(p^2)
&=& Z_A(p^2) p^2 \delta^{ab}\left(\delta_{\mu \nu} -
    \frac{p_\mu p_\nu}{p^2}\right) +{\rm longitud.}\,, \\
{\Gamma_C^{(2)}}^{ab}(p^2)&=& Z_C(p^2) p^2\delta^{ab}\,,\label{eq:frgprops} 
\end{eqnarray} 
as well as the vertices
$\Gamma^{(2,1)},\Gamma^{(0,3)},\Gamma^{(0,4)}$. For the vertices we
refer to the parameterization in section~\ref{sec:sec2} and to
\cite{Pawlowski:2003hq,Pawlowski:2005xe,pawlowski}. The wave function
renormalization functions $Z_A,Z_C$ relate to the dressing functions
$Z(p^2), G(p^2)$ as
\begin{eqnarray} \label{eq:DSEFRGrel}
Z_A(p^2)=\0{1}{Z(p^2)}\,,&\qquad \qquad & Z_C(p^2)=\0{1}{G(p^2)}\,, 
\end{eqnarray} 
whereas $Z_3,\tilde Z_3$ and $\tilde Z_1$ are defined with
$\Gamma_\Lambda^{(2,0)}$, $\Gamma_\Lambda^{(0,2)}$ and
$\Gamma_\Lambda^{(2,1)}$. Note that RG-invariance of the FRG-equation
is automatically guaranteed by the separate RG-invariance of each term
in \eq{eq:FRG}, and hence the invariance of the FRG solution under a
change of the renormalization scale $\mu$. The scaling and decoupling
solutions (at $k=0$) are now adjusted via fine-tuning of $\tilde Z_3$
as in the DSEs. Indeed, the ghost FRG can be mapped into the
corresponding DSE and one can directly use eq.~(\ref{subsub}).
Finally, the truncation is optimized
\cite{Litim:2000ci,Pawlowski:2005xe} by an appropriate choice of the
regulator function $R$. The optimized regulator employed for the
present work has been derived from functional optimization in
\cite{Pawlowski:2005xe}. It has been already shown for the present
truncation in \cite{Pawlowski:2003hq} that optimized regulators lead
to the same analytic expressions for $\kappa$'s and $\alpha(0)$ in the
scaling solution as for the DSE equations, $\kappa_C\approx 0.595353,
\alpha(0)\approx 3$. Other regulators lead to scaling solutions with
slightly varying $\kappa$, $\kappa\in [.539...\,,\,.59535...]$, for a
specific choice see also \cite{Fischer:2004uk}. Note that the above
interval serves as an estimate on the systematic error of the present
FRG and also DSE truncation. Furthermore it has been argued in
\cite{Pawlowski:2004ip} that the lower value relates to finite volume
studies, providing a link to the torus DSE analyses in
\cite{Fischer:2002eq,Fischer:2005ui,Fischer:2007pf}. More details on
the present FRG analysis will be presented in \cite{pawlowski}.

%%%%%%%%%%%%%%%%%%%%%%%%%%%%%%%%%%%%%%%%%%%%%%%%%%%%%%%%%%%%%%%%%%%%%
\begin{figure}[th!]
  \includegraphics[width=0.95\columnwidth]{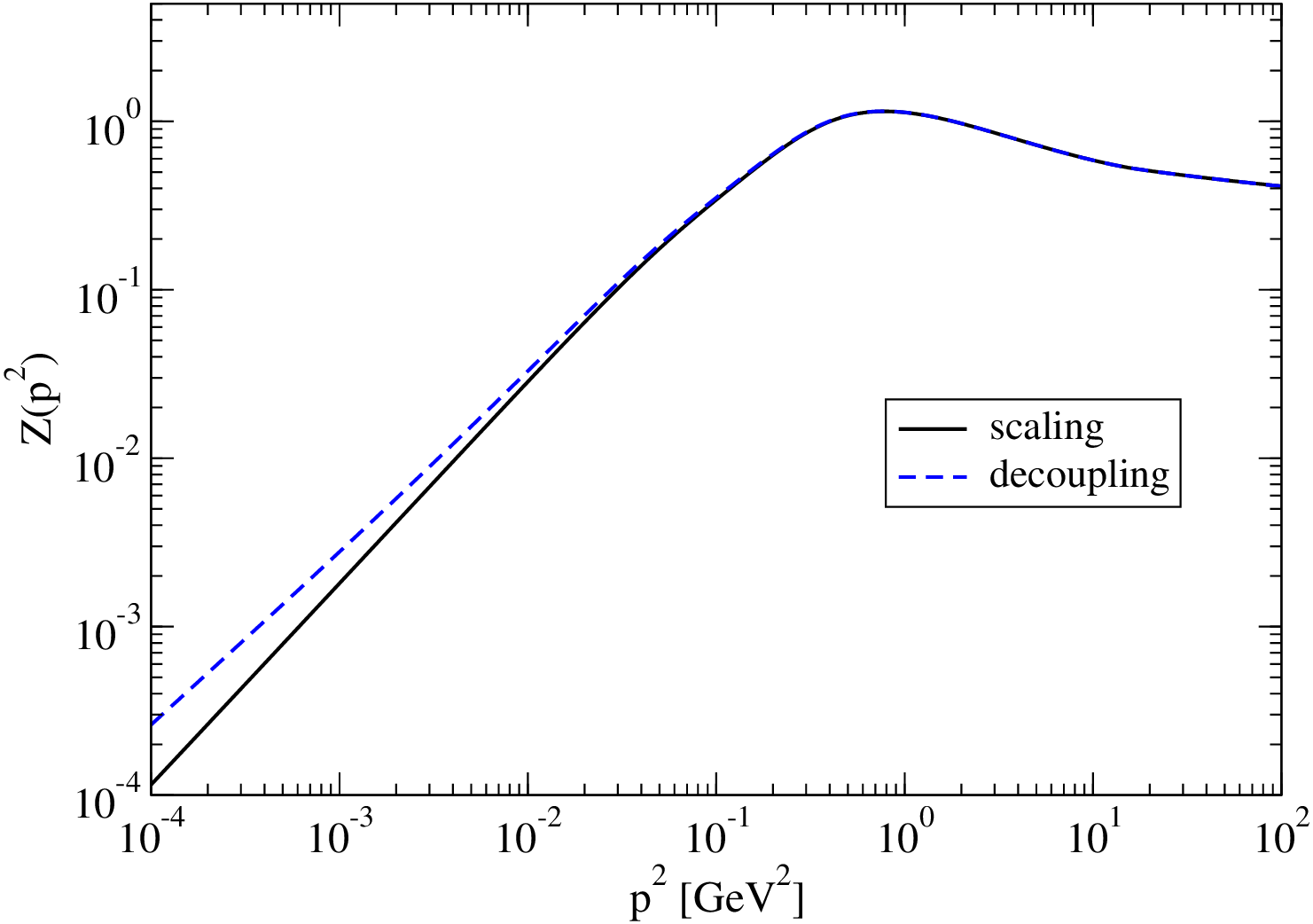}
  \includegraphics[width=0.95\columnwidth]{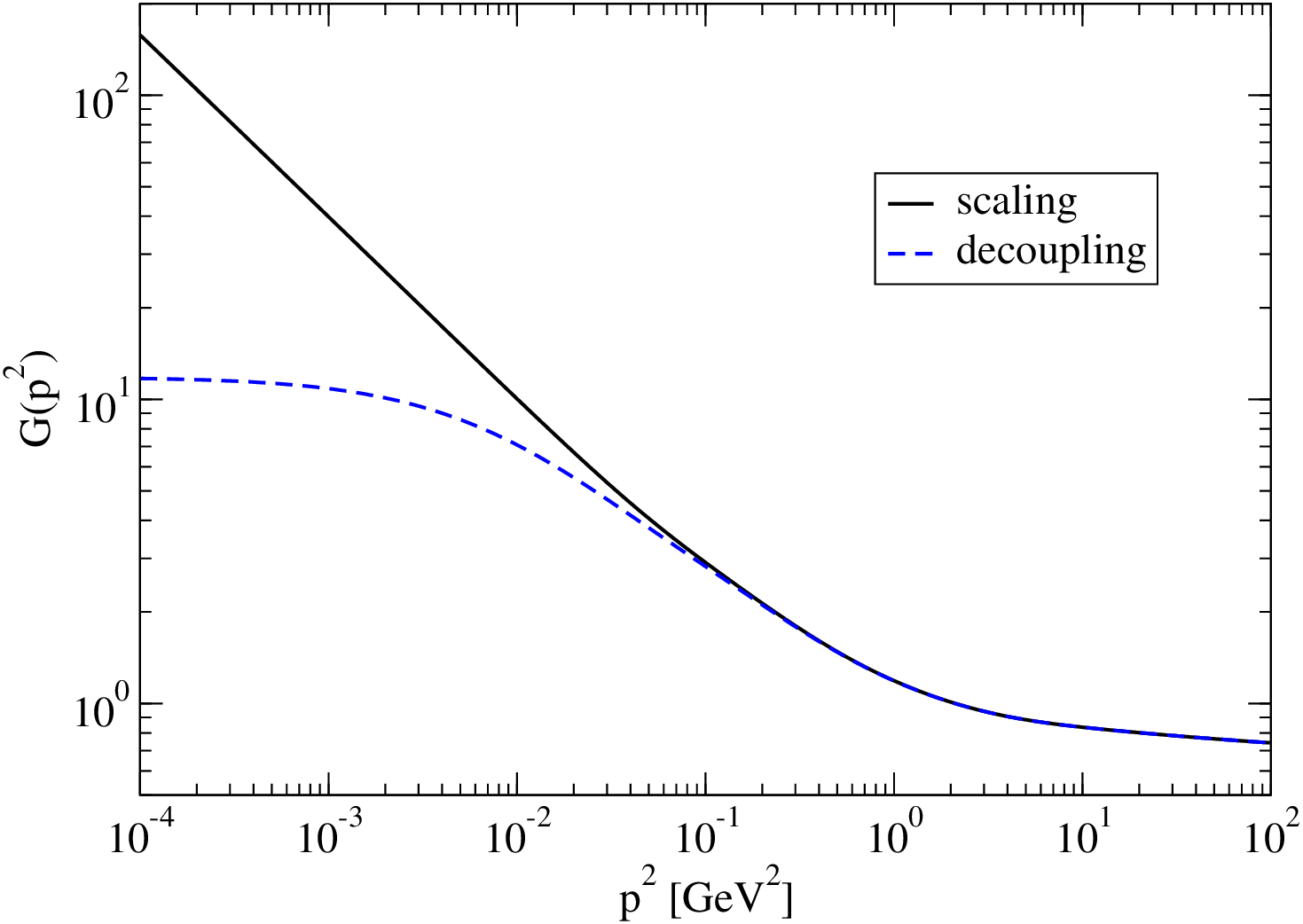}
  \caption{Numerical solutions for the ghost and gluon
    dressing function with two different boundary conditions $G(0)$
    calculated from the FRGs.}
\label{res:sec4}
\end{figure}
\begin{figure}[th!]
  \includegraphics[width=0.95\columnwidth]{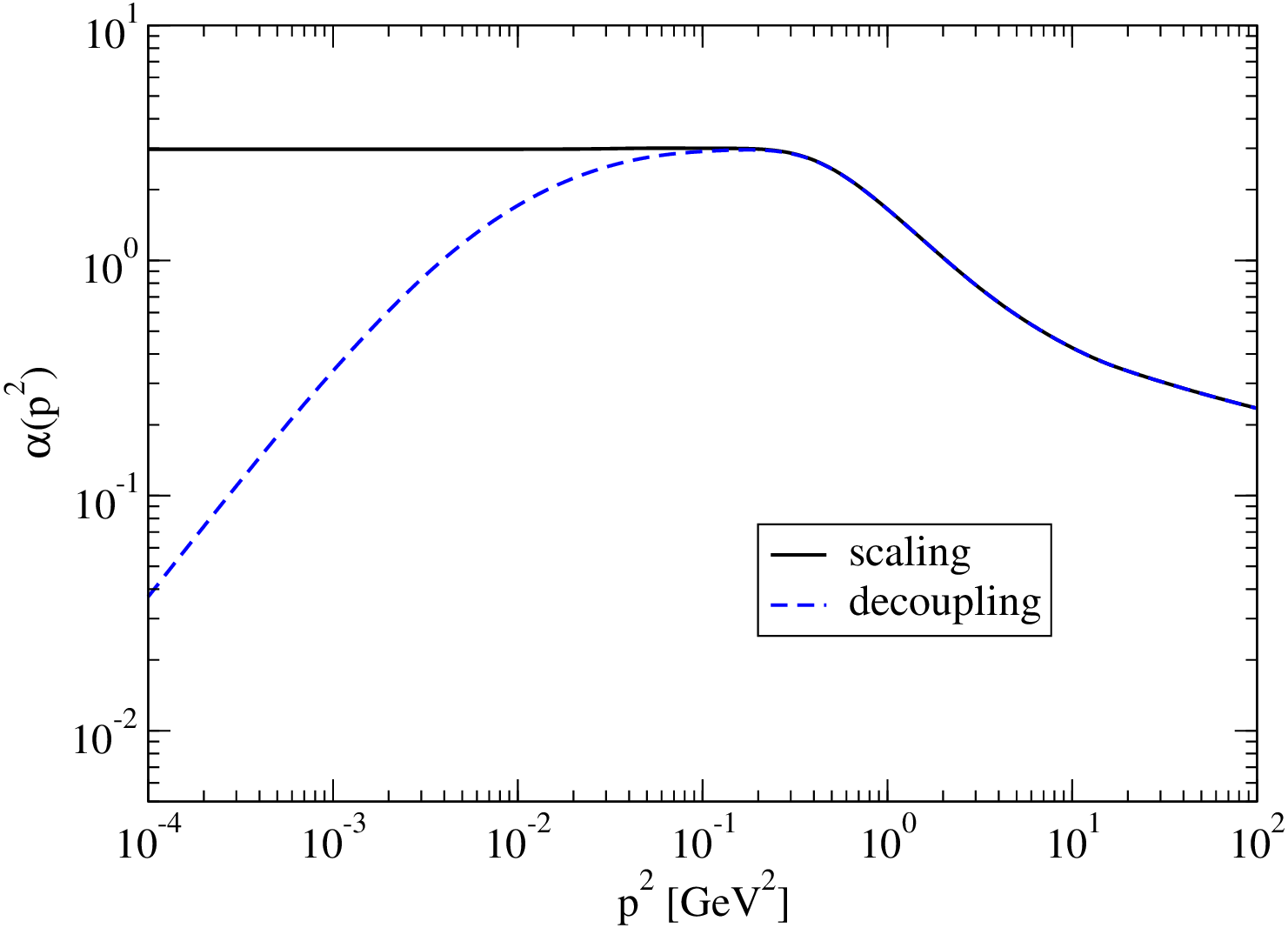}
  \caption{The running coupling from the FRGs, defined as
    $\alpha(p^2)=\alpha(\mu^2)G(p^2)^2 Z(p^2)$.} 
 \label{res:sec4:coupling}
\end{figure}
%%%%%%%%%%%%%%%%%%%%%%%%%%%%%%%%%%%%%%%%%%%%%%%%%%%%%%%%%%%%%%%%%%%%%%

The numerical solutions from the FRGs are presented in
Figs.~\ref{res:sec4} and \ref{res:sec4:coupling}. We show the scaling 
solution and one representative for a decoupling solution. They agree
qualitatively and to a large extent also quantitatively with the ones
from the DSEs presented in the previous section. Therefore both
frameworks support the conclusions discussed there.  Quantitative
differences in the mid-momentum regime can be attributed to the
different diagrammatics used in both approaches.  
In summary we have discussed the equivalence and consistency of the
renormalization procedure for both, DSEs and FRGs. Moreover, the FRG
provides a consistent momentum cut-off regularization of the
corresponding DSE equation via \eq{eq:FRG} and thus allows to
deduce the modified STIs for the DSE in the presence of an ultraviolet
momentum cut-off, see \cite{Pawlowski:2005xe,pawlowski}. A
crucial difference in the present truncation is the tadpole diagram in
the gluon FRG-equation that depends on the full four-gluon vertex.
This incorporates two-loop contributions of the sunset
diagram in the gluon DSE, see Fig.~\ref{fig:DSE-gl}.

%%%%%%%%%%%%%%%%%%%%%%%%%%%%%%%%%%%%%%%%%%%%%%%%%%%%%%%%%%%%%%%%%%%%%%%%%%
\begin{figure}[th!]
\includegraphics[width=\columnwidth]{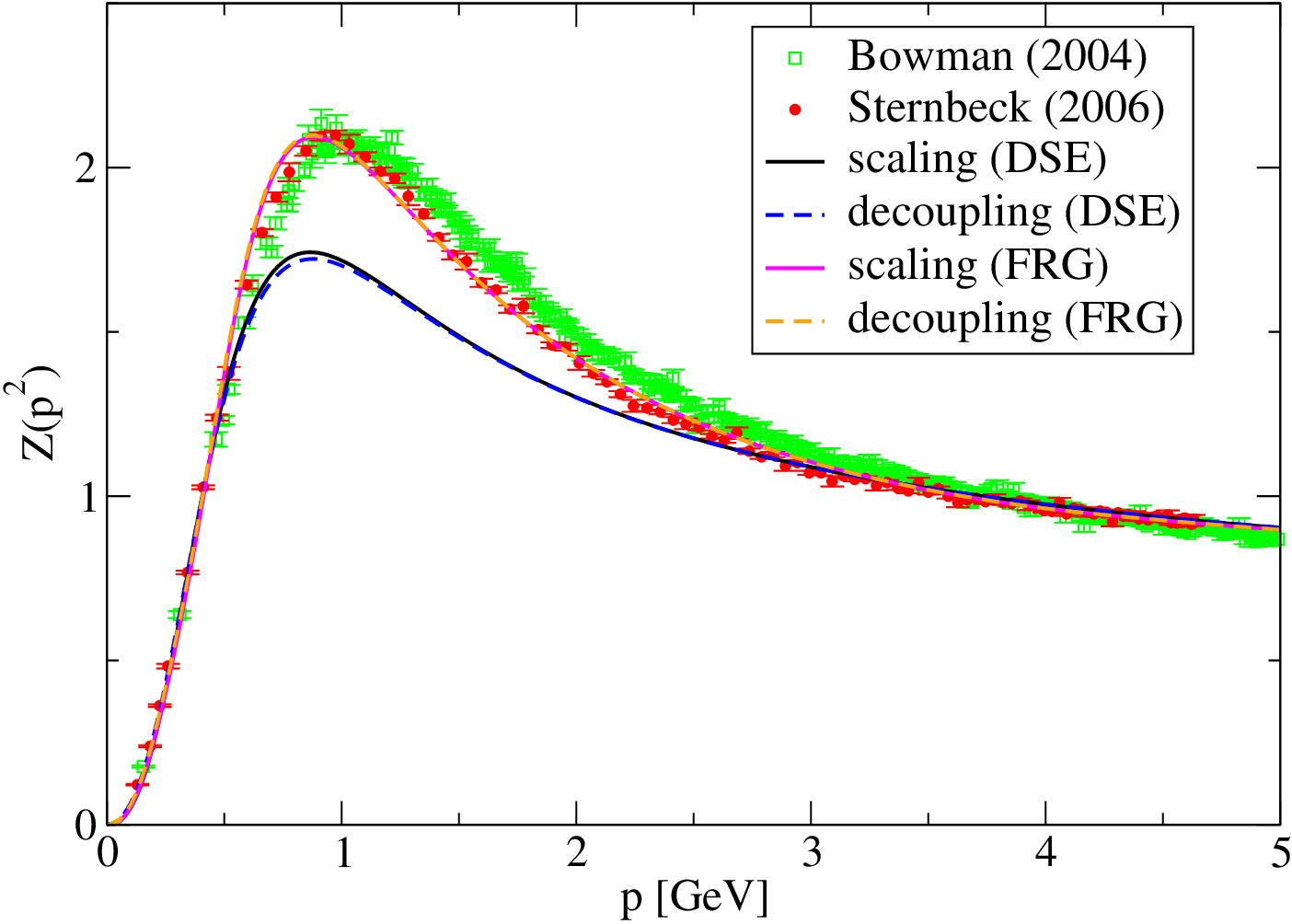}
\includegraphics[width=\columnwidth]{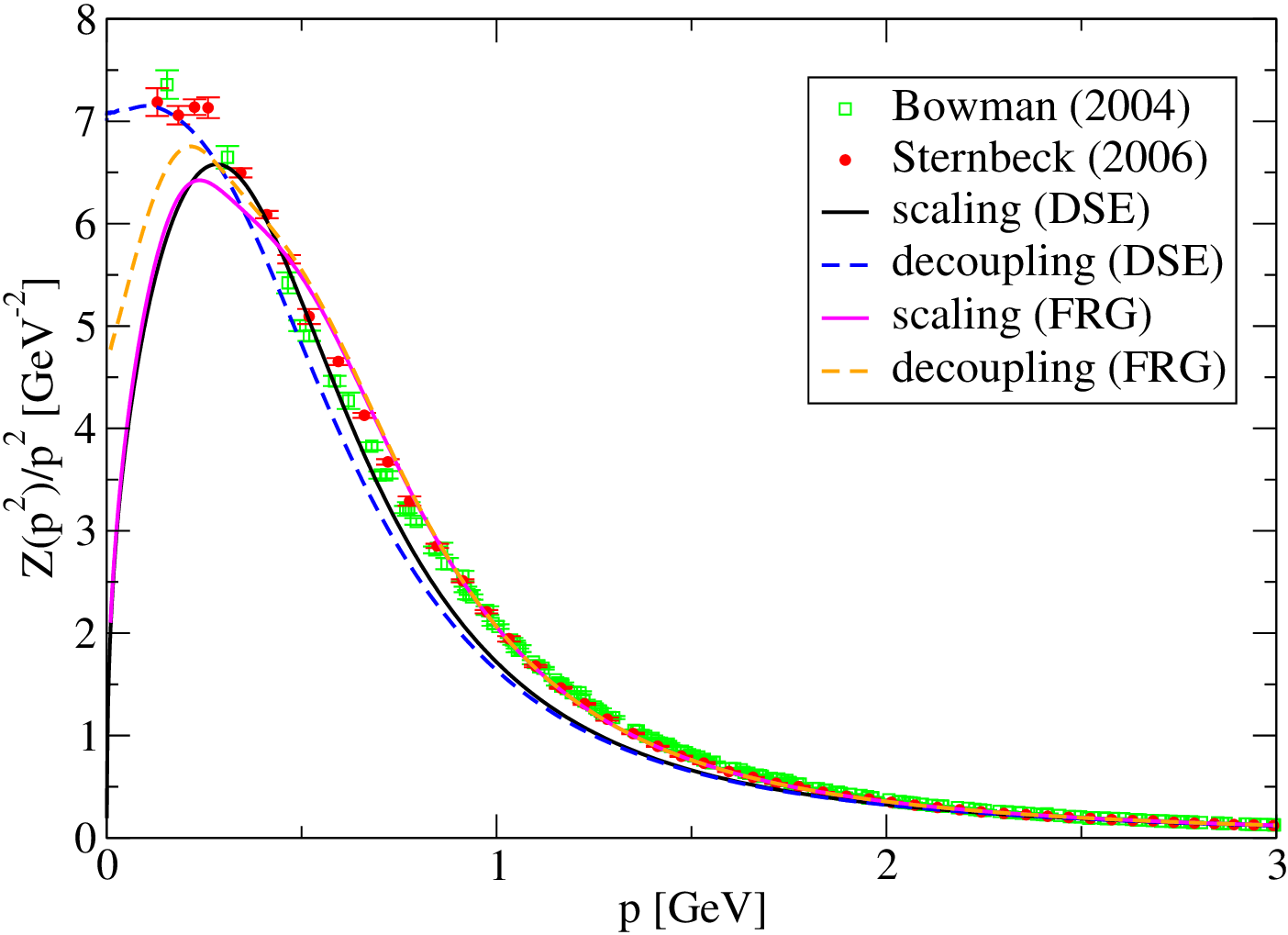}
\includegraphics[width=\columnwidth]{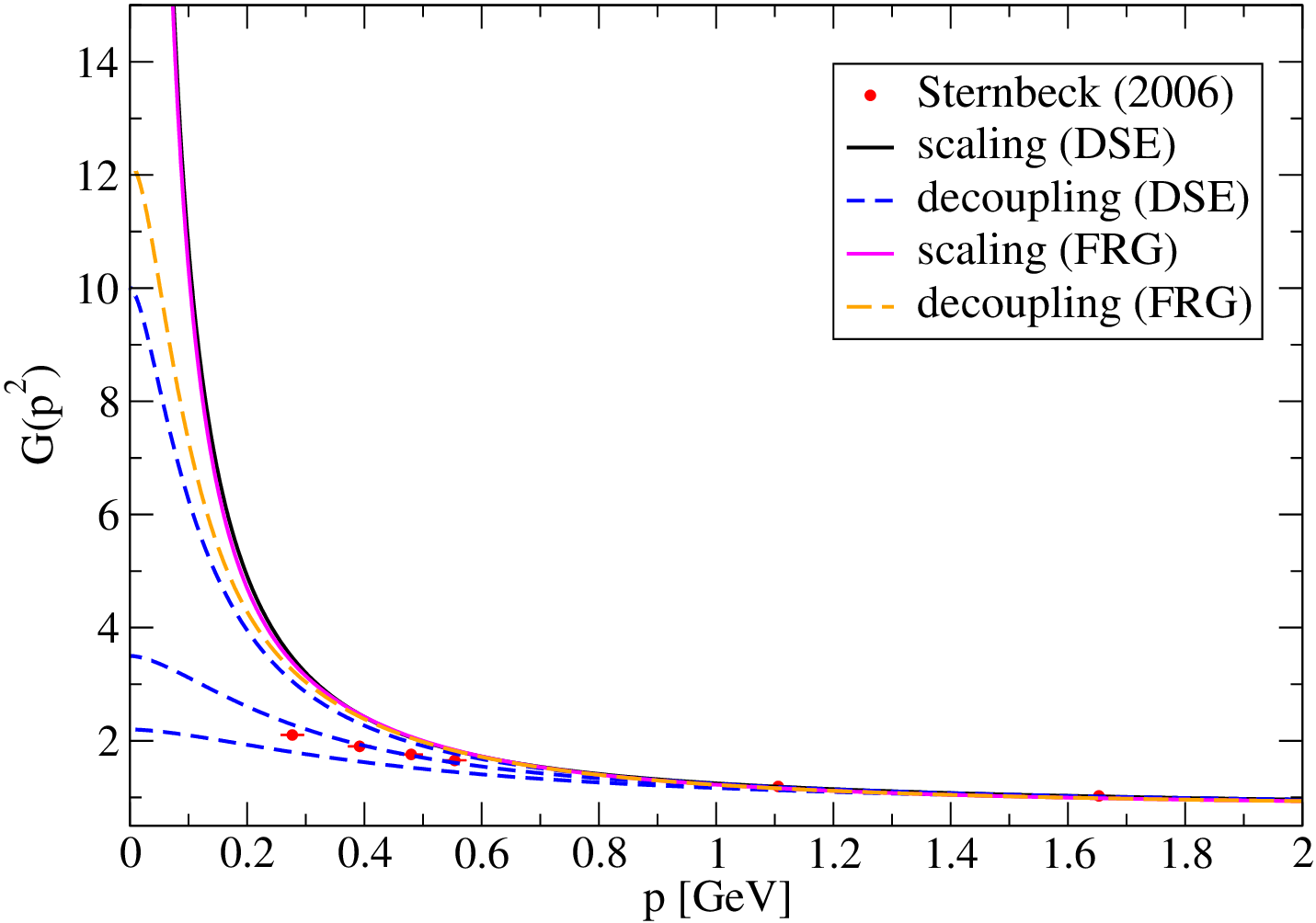}
\caption{Both type of solutions of sections \ref{sec:sec2} and 
\ref{sec:FRG} compared to 
lattice results in minimal Landau gauge from \cite{sternbeck06,Bowman:2004jm}.}
\label{res:lattice}
\end{figure}
%%%%%%%%%%%%%%%%%%%%%%%%%%%%%%%%%%%%%%%%%%%%%%%%%%%%%%%%%%%%%%%%%%%%%%%%%%

\section{Comparison with lattice results}\label{sec:lattice}

In the previous two sections we obtained two different 
types of solutions for the ghost and gluon propagators in the DSE 
and FRG approaches. It is certainly instructive to compare these
results to the ones from lattice calculations. As became apparent 
from a number of works in the past years such a comparison is not
unambiguous. Ideally one strives for a situation where exactly the
same quantities are calculated in the continuum and on the lattice.
However, this is currently not the case for a number of reasons.
First, lattice calculations are necessarily done in a 
finite volume. It is therefore
mandatory to take into account finite volume effects and zero 
mode contributions absent in the infinite volume/continuum limit.
Second, one encounters finite size contributions due to the non-vanishing
lattice spacing. Third, artefacts due to the gauge fixing procedure
are different from the ones in a continuum formulation.
 
Before we discuss these issues further let us compare the continuum
solutions with the lattice results of
refs.~\cite{sternbeck06,Bowman:2004jm} in minimal Landau gauge.  In
the top diagram of fig.~\ref{res:lattice} we display the gluon
dressing function from different approaches. At large momenta, where
perturbation theory sets in, all results are in excellent agreement
with each other. The DSE results as well as the FRG results in the
intermediate regime show only a mild dependence of the type of
solution, i.e. scaling or decoupling does not really matter here, as
expected. As compared to the standard DSE results the dressing
function from the functional RG approach is closer to the lattice
data. From the discussion of the last section this was to be expected,
since the FRG truncation included effects from the gluonic two-loop
diagrams neglected in the DSE-truncation.  Note that such
contributions can be either included directly or phenomenologically by
modifying the three-gluon interaction in the one-loop diagram also
into the DSE framework, see e.g. \cite{Maas:2004se}.

The infrared behavior of the propagator functions for the gluon,
$D(p^2)=Z(p^2)/p^2$, of both solutions are compared in the second
panel of fig.~\ref{res:lattice}. Clearly, the scaling solution
comprises an infrared vanishing propagator, whereas the decoupling
solutions are infrared finite. Changing the boundary condition
$G^{-1}(0,\mu^2)$ from zero to finite values first leads to a finite
but small value for $D(0)$ with the corresponding gluon propagator
still being non-monotonous.  From a certain minimal value of
$G^{-1}(0,\mu^2)$ on, this behavior changes and the gluon becomes a
monotonously decreasing function of momentum.  Such a monotonous
behavior is also seen in the lattice data, which therefore clearly
represent a decoupling type of solution for the gluon.

The third diagram of fig.~\ref{res:lattice} compares results for the
ghost dressing function from the continuum approaches with the lattice
data. Again we find that the lattice results resemble a decoupling
type of solution, i.e. it seems from the plot that the corresponding
value $G(0,\mu^2)$ on the lattice would be finite.

In conclusion, the lattice results support a decoupling solution with 
broken BRST symmetry. The origin of this BRST breaking is
currently not understood. It may be induced by one or more of the three
aforementioned effects.

Finite volume corrections are unambiguously present and are sizeable
even for large volumes \cite{Fischer:2007pf}. However, recent
calculations on very large lattices
\cite{cucchieril7,Cucchieri:2007rg,limghost} have demonstrated that
these corrections alone do not change the solution from a decoupling
one to a scaling one. These results have been obtained on rather
coarse lattices in the minimal Landau gauge\footnote{Note that minimal
  Landau gauge may have ambiguities, that possibly lead to differing
  propagators or at least differing finite volume effects \cite{gc},
  in contrast to the unique definition of the absolute Landau gauge.}.
Correcting for both effects has a significant impact.

Recently, it has been shown \cite{lvs} that the gluon propagator, 
and in particular its value at zero momentum $D(0)$, are affected 
by discretization artifacts which originate from different 
possibilities to define the gauge field on the lattice. 
The consequences of such discretization artifacts are not yet 
studied in detail, but evidence has been found that is in favor 
of a scaling solution on very fine, large lattices \cite{lvs}.

Finally, the question of adequate gauge-fixing is of relevance. As
discussed in the beginning the basic scenario is established in the
absolute Landau gauge, and therefore the lattice calculations should
be done in this gauge. However, the numerical implementation of this
gauge is expensive, and hence most calculations so far have been done
in the minimal Landau gauge. Nonetheless, it is known that at least
the ghost propagator differs in both gauges, but the extent and
volume-dependence is not known sufficiently well in four dimensions
\cite{Bogolubsky,Cucchieri:1997ns}. Again, recently \cite{gc} evidence
has been found which suggest that the absolute Landau gauge produces
the scaling solution. However, only in three and two dimensions
support for this has been found yet and only
indications are so far available in four dimensions \cite{Bogolubsky,Bakeev:2003rr,gc}, mainly due
to the computing time required in four dimensions.
Extrapolations in the number of dimensions $d$ make it nonetheless
likely that also in four dimensions in the absolute Landau gauge the
propagators are scaling.

Hence both effects are in favor of a scaling solution. Nonetheless,
the interplay of these effects and finite volume effects has not yet
been fully investigated, and the situation in lattice gauge theory is
not yet unambiguously resolved.

By comparison of the lattice and the functional results it seems
furthermore tempting to conjecture that the scaling solution is in
fact obtained in absolute Landau gauge, and the absolute Landau gauge
is implemented in functional calculations by the implementation of the
boundary condition $G(0,\mu^2)^{-1}=0$.  The good agreement of the
massive solution to the lattice results in minimal Landau gauge, see
Fig.~\ref{res:lattice}, suggest further that the other boundary
condition $G(0,\mu^2)^{-1}\neq 0<\infty$ seems to implement the
minimal Landau gauge. However, final justification of this idea
requires a clarification of the role of discretization artifacts
\cite{lvs} and other subtleties in defining the minimal Landau gauge
\cite{gc}.

Alternatively, one could check which solution satisfies the condition
for absolute Landau gauge.  One is the positivity of the ghost
dressing function, which is given in both cases. The other is the
requirement of minimizing the trace of the gluon propagator,
equivalent to the condition \pref{eq:abslancon} \cite{gc},
\begin{equation} 
{\cal F}(D)=c \int p Z(p) dp\label{momdesc},
\end{equation} 
\no with a positive constant $c$ \cite{gc}. As is visible from
Fig.~\ref{res:sec3} and \ref{res:lattice}, at low momenta the scaling
solution is smaller for both the DSE and FRG results. At mid-momentum,
both solutions are similar from the FRG, while from the DSEs the 
massive solution is somewhat smaller. This already shows that the
question whether \pref{momdesc} is minimal for scaling or decoupling 
depends on the details of the truncation and therefore cannot be 
answered from the results plotted in Fig.~\ref{res:sec3} and 
\ref{res:lattice}.  

Finally, we would like to mention two general issues.
First, we expect that the general structure 
of the interplay between gauge fixing and the infrared behavior of 
Green's functions is general and also present in other gauges. 
Clear indications for such a picture are given in \cite{Reinhardt:2008ij} 
for the case of Coulomb gauge. Second, we wish to emphasize that our
discussion is independent of the number of colors, $N_c$, and therefore
valid for SU($N_c$)-Yang-Mills theory. In the DSE/FRG framework an 
indirect and very mild dependence of the ghost and gluon dressing 
functions on $N_c$ is induced by sub-leading components of the four-gluon 
vertex \cite{Kellermann:2008iw}, which only have an impact
in the mid-momentum region. This mild dependence is also found in
lattice calculations comparing the dressing functions for the cases
of SU(2) and SU(3) \cite{Cucchieri:2007zm,Sternbeck:2007ug}. 

In fact, based on the structure of the DSEs it has been conjectured 
\cite{Maas:2005ym} that the scenario presented here
can be extended to any (semi-)simple Lie-group without any
qualitative change. First support for this
conjecture in the case of the exceptional Lie-group G$_2$ in
lattice calculations has been presented in \cite{Maas:2007af}.

\section{Gluon confinement}\label{sec:selec}

\begin{figure}[th!]
 \includegraphics[width=\columnwidth]{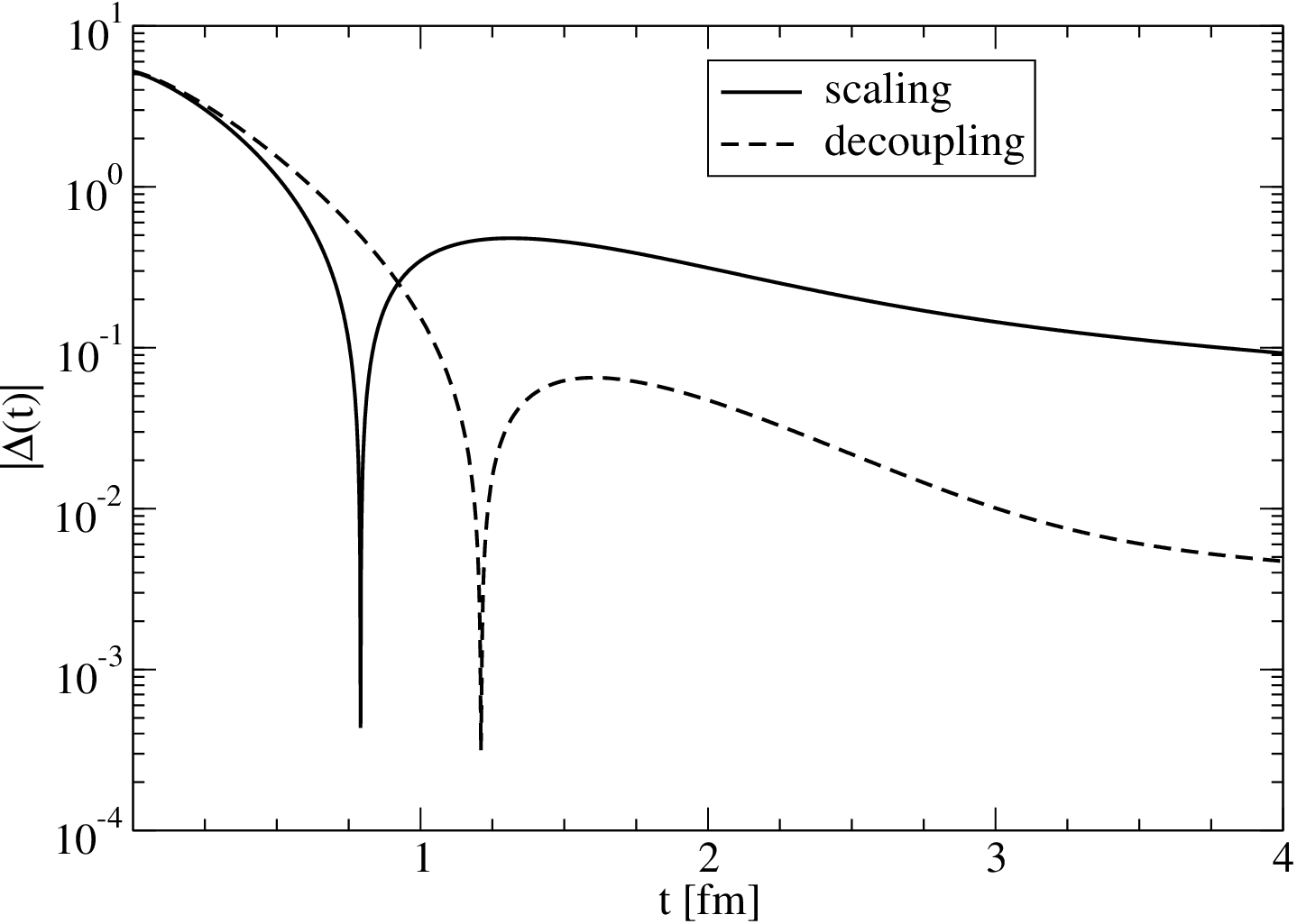}
  \caption{The absolute value of the Schwinger function $\Delta(t)$ plotted 
  against time for both, the decoupling and scaling type of solutions. 
  (The latter result with a slightly different scale has previously been 
  published in \cite{Alkofer:2003jj}.)}
\label{res:FFT}
\end{figure}

Finally we wish to investigate the issue of positivity violations in
our two types of solutions.  The confinement of gluons by the absence
of a spectral representation of its propagator has been addressed e.g.
in \cite{gzwanziger2,von Smekal:1997vx,Alkofer:2003jj} with violation
of positivity being a sufficient criterion for the absence of gluons
from the physical part of the state space of QCD. For the scaling type
of solutions with an infrared vanishing gluon propagator these
violations can be shown analytically \cite{gzwanziger2}.  This is not
so for the decoupling type of solutions. Thus to investigate this
question one has to determine the Schwinger function
\begin{equation}
  \Delta(t)\; =\; \int d^3x \int \frac{d^4p}{(2\pi)^4} \exp(i p \cdot x)
  D(p^2),
\end{equation}
numerically. Here $D(p^2)=Z(p^2)/p^2$ is the gluon propagator
function.  Our results for both type of solutions are shown in
Fig.~\ref{res:FFT}. Apart from some variations in the scale this type
of behavior can be seen for all our decoupling solutions, i.e.  it is
independent of the value of $G(0)$. This statement, however, may
depend on the truncation scheme.

From Fig.~\ref{res:FFT} it is plainly visible that, despite
the differences in momentum space, the Schwinger function for both
solutions is very similar. In particular, in both cases positivity is
violated, and gluons are confined. This occurs for both cases at about
the same scale of 1 fm, typical for the size of bound states.

Furthermore, both results for the Schwinger function
are in qualitative agreement with corresponding results from
lattice calculations \cite{Bowman:2007du,Cucchieri:2004mf}.  Hence,
despite the completely different status with respect to the Kugo-Ojima
framework, both solutions do not describe propagating gluons.

In particular, the gluon is {\it not} 
characterized by a pole mass and an exponential decrease of 
a (positive) Schwinger function at large times. This implies 
that the mass parameter defined in \pref{masseq} is also not a 
pole mass in the ordinary sense but at best a screening mass. 
In addition, a gauge-independent mass in a sense defined 
in \cite{Nielsen:1975fs} cannot be constructed.
Hence, irrespective of its infrared properties, the gluon is 
never an ordinary massive particle.

\section{Summary and discussion \label{sec:sec6}}

We close with summarizing again some of our main results and
conclusions. More detailed discussions have been provided in the related 
chapters. 

We have obtained, in various truncations of DSEs and FRGs, a
one-parameter family of solutions for the ghost and gluon dressing
functions of Landau gauge Yang-Mills theory. We argued that
Slavnov-Taylor identities cannot be used to discriminate between these,
contrary to what has been claimed in \cite{Boucaud:2008ky}.
In particular, transversality is not indicative in this respect. 
Our numerical solutions for both, scaling and decoupling, are 
transverse on the level of numerical accuracy. 
However, global symmetries are indicative: exactly one member of this
family, the only one exhibiting scaling behavior, is consistent with
the existence of a globally well-defined BRST charge. The remaining 
solutions are of a decoupling type and are not BRST-symmetric.

The appearance of the above mentioned one-parameter family of 
solutions in the DSEs and FRGs can be achieved by implementing 
boundary conditions \cite{Lerche:2002ep,Boucaud:2008ky}.  
The scaling solution is in accordance with the Kugo-Ojima and 
Gribov-Zwanziger confinement scenarios and solely generated
by the gauge fixing sector of the theory. In case of the decoupling
solution the infrared dynamics is qualitatively different. 
We have argued that the latter type of dynamics
cannot be color-confining and preserving BRST symmetry at the same
time. Both solutions, however, satisfy the quark confinement criterion
put forward in \cite{Braun:2007bx}. 

Finally, we would like to emphasize once more that neither solution
corresponds to an ordinary massive gluon with a pole mass. Instead, in
both cases the Schwinger function shows the presence of positivity
violation and thus the gluon is not a free particle. 

In conclusion, each member of the one-parameter family of solutions
is confining.\\[1mm]

{\bf Acknowledgments}\\
We thank Reinhard Alkofer, Jens Braun, Joannis Papavassiliou, Olivier
Pene, Kai Schwenzer and Lorenz von Smekal for useful discussions.
A.~M. was supported by the FWF under grant number P20330, C.~F by the
Helmholtz-University Young Investigator Grant number VH-NG-332
and J.~M.~P. by the ExtreMe Matter Institute (EMMI).\\[1mm]

\begin{appendix}

  \section{Explicit expressions for the dressing of the three-gluon
    vertex and the integral kernels \label{kernels}}

The ansatz for the three-gluon vertex used in our truncation 
includes the dressing function \cite{Fischer:2002eq}
\begin{eqnarray}
\Gamma^{3g}(p,q)  = \frac{1}{Z_1}
\frac{[G(y+\Lambda_{dec}^2) \,G(z+\Lambda_{dec}^2)]^{(1-a/\delta-2a)}}
{[Z(y+\Lambda_{dec}^2)\,Z(z+\Lambda_{dec}^2)]^{(1+a)}}
\label{3g-vertex}
\end{eqnarray}
with $y=q^2$, $z=(p-q)^2$, the anomalous
dimension $\delta=-9/44$ of the ghost dressing function and the
renormalization constant $Z_1$ for the three-gluon vertex. The value
of $a$ is a free parameter which regulates the interaction strength of
the vertex. In \cite{Fischer:2002eq} $a=3\delta$ has been chosen. The scale
$\Lambda_{dec} \approx 500$ MeV is irrelevant for the scaling type of solutions
since the gluon-loop including the three-gluon vertex becomes sub-leading at 
such small scales compared to the ghost loop. For decoupling type of solutions
this scale implements decoupling within the three-gluon vertex. All results 
in this paper are qualitatively independent of $\Lambda_{dec}$. Quantitative 
changes in the decoupling solutions are at the percent level when varying 
this scale within reasonable bounds.

The kernels $\bar{K}$, $\bar{M}$ and $\bar{Q}$ for the
integral equations (\ref{gluon-dse2}) and (\ref{ghost-dse2}) are given
by
\begin{widetext}
\begin{eqnarray}
  \bar{M}(p^2,q^2,(p-q)^2) &=& M(p^2,q^2,(p-q)^2) \nonumber\\
  &&+\left( (p-q)^2 \left[ - \frac{1}{2q^2} - \frac{1}{2p^2} \right]
    +  \frac{(p-q)^4}{4p^2q^2} - \frac{1}{2} 
    + \frac{p^2}{4q^2} 
    + \frac{q^2}{4p^2}  \right)
  f_{UV}(p^2,q^2;k^2)  \,,\\
%\end{eqnarray}
%\end{widetext}
%\begin{widetext}
%\begin{eqnarray}
  \bar{Q}(p^2,q^2,(p-q)^2) &=& Q(p^2,q^2,(p-q)^2)   \nonumber\\
  &&\hspace*{-1cm}+ \left( \frac{1}{(p-q)^4} \left[
      \frac{39 q^8}{68 p^4}
      + \frac{236 q^6}{51p^2} 
      - \frac{1021 q^4}{102} 
      + \frac{66 p^2q^2}{17} 
      + \frac{181 p^4}{204} 
      + \frac{2 p^6}{51 q^2}  
    \right]\right. \nonumber\\
  &&\hspace*{-1cm}+ \frac{1}{(p-q)^2} \left[ 
    + \frac{65 q^6}{17 p^4}
    - \frac{9923 q^4}{408 p^2} 
    - \frac{365 q^2}{51}  
    + \frac{35 p^2}{204}  
    - \frac{30 p^4}{17 q^2} 
    - \frac{67 p^6}{408 q^4} 
  \right]\nonumber\\
  &&\hspace*{-1cm}+ (p-q)^2 \left[ 
    \frac{5347}{204 p^2} 
    + \frac{674}{51 q^2} 
    + \frac{338 q^2}{17 p^4} 
    + \frac{305 p^2}{68 q^4} 
  \right]\nonumber\\
  &&\hspace*{-1cm}+ (p-q)^4 \left[ 
    - \frac{689}{51 p^2 q^2}
    - \frac{485}{102 q^4} 
    - \frac{611}{68 p^4}\right]
  + (p-q)^6 \left[ 
    +\frac{557}{408 p^2q^4} 
    +\frac{13}{17 p^4 q^2}  \right]
  + (p-q)^8  \frac{13}{68 p^4q^4} \nonumber\\
  &&\hspace*{-1cm}\left.
    + \frac{250}{17} 
    + \frac{287 q^2}{51 p^2} 
    - \frac{2 p^2}{51 q^2} 
    - \frac{229 p^4}{204 q^4} 
    - \frac{65 q^4}{4 p^4}   
  \right)
  f_{UV}(p^2,q^2;k^2)\\
  \bar{K}(p^2,q^2,(p-q)^2) &=& K(p^2,q^2,(p-q)^2) \nonumber\\
  &&+\left( \frac{1}{(p-q)^4} \left[ \frac{p^2 q^2}{4} - \frac{p^4}{2} +
      \frac{p^6}{4q^2}\right] - \frac{1}{(p-q)^2} \left[\frac{p^2}{2} +
      \frac{p^4}{2q^2} \right]
    +  \frac{p^2}{4q^2}\right)f_{UV}(p^2,q^2;k^2) \,,
\end{eqnarray}
\end{widetext}
with the original kernels  
$K$, $M$ and $Q$ 
of the truncation scheme of ref.~\cite{Fischer:2002eq,Fischer:2003zc}:
\begin{widetext}
\begin{eqnarray}
  K(p^2,q^2,(p-q)^2) &=& \frac{1}{(p-q)^4}\left(-
    \frac{(p^2-q^2)^2}{4}\right) + 
  \frac{1}{(p-q)^2}\left(\frac{p^2+q^2}{2}\right)-\frac{1}{4} \,,
  \nonumber\\
  M(p^2,q^2,(p-q)^2) &=& \frac{1}{(p-q)^2} \left( \frac{-1}{4}p^2 + 
    \frac{q^2}{2} - \frac{1}{4}\frac{q^4}{p^2}\right)
  +\frac{1}{2} + \frac{1}{2}\frac{q^2}{p^2} - 
  \frac{1}{4}\frac{(p-q)^2}{p^2} \,, \nonumber\\
  Q(p^2,q^2,(p-q)^2) &=& \frac{1}{(p-q)^4} 
  \left( \frac{1}{8}\frac{p^6}{q^2} + p^4 -\frac{18}{8}p^2q^2 + q^4
    +\frac{1}{8}\frac{q^6}{p^2} \right)\nonumber\\
  && +\frac{1}{(p-q)^2} \left( \frac{p^4}{q^2} - {4}p^2-
    {4}q^2+\frac{q^4}{p^2}\right)\nonumber\\
  && - \left( \frac{9}{4}\frac{p^2}{q^2}+{4}+
    \frac{9}{4}\frac{q^2}{p^2} \right) 
  + (p-q)^2\left(\frac{1}{p^2}+\frac{1}{q^2}\right) + (p-q)^4
  \frac{1}{8p^2q^2} \,.
  \hspace*{0.5cm} \label{new_kernels}
\end{eqnarray}
\end{widetext}

\section{Multiplicative renormalizability of the ghost DSE \label{MR}}

Multiplicative renormalizability of the ghost and gluon dressing
functions implies the relations \begin{eqnarray} G(p^2,\mu^2)
  \widetilde{Z}_3(\mu^2,\Lambda^2) &=&
  G^0(p^2,\Lambda^2) \nonumber\\
  Z(p^2,\mu^2) Z_3(\mu^2,\Lambda^2) &=&
  Z^0(p^2,\Lambda^2) \nonumber\\
  g(\mu^2) Z_g(\mu^2,\Lambda^2) &=&
  g^0(\Lambda^2) \nonumber\\
  \Gamma^{(2,1)}_\mu(p,q,\mu^2) \widetilde{Z}_1^{-1}(\mu^2,\Lambda^2)&=&
  \Gamma_\mu^{(2,1)0}(p,q,\Lambda^2) \label{eq1} \end{eqnarray} between the
bare dressing functions depending on a UV-cutoff $\Lambda$ and denoted
with superscript zero and the renormalized dressing functions depending
on the renormalization point $\mu^2$.

Note that the right hand sides of eqs.\ (\ref{eq1}) are independent of
the renormalization point. Thus a finite re-normalization, i.\ e., a
change in the renormalization point from $\mu^2$ to $\nu^2$ is
described by
\begin{eqnarray}
G(p^2,\nu^2)  &=& G(p^2,\mu^2) 
\frac{\widetilde{Z}_3(\mu^2,\Lambda^2)}{\widetilde{Z}_3(\nu^2,\Lambda^2)} 
							\nonumber\\
Z(p^2,\nu^2)  &=& Z(p^2,\mu^2) 
\frac{{Z}_3(\mu^2,\Lambda^2)}{{Z}_3(\nu^2,\Lambda^2)} 	\nonumber\\
g(\nu^2)  &=& g(\mu^2) 
\frac{{Z}_g(\mu^2,\Lambda^2)}{{Z}_g(\nu^2,\Lambda^2)} 	\nonumber\\
\Gamma^{(2,1)}_\mu(p,q,\nu^2) &=& \Gamma^{(2,1)}_\mu(p,q,\mu^2) 
\frac{\widetilde{Z}_1(\nu^2,\Lambda^2)}{\widetilde{Z}_1(\mu^2,\Lambda^2)}.
 \label{eq2}
\end{eqnarray}
Furthermore we need the identity
\begin{equation}
\widetilde{Z}_1 = Z_g Z_3^{1/2} \widetilde{Z}_3.
\end{equation}

Now let us consider the Dyson-Schwinger equation for the ghost
propagator
\begin{widetext}
\begin{equation}
\frac{1}{G(p^2,\mu^2)} = \widetilde{Z}_3(\mu^2,\Lambda^2) 
- \widetilde{Z}_1(\mu^2,\Lambda^2) g^2(\mu^2)
\int^\Lambda \frac{d^4q}{(2\pi)^4} \frac{G(q^2,\mu^2)}{q^2} 
\frac{Z(k^2,\mu^2)}{k^2} \gamma_\mu P^T_{\mu \nu}(k) \Gamma^{(2,1)}_\nu(p,q,\mu^2)
\label{ghost1}
\end{equation} 
\end{widetext}
where $P^T_{\mu \nu}(k) = \left( \delta_{\mu \nu} - \frac{k_\mu
    k_\nu}{k^2}\right)$ denotes the transverse projector and $k_\mu =
q_\mu - p_\mu$. This equation is invariant under a change of the
renormalization point, since we have 
\begin{widetext}
\begin{eqnarray} 
\frac{1}{G(p^2,\nu^2)} &=&
\frac{1}{G(p^2,\mu^2)}
\frac{\widetilde{Z}_3(\nu^2,\Lambda^2)}{\widetilde{Z}_3(\mu^2,\Lambda^2)}\\
&=& \widetilde{Z}_3(\nu^2,\Lambda^2) -
\frac{\widetilde{Z}_3(\nu^2,\Lambda^2)}{\widetilde{Z}_3(\mu^2,\Lambda^2)}
\widetilde{Z}_1(\mu^2,\Lambda^2)
\frac{g^2(\nu^2){Z}_g^2(\nu^2,\Lambda^2)}{{Z}_g^2(\mu^2,\Lambda^2)} \nonumber\\
&&\hspace*{5mm} \int^\Lambda \frac{d^4q}{(2\pi)^4} \frac{G(q^2,\nu^2)
  \widetilde{Z}_3(\nu^2,\Lambda^2)}
{q^2\widetilde{Z}_3(\mu^2,\Lambda^2)}
\frac{Z(k^2,\nu^2){Z}_3(\nu^2,\Lambda^2)} {k^2{Z}_3(\mu^2,\Lambda^2)}
\gamma_\mu P^T_{\mu \nu}(k) \Gamma^{(2,1)}_\nu(p,q,\nu^2)
\frac{\widetilde{Z}_1(\mu^2,\Lambda^2)}{\widetilde{Z}_1(\nu^2,\Lambda^2)}\\
&=& \widetilde{Z}_3(\nu^2,\Lambda^2) -
\widetilde{Z}_1(\nu^2,\Lambda^2)g^2(\nu^2) \int^\Lambda
\frac{d^4q}{(2\pi)^4} \frac{G(q^2,\nu^2) }{q^2}
\frac{Z(k^2,\nu^2)}{k^2} \gamma_\mu P^T_{\mu \nu}(k)
\Gamma^{(2,1)}_\nu(p,q,\nu^2) \nonumber\\
&&\hspace*{5mm}
\frac{\widetilde{Z}_1^2(\mu^2,\Lambda^2)}{\widetilde{Z}_1^2(\nu^2,\Lambda^2)}
\frac{{Z}_g^2(\nu^2,\Lambda^2)
  \widetilde{Z}_3^2(\nu^2,\Lambda^2){Z}_3(\nu^2,\Lambda^2)}
{{Z}_g^2(\mu^2,\Lambda^2)\widetilde{Z}_3^2(\mu^2,\Lambda^2){Z}_3(\mu^2,\Lambda^2)}\\
&=& \widetilde{Z}_3(\nu^2,\Lambda^2) -
\widetilde{Z}_1(\nu^2,\Lambda^2)g^2(\nu^2) \int^\Lambda
\frac{d^4q}{(2\pi)^4} \gamma_\mu \frac{G(q^2,\nu^2) }{q^2}
\frac{Z(k^2,\mu^2)}{k^2} P^T_{\mu \nu}(k) \Gamma^{(2,1)}_\nu(p,q,\nu^2)
\label{ghost2} 
\end{eqnarray} 
\end{widetext}
This demonstrates explicitly that a solution $G(p^2,\mu^2)$ of eq.\
(\ref{ghost1}) and a solution $G(p^2,\nu^2)$ of eq.\ (\ref{ghost2})
are uniquely related by a multiplicative factor
$\frac{\widetilde{Z}_3(\mu^2,\Lambda^2)}{\widetilde{Z}_3(\nu^2,\Lambda^2)}$.
Thus a change of $g^2$, corresponding to $g(\mu^2) \rightarrow
g(\nu^2)$ with $\mu^2 \neq \nu^2$ must not generate a qualitatively
different solution.

\begin{figure}[t]
\includegraphics[width=\columnwidth]{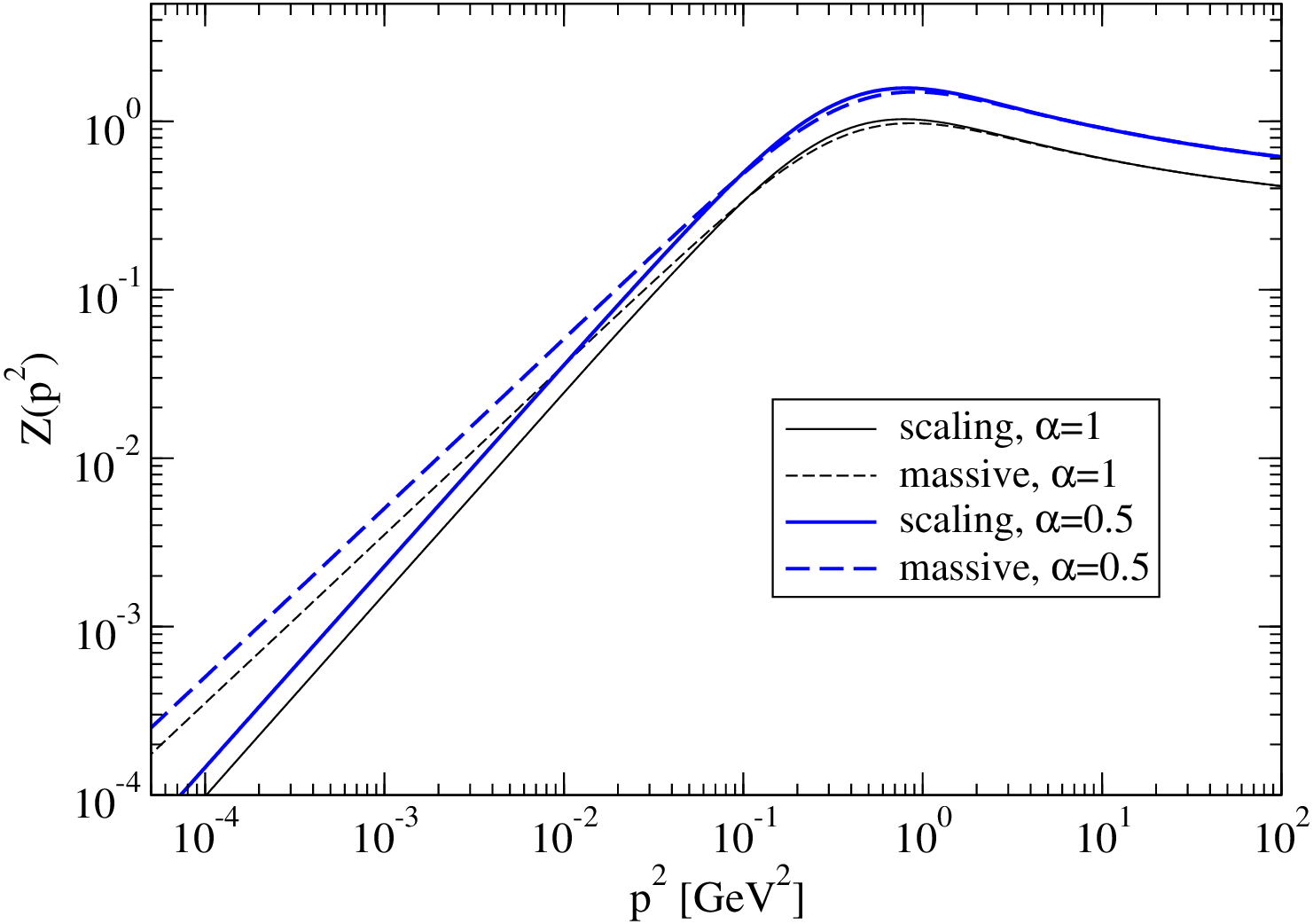}
\includegraphics[width=\columnwidth]{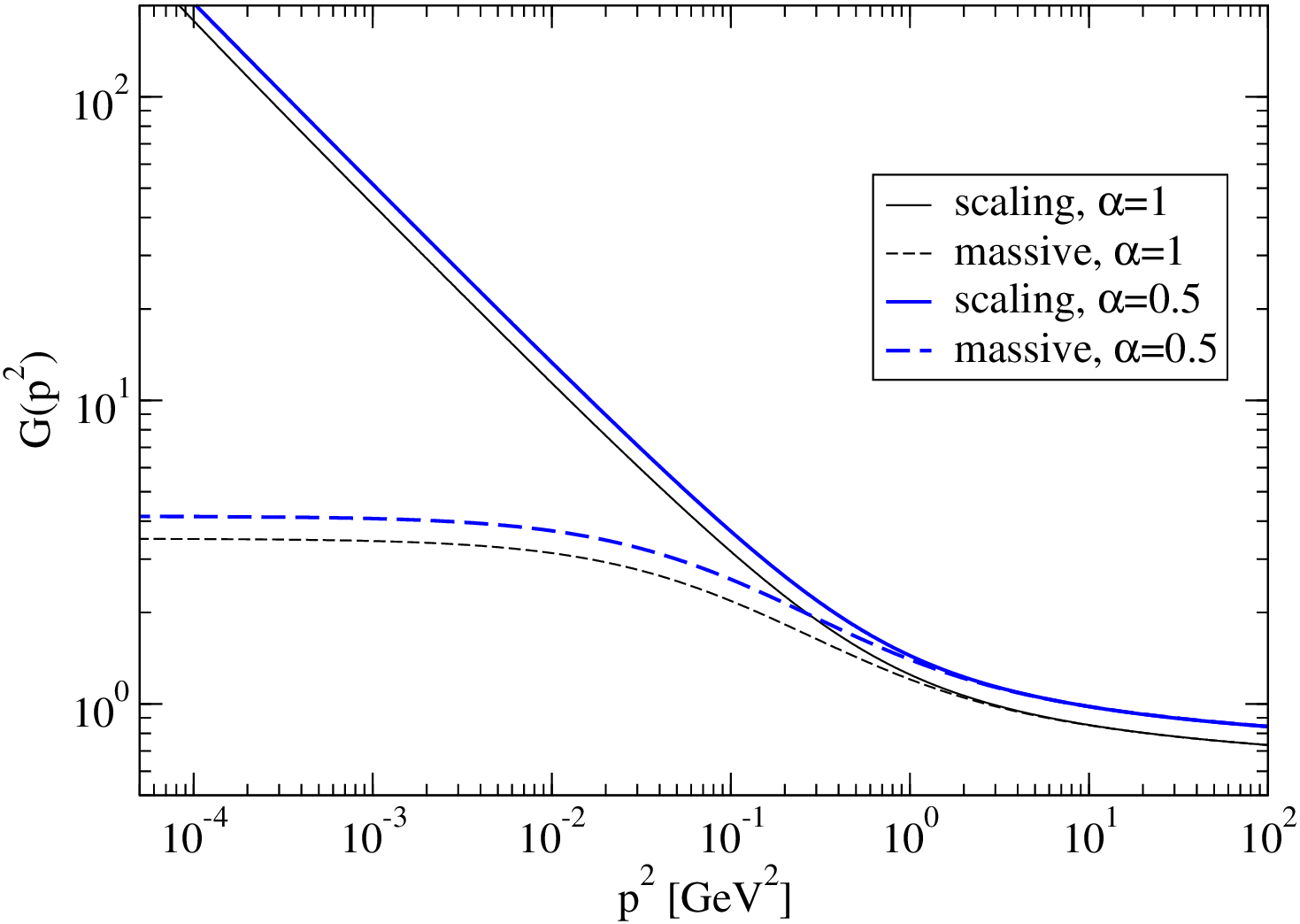}
\includegraphics[width=\columnwidth]{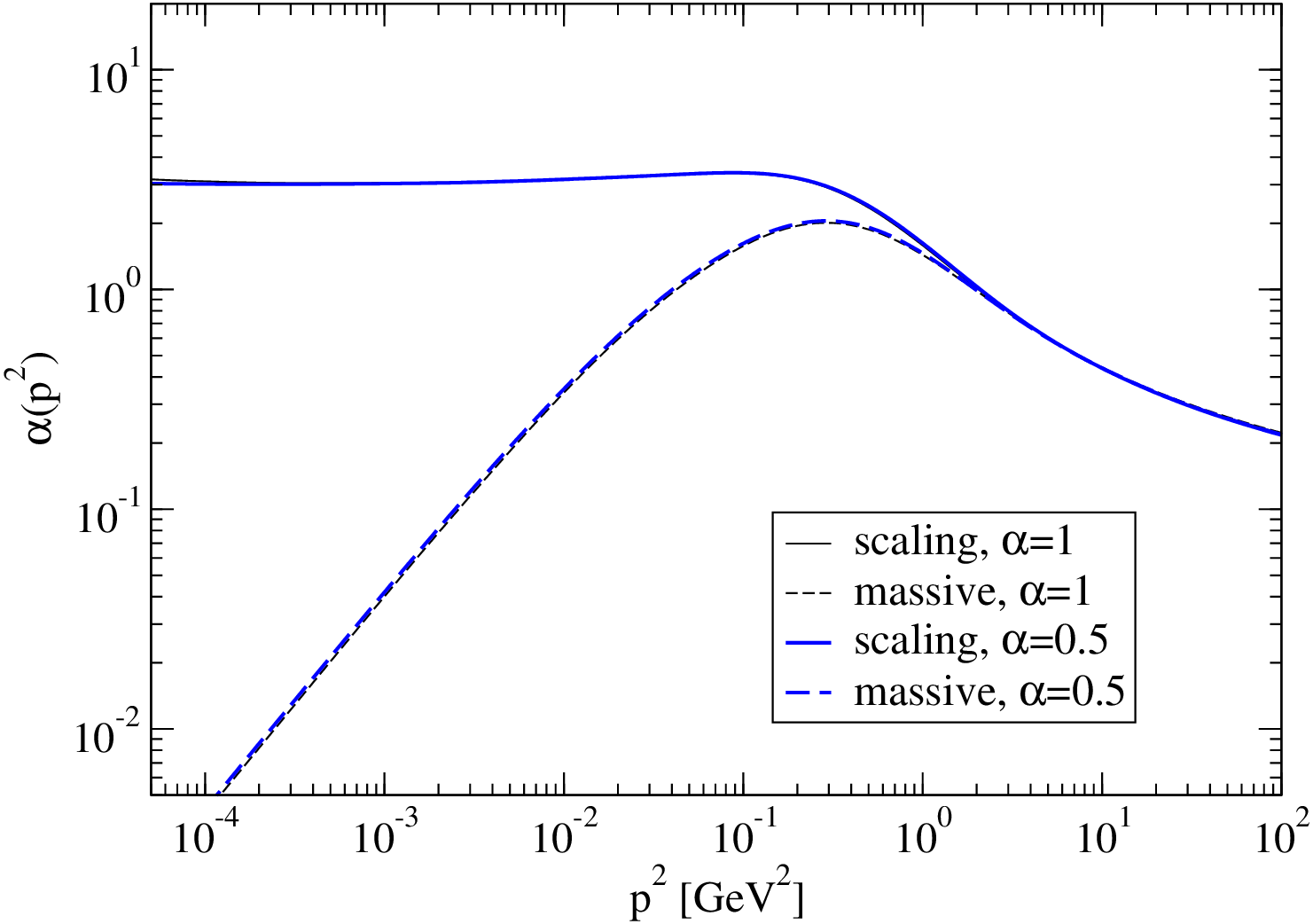}
\caption{Numerical results for the ghost and gluon dressing function
  as well as the running coupling for two different renormalization
  points.}
\label{mrfig}
\end{figure}

Since our truncation respects multiplicative renormalizability this
requirement is certainly respected. This is explicitly demonstrated in
Fig.~\ref{mrfig}, where we compare our results for $\alpha(\mu^2)=1$
and $\alpha(\mu^2)=0.5$. The resulting running coupling is clearly an
RG-invariant, whereas the ghost and gluon dressing functions are
related by finite re-normalization factors according to
eq.~(\ref{eq2}), as they should.

\end{appendix}


\newpage
\begin{thebibliography}{99}

%%%%% DSEs 1 %%%%%

%\cite{von Smekal:1997vx}
\bibitem{von Smekal:1997vx}
  L.~von Smekal, A.~Hauck and R.~Alkofer,
  %``A solution to coupled Dyson-Schwinger equations for gluons and ghosts  in
  %Landau gauge,''
  Annals Phys.\  {\bf 267} (1998) 1
  [Erratum-ibid.\  {\bf 269} (1998) 182]
  [arXiv:hep-ph/9707327];
  %%CITATION = APNYA,267,1;%%
%\cite{von Smekal:1997is}
%\bibitem{von Smekal:1997is}
  L.~von Smekal, R.~Alkofer and A.~Hauck,
  %``The infrared behavior of gluon and ghost propagators in Landau gauge
  %QCD,''
  Phys.\ Rev.\ Lett.\  {\bf 79} (1997) 3591
  [arXiv:hep-ph/9705242].
  %%CITATION = PRLTA,79,3591;%%


\bibitem{Alkofer:2000wg}
  R.\ Alkofer and L.\ von Smekal,
  %``The infrared behavior of QCD Green's functions: Confinement, dynamical
  %symmetry breaking, and hadrons as relativistic bound states,''
  Phys.\ Rept.\ {\bf 353} (2001) 281
  [arXiv:hep-ph/0007355].
  %%CITATION = PRPLC,353,281;%%

\bibitem{Watson:2001yv}
  P.~Watson and R.~Alkofer,
  %``Verifying the Kugo-Ojima confinement criterion in Landau gauge QCD,''
  PRL {\bf 86} (2001) 5239
  [hep-ph/0102332].
  %%CITATION = HEP-PH 0102332;%%

\bibitem{Lerche:2002ep}
  C.~Lerche and L.~von Smekal,
  %``On the infrared exponent for gluon and ghost propagation in Landau  gauge
  %QCD,''
  Phys.\ Rev.\  D {\bf 65}, 125006 (2002)
  [arXiv:hep-ph/0202194].
  %%CITATION = PHRVA,D65,125006;%%

\bibitem{Fischer:2002eq}
  C.~S.~Fischer and R.~Alkofer,
  %``Infrared exponents and running coupling of SU(N) Yang-Mills theories,''
  Phys.\ Lett.\  B {\bf 536} (2002) 177
  [arXiv:hep-ph/0202202];
  %%CITATION = PHLTA,B536,177;%%
  C.~S.~Fischer, R.~Alkofer and H.~Reinhardt,
  %``The elusiveness of infrared critical exponents in Landau gauge  Yang-Mills
  %theories,''
  Phys.\ Rev.\  D {\bf 65} (2002) 094008
  [arXiv:hep-ph/0202195].
  %%CITATION = PHRVA,D65,094008;%%

\bibitem{Alkofer:2004it}
  R.\ Alkofer, C.\ S.\ Fischer and F.\ J.\ Llanes-Estrada,
  %``Vertex functions and infrared fixed point in Landau gauge SU(N)  Yang-Mills
  %theory,''
  Phys.\ Lett.\ B {\bf 611}, 279 (2005)
  [arXiv:hep-th/0412330].
  %%CITATION = PHLTA,B611,279;%%

\bibitem{Schleifenbaum:2004id}
  W.\ Schleifenbaum, A.\ Maas, J.\ Wambach and R.\ Alkofer,
  %``Infrared behaviour of the ghost gluon vertex in Landau gauge Yang-Mills
  %theory,''
  Phys.\ Rev.\ D {\bf 72}, 014017 (2005)
  [arXiv:hep-ph/0411052].
  %%CITATION = PHRVA,D72,014017;%%

\bibitem{Fischer:2006ub}
  C.\ S.\ Fischer,
  %``Infrared properties of QCD from Dyson-Schwinger equations,''
  J.\ Phys.\ G {\bf 32}, R253 (2006)
  [arXiv:hep-ph/0605173].
  %%CITATION = JPHGB,G32,R253;%%

%\cite{Schleifenbaum:2006bq}
\bibitem{Schleifenbaum:2006bq}
  W.~Schleifenbaum, M.~Leder and H.~Reinhardt,
  %``Infrared analysis of propagators and vertices of Yang-Mills theory in
  %Landau and Coulomb gauge,''
  Phys.\ Rev.\  D {\bf 73} (2006) 125019
  [arXiv:hep-th/0605115].
  %%CITATION = PHRVA,D73,125019;%%

\bibitem{Fischer:2006vf}
  C.\ S.\ Fischer and J.\ M.\ Pawlowski,
  %``Uniqueness of infrared asymptotics in Landau gauge Yang-Mills theory,''
  Phys.\ Rev.\ D {\bf 75}, 025012 (2007)
  [arXiv:hep-th/0609009]; 
  %%CITATION = PHRVA,D75,025012;%%
%\cite{Fischer:2009tn}
%\bibitem{Fischer:2009tn}
%  C.~S.~Fischer and J.~M.~Pawlowski,
  %``Uniqueness of infrared asymptotics in Landau gauge Yang-Mills theory II,''
 Phys.\ Rev.\ D in press, arXiv:0903.2193 [hep-th].
  %%CITATION = ARXIV:0903.2193;%%



\bibitem{Alkofer:2008jy}
  R.\ Alkofer, M.\ Q.\ Huber and K.\ Schwenzer,
  %``On infrared singularities in Landau gauge Yang-Mills theory,''
  arXiv:0801.2762 [hep-th]; 
  %%CITATION = ARXIV:0801.2762;%%
%\cite{Huber:2009wh}
%\bibitem{Huber:2009wh}
%  M.~Q.~Huber, K.~Schwenzer and R.~Alkofer,
  %``On the infrared scaling solution of SU(N) Yang-Mills theories in the
  %maximally Abelian gauge,''
  arXiv:0904.1873 [hep-th].
  %%CITATION = ARXIV:0904.1873;%%



\bibitem{Kellermann:2008iw}
  C.~Kellermann and C.~S.~Fischer,
  %``The running coupling from the four-gluon vertex in Landau gauge Yang-Mills
  %theory,''
  Phys.\ Rev.\  D {\bf 78}, 025015 (2008)
  [arXiv:0801.2697 [hep-ph]].

%\cite{Alkofer:2008dt}
\bibitem{Alkofer:2008dt}
  R.~Alkofer, M.~Q.~Huber and K.~Schwenzer,
  %``Infrared Behavior of Three-Point Functions in Landau Gauge Yang-Mills
  %Theory,''
  arXiv:0812.4045 [hep-ph].
  %%CITATION = ARXIV:0812.4045;%%



%%%%% DSEs 2 %%%%%
%\cite{Cornwall:1981zr}
\bibitem{Cornwall:1981zr}
  J.~M.~Cornwall,
  %``Dynamical Mass Generation In Continuum QCD,''
  Phys.\ Rev.\  D {\bf 26} (1982) 1453.
  %%CITATION = PHRVA,D26,1453;%%

\bibitem{Aguilar:2007nf}
  A.\ C.\ Aguilar and J.\ Papavassiliou,
  %``Infrared finite ghost propagator in the Feynman gauge,''
  arXiv:0712.0780 [hep-ph].
  %%CITATION = ARXIV:0712.0780;%%
  
%\cite{Binosi:2007pi}
\bibitem{Binosi:2007pi}
  D.~Binosi and J.~Papavassiliou,
  %``Gauge-invariant truncation scheme for the Schwinger-Dyson equations of
  %QCD,''
  Phys.\ Rev.\  D {\bf 77} (2008) 061702
  [arXiv:0712.2707 [hep-ph]].
  %%CITATION = PHRVA,D77,061702;%%

\bibitem{Aguilar:2008xm}  
  A.~C.~Aguilar, D.~Binosi and J.~Papavassiliou,
  %``Gluon and ghost propagators in the Landau gauge: Deriving lattice results
  %from Schwinger-Dyson equations,''
  Phys.\ Rev.\  D {\bf 78}, 025010 (2008)
  [arXiv:0802.1870 [hep-ph]].
  %%CITATION = ARXIV:0802.1870;%%

%%%%% DSEs 3 %%%%%

\bibitem{Boucaud:2006if}
  Ph.\ Boucaud, Th.\ Bruntjen, J.\ P.\ Leroy, A.\ Le Yaouanc, A.\ Y.\ Lokhov,
  J.\ Micheli, O.\ Pene and J.\ Rodriguez-Quintero,
  %``Is the QCD ghost dressing function finite at zero momentum?,''
  JHEP {\bf 0606}, 001 (2006)
  [arXiv:hep-ph/0604 056].
  %%CITATION = JHEPA,0606,001;%%

\bibitem{Boucaud:2008ji}
  Ph.~Boucaud, J.~P.~Leroy, A.~L.~Yaouanc, J.~Micheli, O.~Pene and J.~Rodriguez-Quintero,
  %``IR finiteness of the ghost dressing function from numerical resolution of
  %the ghost SD equation,''
  JHEP {\bf 0806}, 012 (2008)
  [arXiv:0801.2721 [hep-ph]].
  %%CITATION = JHEPA,0806,012;%%

%\cite{Boucaud:2008ky}
\bibitem{Boucaud:2008ky}  
  Ph.~Boucaud, J.~P.~Leroy, A.~Le Yaouanc, J.~Micheli, O.~Pene and J.~Rodriguez-Quintero,
  %``On the IR behaviour of the Landau-gauge ghost propagator,''
  JHEP {\bf 0806}, 099 (2008)
  [arXiv:0803.2161 [hep-ph]].
  %%CITATION = JHEPA,0806,099;%%

%%%%% DSEs 4 %%%%%

%\cite{Bloch:2002eq}
\bibitem{Bloch:2002eq}
  J.~C.~R.~Bloch,
  %``Multiplicative renormalizability and quark propagator,''
  Phys.\ Rev.\  D {\bf 66} (2002) 034032
  [arXiv:hep-ph/0202073].
  %%CITATION = PHRVA,D66,034032;%%




%%%%% RG %%%%%


%\cite{Wetterich:1992yh}
\bibitem{Wetterich:1992yh}
  C.~Wetterich,
   %``Exact evolution equation for the effective potential,''
  %
  Phys.\ Lett.\ B {\bf 301} (1993) 90.
  %%CITATION = PHLTA,B301,90;%%


%\cite{Litim:1998nf}
\bibitem{Litim:1998nf} D.~F.~Litim and J.~M.~Pawlowski, in {\it The Exact
  Renormalization Group}, Eds.~Krasnitz et al, World Sci (1999) 168. 
   %``On gauge invariant Wilsonian flows,''
  %
[hep-th/9901063].
  %%CITATION = HEP-TH 9901063;%%



%\cite{Pawlowski:2005xe}
\bibitem{Pawlowski:2005xe}
  J.~M.~Pawlowski,
  %``Aspects of the functional renormalisation group,''
  Annals Phys.\  {\bf 322} (2007) 2831. 
  [arXiv:hep-th/0512261].
  %%CITATION = APNYA,322,2831;%%

%\cite{Gies:2006wv}
\bibitem{Gies:2006wv}
  H.~Gies,
  %``Introduction to the functional RG and applications to gauge theories,''
  arXiv:hep-ph/0611146.
  %%CITATION = HEP-PH 0611146;%%

%\cite{Ellwanger:1994iz}
\bibitem{Ellwanger:1994iz}
  U.~Ellwanger,
%   ``Flow equations and BRS invariance for Yang-Mills theories,''
  %
  Phys.\ Lett.\ B {\bf 335} (1994) 364 
  [arXiv:hep-th/9402077].
  %%CITATION = HEP-TH 9402077;%%


%\cite{Bonini:1994kp}
\bibitem{Bonini:1994kp}
  M.~Bonini, M.~D'Attanasio and G.~Marchesini,
  %``BRS symmetry for Yang-Mills theory with exact renormalization group,''
  Nucl.\ Phys.\  B {\bf 437} (1995) 163
  [arXiv:hep-th/9410138].
  %%CITATION = NUPHA,B437,163;%%




%\cite{Ellwanger:1995qf}
\bibitem{Ellwanger:1995qf}
  U.~Ellwanger, M.~Hirsch and A.~Weber,
%   ``Flow equations for the relevant part of the pure Yang-Mills action,''
  %
  Z.\ Phys.\ C {\bf 69} (1996) 687 
  [arXiv:hep-th/9506019]; 
  %%CITATION = HEP-TH 9506019;%%
%\cite{Ellwanger:1996wy}
%\bibitem{Ellwanger:1996wy}
%  U.~Ellwanger, M.~Hirsch and A.~Weber,
%   ``The heavy quark potential from Wilson's exact renormalization group,''
  %
  Eur.\ Phys.\ J.\ C {\bf 1} (1998) 563 
  [arXiv:hep-ph/9606468]; 
  %%CITATION = HEP-PH 9606468;%%
%\cite{Ellwanger:1997tp}
%\bibitem{Ellwanger:1997tp}
  U.~Ellwanger, 
  %``The running gauge coupling in the exact renormalization group approach,''
  Z.\ Phys.\ C {\bf 76} (1997) 721 
  [arXiv:hep-ph/9702309]; 
  %%CITATION = HEP-PH 9702309;%%
%\cite{Bergerhoff:1997cv}
%\bibitem{Bergerhoff:1997cv}
  B.~Bergerhoff and C.~Wetterich,
%   ``Effective quark interactions and QCD propagators,''
  %
  Phys.\ Rev.\ D {\bf 57} (1998) 1591 
  [arXiv:hep-ph/9708425].
  %%CITATION = HEP-PH 9708425;%%

%\cite{Pawlowski:2003hq}
\bibitem{Pawlowski:2003hq}
  J.~M.~Pawlowski, D.~F.~Litim, S.~Nedelko and L.~von Smekal,
   %``Infrared behaviour and fixed points in Landau gauge QCD,''
  %
  Phys.\ Rev.\ Lett.\  {\bf 93} (2004) 152002 
  [arXiv:hep-th/0312324].
  %%CITATION = HEP-TH 0312324;%%

%\cite{Pawlowski:2004ip}
\bibitem{Pawlowski:2004ip}
  J.~M.~Pawlowski, D.~F.~Litim, S.~Nedelko and L.~von Smekal,
  %%``Signatures of confinement in Landau gauge QCD,''
  AIP Conf.\ Proc.\  {\bf 756} (2005) 278 
  [arXiv:hep-th/0412326].
  %%CITATION = HEP-TH 0412326;%%

%\cite{Kato:2004ry}
\bibitem{Kato:2004ry}
  J.~Kato,
  %``Infrared non-perturbative propagators of gluon and ghost via exact
  %renormalization group,''
  arXiv:hep-th/0401068.
  %%CITATION = HEP-TH/0401068;%%



%\cite{Fischer:2004uk}
\bibitem{Fischer:2004uk}
  C.~S.~Fischer and H.~Gies,
%   ``Renormalization flow of Yang-Mills propagators,''
  %
  JHEP {\bf 0410} (2004) 048 
  [arXiv:hep-ph/0408089].
  %%CITATION = HEP-PH 0408089;%%

%\cite{Braun:2007bx}
\bibitem{Braun:2007bx}
  J.~Braun, H.~Gies and J.~M.~Pawlowski,
  %``Quark Confinement from Color Confinement,''
  arXiv:0708.2413 [hep-th].
  %%CITATION = ARXIV:0708.2413;%%


%\cite{Litim:2000ci}
\bibitem{Litim:2000ci}
  D.~F.~Litim,
   %``Optimisation of the exact renormalisation group,''
  %
  Phys.\ Lett.\ B {\bf 486} (2000) 92  
  [arXiv:hep-th/0005245].
  %%CITATION = HEP-TH 0005245;%%
%\cite{Litim:2001up}
%\bibitem{Litim:2001up}
%  D.~F.~Litim,
   %``Optimised renormalisation group flows,''
  %
  Phys.\ Rev.\ D {\bf 64} (2001) 105007  
  [hep-th/0103195].
  %%CITATION = HEP-TH 0103195;%%
%\cite{Litim:2001fd}
%\bibitem{Litim:2001fd}
%  D.~F.~Litim,
   %``Mind the gap,''
  %
  Int.\ J.\ Mod.\ Phys.\ A {\bf 16} (2001) 2081 
  [arXiv:hep-th/0104221].
  %%CITATION = HEP-TH 0104221;%%



%%%%% SQ %%%%%

\bibitem{Zwanziger:2003cf}
  D.\ Zwanziger,
  %``Non-perturbative Faddeev-Popov formula and infrared limit of QCD,''
  Phys.\ Rev.\ D {\bf 69}, 016002 (2004)
  [arXiv: hep-ph/0303028].
  %%CITATION = HEP-PH 0303028;%%

\bibitem{gzwanziger}
   D.\ Zwanziger,
  %``Non-perturbative Landau gauge and infrared critical exponents in QCD,''
  Phys.\ Rev.\ D {\bf 65}, 094039 (2002) [arXiv: hep-th/0109224];
  %%CITATION = HEP-TH 0109224;%%
  % D.\ Zwanziger,
  %``Time-independent stochastic quantization, DS equations, and infrared
  %critical exponents in QCD,''
  Phys.\ Rev.\ D {\bf 67}, 105001 (2003) [arXiv: hep-th/0206053].
  %%CITATION = HEP-TH 0206053;%%

%%%%% EA %%%%%

\bibitem{Dudal:2005na}
  D.\ Dudal, R.\ F.\ Sobreiro, S.\ P.\ Sorella and H.\ Verschelde,
  %``The Gribov parameter and the dimension two gluon condensate in  Euclidean
  %Yang-Mills theories in the Landau gauge,''
  Phys.\ Rev.\ D {\bf 72}, 014016 (2005)
  [arXiv:hep-th/0502183].
  %%CITATION = PHRVA,D72,014016;%%

\bibitem{Dudal:2007cw}
  D.\ Dudal, S.\ P.\ Sorella, N.\ Vandersickel and H.\ Verschelde,
  %``New features of the gluon and ghost propagator in the infrared region from
  %the Gribov-Zwanziger approach,''
  arXiv:0711.4496 [hep-th].
  %%CITATION = ARXIV:0711.4496;%%

\bibitem{Capri:2007ix}
  M.~A.~L.~Capri, D.~Dudal, V.~E.~R.~Lemes, R.~F.~Sobreiro, S.~P.~Sorella, R.~Thibes and H.~Verschelde,
  %``The Gribov-Zwanziger action in the presence of the gauge invariant,
  %nonlocal mass operator $Tr \int d^4x F_{\mu\nu} (D^2)^{-1} F_{\mu\nu}$ in the
  %Landau gauge,''
  Eur.\ Phys.\ J.\  C {\bf 52}, 459 (2007)
  [arXiv:0705.3591 [hep-th]].
  %%CITATION = EPHJA,C52,459;%%

%\cite{Dudal:2008sp}
\bibitem{Dudal:2008sp}
  D.~Dudal, J.~A.~Gracey, S.~P.~Sorella, N.~Vandersickel and H.~Verschelde,
  %``A refinement of the Gribov-Zwanziger approach in the Landau gauge: infrared
  %propagators in harmony with the lattice results,''
  arXiv:0806.4348 [hep-th].
  %%CITATION = ARXIV:0806.4348;%%
%%%%% Lattice %%%%%

\bibitem{Oliveira:2007dy}
  O.\ Oliveira and P.\ J.\ Silva,
  %``Infrared gluon and ghost propagators exponents from lattice QCD,''
  arXiv:0705.0964 [hep-lat];
  %%CITATION = ARXIV:0705.0964;%% 
%\cite{Oliveira:2008uf}
%\bibitem{Oliveira:2008uf}
  %O.~Oliveira and P.~J.~Silva,
  %``Does The Lattice Zero Momentum Gluon Propagator for Pure Gauge SU(3)
  %Yang-Mills Theory Vanishes in the Infinite Volume Limit?,''
  arXiv:0809.0258 [hep-lat].
  %%CITATION = ARXIV:0809.0258;%%

\bibitem{4d}
%\bibitem{Bloch:2003sk}
  J.\ C.\ R.\ Bloch, A.\ Cucchieri, K.\ Langfeld and T.\ Mendes,
  %``Propagators and running coupling from SU(2) lattice gauge theory,''
  Nucl.\ Phys.\ B {\bf 687}, 76 (2004)
  [arXiv:hep-lat/0312036];
  %%CITATION = NUPHA,B687,76;%%
  
\bibitem{4d2}
  A.\ Sternbeck, E.\ M.\ Ilgenfritz, M.\ Mueller-Preussker and A.\ Schiller,
  %``Going infrared in SU(3) Landau gauge gluodynamics,''
  Phys.\ Rev.\ D {\bf 72}, 014507 (2005)
  [arXiv: hep-lat/0506007];
  %%CITATION = PHRVA,D72,014507;%% 

\bibitem{sternbeck06} 
  A.\ Sternbeck, E.\ M.\ Ilgenfritz, M.\ Muller-Preussker, A.\ Schiller and I.\ L.\ Bogolubsky,
  %``Lattice study of the infrared behavior of QCD Green's functions in Landau
  %gauge,''
  PoS {\bf LAT2006}, 076 (2006)
  [arXiv:hep-lat/0610053].
  %%CITATION = POSCI,LAT2006,076;%%

\bibitem{Cucchieri:2006xi}  
  A.~Cucchieri and T.~Mendes,
  %``Propagators, running coupling and condensates in lattice QCD,''
  Braz.\ J.\ Phys.\  {\bf 37}, 484 (2007)
  [arXiv:hep-ph/0605224].
  %%CITATION = BJPHE,37,484;%%

\bibitem{cucchieril7}
  I.\ L.\ Bogolubsky, E.\ M.\ Ilgenfritz, M.\ Muller-Preussker and A.\ Sternbeck,
  %``The Landau gauge gluon and ghost propagators in 4D SU(3) gluodynamics in
  %large lattice volumes,''
  PoS (LATTICE-2007) 290, arXiv: 0710.1968 [hep-lat];
  %%CITATION = ARXIV:0710.1968;%%
  A.\ Cucchieri and T.\ Mendes,
  %``What's up with IR gluon and ghost propagators in Landau gauge? A puzzling
  %answer from huge lattices,''
  PoS (LATTICE 2007) 297, arXiv:0710.0412 [hep-lat];
  %%CITATION = ARXIV:0710.0412;%%
  A.\ Sternbeck, L.\ von Smekal, D.\ B.\ Leinweber and A.\ G.\ Williams,
  %``Comparing SU(2) to SU(3) gluodynamics on large lattices,''
  PoS (LATTICE 2007) 340, arXiv:0710.1982 [hep-lat].
  %%CITATION = ARXIV:0710.1982;%% 

\bibitem{Cucchieri:2007rg}  
  A.~Cucchieri and T.~Mendes,
  %``Constraints on the IR behavior of the gluon propagator in Yang-Mills
  %theories,''
  Phys.\ Rev.\ Lett.\  {\bf 100}, 241601 (2008)
  [arXiv:0712.3517 [hep-lat]].
  %%CITATION = PRLTA,100,241601;%%

\bibitem{Bogolubsky}
  % \bibitem{Bogolubsky:2005wf}
  I.\ L.\ Bogolubsky, G.\ Burgio, M.\ Muller-Preussker and V.\ K.\ Mitrjushkin,
  %``Landau gauge ghost and gluon propagators in SU(2) lattice gauge theory:
  %Gribov ambiguity revisited,''
  Phys.\ Rev.\ D {\bf 74}, 034503 (2006)
  [arXiv:hep-lat/0511056];
  %%CITATION = PHRVA,D74,034503;%%
  I.\ L.\ Bogolubsky {\it et al.},
  % V.\ G.\ Bornyakov, G.\ Burgio, E.\ M.\ Ilgenfritz, M.\ Muller-Preussker and V.\ K.\ Mitrjushkin,
  %``Improved Landau gauge fixing and the suppression of finite-volume effects
  %of the lattice gluon propagator,''
  Phys.\ Rev.\  D {\bf 77}, 014504 (2008)
  [Erratum-ibid.\  D {\bf 77}, 039902 (2008)]
  [arXiv:0707.3611 [hep-lat]].
  %%CITATION = PHRVA,D77,014504;%%

\bibitem{limghost}  
  A.~Cucchieri and T.~Mendes,
  %``Constraints on the IR behavior of the ghost propagator in Yang-Mills
  %theories,''
  arXiv:0804.2371 [hep-lat].
  %%CITATION = ARXIV:0804.2371;%%

\bibitem{gggvertex}
%\bibitem{Parrinello:1994wd}
  C.\ Parrinello,
  %``Exploratory study of the three gluon vertex on the lattice,''
  Phys.\ Rev.\  D {\bf 50} (1994) 4247, [arXiv:hep-lat/9405024];
  %%CITATION = PHRVA,D50,4247;%%
%\bibitem{Boucaud:1998bq}
  P.\ Boucaud, J.\ P.\ Leroy, J.\ Micheli, O.\ Pene and C.\ Roiesnel,
  %``Lattice calculation of alpha(s) in momentum scheme,''
  JHEP {\bf 9810}, 017 (1998), [arXiv:hep-ph/9810322].
  %%CITATION = JHEPA,9810,017;%%

%\cite{Ilgenfritz:2006he}
\bibitem{Ilgenfritz:2006he}
  E.~M.~Ilgenfritz, M.~Muller-Preussker, A.~Sternbeck, A.~Schiller and I.~L.~Bogolubsky,
  %``Landau gauge gluon and ghost propagators from lattice QCD,''
  Braz.\ J.\ Phys.\  {\bf 37}, 193 (2007)
  [arXiv:hep-lat/0609043].
  %%CITATION = BJPHE,37,193;%%

\bibitem{Cucchieri:2008qm}
  A.~Cucchieri, A.~Maas and T.~Mendes,
  %``Three-point vertices in Landau-gauge Yang-Mills theory,''
  Phys.\ Rev.\  D {\bf 77}, 094510 (2008)
  [arXiv:0803.1798 [hep-lat]].
  %%CITATION = PHRVA,D77,094510;%%

%%%%% Other %%%%%

\bibitem{Kugo}
T.\ Kugo and I.\ Ojima,
%``Local Covariant Operator Formalism Of Nonabelian Gauge Theories And Quark Confinement Problem,''
Prog.\ Theor.\ Phys.\ Suppl.\ {\bf 66}, 1 (1979) [Erratum Prog.\ Theor.\ Phys.\ {\bf 71}, 1121 (1984)];
%%CITATION = PTPSA,66,1;%%
T.\ Kugo,
%``The Universal Renormalization Factors Z_1/Z_3 and Color Confinement Condition in Non-Abelian Gauge Theory,''
arXiv:hep-th/9511033.
%%CITATION = HEP-TH 9511033;%%

\bibitem{lvs}
  A.~Sternbeck and L.~von Smekal,
  arXiv:0811.4300 [hep-lat].
  %%CITATION = ARXIV:0811.4300;%%

%\cite{Bakeev:2003rr}
\bibitem{Bakeev:2003rr}
  T.~D.~Bakeev, E.~M.~Ilgenfritz, V.~K.~Mitrjushkin and M.~Mueller-Preussker,
  %``On practical problems to compute the ghost propagator in SU(2) lattice
  %gauge theory,''
  Phys.\ Rev.\  D {\bf 69} (2004) 074507
  [arXiv:hep-lat/0311041];
  %%CITATION = PHRVA,D69,074507;%%
%\cite{Sternbeck:2005vs}
%\bibitem{Sternbeck:2005vs}
  A.~Sternbeck, E.~M.~Ilgenfritz and M.~Muller-Preussker,
  %``Spectral properties of the Landau gauge Faddeev-Popov operator in  lattice
  %gluodynamics,''
  Phys.\ Rev.\  D {\bf 73}, 014502 (2006)
  [arXiv:hep-lat/0510109].
  %%CITATION = PHRVA,D73,014502;%%


\bibitem{gc}
A.~Maas, arXiv:0808.3047 [hep-lat].
  %%CITATION = ARXIV:0808.3047;%%

\bibitem{Gribov}
  %\bibitem{Gribov:1977wm}
  V.\ N.\ Gribov,
  %``Quantization of non-Abelian gauge theories,''
  Nucl.\ Phys.\ B {\bf 139}, 1 (1978).
  %%CITATION = PHRVA,139,B184;%%

\bibitem{Singer:dk}
  I.\ M.\ Singer,
  % ``Some Remarks On The Gribov Ambiguity,''
  Commun.\ Math.\ Phys.\ {\bf 60} (1978) 7.
  %% CITATION = CMPHA,60,7;%%

\bibitem{Neuberger:1986xz}
  H.~Neuberger,
  %``NONPERTURBATIVE BRS INVARIANCE AND THE GRIBOV PROBLEM,''
  Phys.\ Lett.\  B {\bf 183}, 337 (1987).
  %%CITATION = PHLTA,B183,337;%%

%\cite{vonSmekal:2007ns}
\bibitem{vonSmekal:2007ns}
  L.~von Smekal, D.~Mehta, A.~Sternbeck and A.~G.~Williams,
  %``Modified Lattice Landau Gauge,''
  PoS {\bf LAT2007} (2007) 382
  [arXiv:0710.2410 [hep-lat]]; 
  %%CITATION = POSCI,LAT2007,382;%%
%\cite{vonSmekal:2008es}
%\bibitem{vonSmekal:2008es}
  L.~von Smekal, A.~Jorkowski, D.~Mehta and A.~Sternbeck,
  %``Lattice Landau gauge via Stereographic Projection,''
  PoS C {\bf ONFINEMENT8} (2008) 048
  [arXiv:0812.2992 [hep-th]].
  %%CITATION = POSCI,CONFINEMENT8,048;%%


\bibitem{vanBaal:1997gu}
P.~van Baal,
%``Gribov ambiguities and the fundamental domain,''
arXiv:hep-th/9711070;
%%CITATION = HEP-TH 9711070;%%
.~Semenov-Tyan-Shanskii and V.~Franke, Zap.\ Nauch.\ Sem.\ Leningrad. Otdeleniya Matematicheskogo Institutia in V.~A.~Stekolov, AN SSSR, Vol.\ 120 (1982), 159 (In English translation: New York, Plenum Press 1986);
G.~Dell'Antonio, D.~Zwanziger, Proceedings of the NATO Advanced Workshop on Probabilistic Methods in Quantum Field Theory and Quantum Gravity, Cargese. New York, Plenum Press (1986), 21.

%\cite{Tissier:2008nw}
\bibitem{Tissier:2008nw}
  M.~Tissier and N.~Wschebor,
  %``Gauged supersymmetries in Yang-Mills theory,''
  arXiv:0809.1880 [hep-th].
  %%CITATION = ARXIV:0809.1880;%%


\bibitem{Huber:2007kc}
  M.~Q.~Huber, R.~Alkofer, C.~S.~Fischer and K.~Schwenzer,
  %``The infrared behavior of Landau gauge Yang-Mills theory in d=2, 3 and 4
  %dimensions,''
  Phys.\ Lett.\  B {\bf 659} (2008) 434
  [arXiv:0705.3809 [hep-ph]].
  %%CITATION = PHLTA,B659,434;%%


\bibitem{pawlowski} J.~M.~Pawlowski, work in preparation.

\bibitem{nedelkopawlowski} S.~Nedelko and J.~M.~Pawlowski, work in preparation.

%\cite{Fischer:2004nq}
\bibitem{Fischer:2004nq}
  C.~S.~Fischer, R.~Alkofer, T.~Dahm and P.~Maris,
  %``Dynamical chiral symmetry breaking in unquenched QED(3),''
  Phys.\ Rev.\  D {\bf 70} (2004) 073007
  [arXiv:hep-ph/0407104].
  %%CITATION = PHRVA,D70,073007;%%

\bibitem{Ball:1980ax}
  J.~S.~Ball and T.~W.~Chiu,
  %``Analytic Properties Of The Vertex Function In Gauge Theories. 2,''
  Phys.\ Rev.\  D {\bf 22}, 2550 (1980)
  [Erratum-ibid.\  D {\bf 23}, 3085 (1981)].
  %%CITATION = PHRVA,D22,2550;%%

\bibitem{BarGadda:1979cz}
  U.~Bar-Gadda,
  %``Infrared Behavior Of The Effective Coupling In Quantum Chromodynamics: A
  %Nonperturbative Approach,''
  Nucl.\ Phys.\  B {\bf 163}, 312 (1980).
  %%CITATION = NUPHA,B163,312;%%

%\cite{Fischer:2005en}
\bibitem{Fischer:2005en}
  C.~S.~Fischer, P.~Watson and W.~Cassing,
  %``Probing unquenching effects in the gluon polarisation in light mesons,''
  Phys.\ Rev.\  D {\bf 72}, 094025 (2005)
  [arXiv:hep-ph/0509213].
  %%CITATION = PHRVA,D72,094025;%%

%\cite{Fischer:2003zc}
\bibitem{Fischer:2003zc}
  C.~S.~Fischer, PhD thesis,
  %``Non-perturbative propagators, running coupling and dynamical mass
  %generation in ghost - antighost symmetric gauges in QCD,''
  arXiv:hep-ph/0304233.
  %%CITATION = HEP-PH/0304233;%%

\bibitem{gzwanziger2}
  %\bibitem{Zwanziger:1990by}
  D.\ Zwanziger,
  %``Vanishing color magnetization in lattice Landau and Coulomb gauges,''
  Phys.\ Lett.\ B {\bf 257}, 168 (1991);
  %%CITATION = PHLTA,B257,168;%%
  % D.\ Zwanziger,
  %``Vanishing of zero momentum lattice gluon propagator and color
  %confinement,''
  Nucl.\ Phys.\ B {\bf 364}, 127 (1991);
  %%CITATION = NUPHA,B364,127;%%
  % D.\ Zwanziger,
  %``Fundamental modular region, Boltzmann factor and area law in lattice gauge
  %theory,''
  Nucl.\ Phys.\ B {\bf 412}, 657 (1994);
  %%CITATION = NUPHA,B412,657;%%
  % D.\ Zwanziger,
  %``Non-perturbative Landau gauge and infrared critical exponents in QCD,''
  Phys.\ Rev.\ D {\bf 65}, 094039 (2002) [arXiv: hep-th/0109224].
  %%CITATION = HEP-TH 0109224;%%
  
\bibitem{Cucchieri:2007ta}
  A.~Cucchieri, A.~Maas and T.~Mendes,
  %``Infrared properties of propagators in Landau-gauge pure Yang-Mills   theory
  %at finite temperature,''
  Phys.\ Rev.\  D {\bf 75} (2007) 076003
  [arXiv:hep-lat/0702022].
  %%CITATION = PHRVA,D75,076003;%%



%%%%%%%%%%%%%%%%%% Curci-Ferrari %%%%%%%%%%%%%%%%%%%%%%%%

%\cite{Kalloniatis:2005if}
\bibitem{Kalloniatis:2005if}
  A.~C.~Kalloniatis, L.~von Smekal and A.~G.~Williams,
  %``Curci-Ferrari mass and the Neuberger problem,''
  Phys.\ Lett.\  B {\bf 609} (2005) 424
  [arXiv:hep-lat/0501016].
  %%CITATION = PHLTA,B609,424;%%


%\cite{vonSmekal:2008en}
\bibitem{vonSmekal:2008en}
  L.~von Smekal, M.~Ghiotti and A.~G.~Williams,
  %``Decontracted double BRST on the lattice,''
  arXiv:0807.0480 [hep-th]; 
  %%CITATION = ARXIV:0807.0480;%%
%\cite{Ghiotti:2006pm}
%\bibitem{Ghiotti:2006pm}
%  M.~Ghiotti, L.~von Smekal and A.~G.~Williams,
  %``Extended double lattice BRST, Curci-Ferrari mass and the Neuberger
  %problem,''
  AIP Conf.\ Proc.\  {\bf 892} (2007) 180
  [arXiv:hep-th/0611058].
  %%CITATION = APCPC,892,180;%%

%\cite{Zwanziger:1997}
\bibitem{Zwanziger:1997}
D.~Zwanziger,
%``Lattice Coulomb hamiltonian and static color-Coulomb field,''
Nucl.\ Phys.\ B {\bf 485} (1997) 185
[arXiv:hep-th/9603203].  See especially the statement following eq. 
(6.9) and Appendix C.
%%CITATION = HEP-TH 9603203;%%



%\cite{Fischer:2005ui}
\bibitem{Fischer:2005ui}
  C.~S.~Fischer, B.~Gr\"uter and R.~Alkofer,
  %``Solving coupled Dyson-Schwinger equations on a compact manifold,''
  Annals Phys.\  {\bf 321}, 1918 (2006)
  [arXiv:hep-ph/0506053].
  %%CITATION = APNYA,321,1918;%%  


%\cite{Fischer:2007pf}
\bibitem{Fischer:2007pf}
  C.~S.~Fischer, A.~Maas, J.~M.~Pawlowski and L.~von Smekal,
  %``Large volume behaviour of Yang-Mills propagators,''
  Annals Phys.\  {\bf 322} (2007) 2916
  [arXiv:hep-ph/0701050].
  %%CITATION = APNYA,322,2916;%%

%\cite{Zwanziger:1992qr}
\bibitem{Zwanziger:1992qr}
  D.~Zwanziger,
  %``Renormalizability of the critical limit of lattice gauge theory by BRS
  %invariance,''
  Nucl.\ Phys.\  B {\bf 399} (1993) 477.
  %%CITATION = NUPHA,B399,477;%%

%\cite{Gracey:2005cx}
\bibitem{Gracey:2005cx}
  J.~A.~Gracey,
  %``Two loop correction to the Gribov mass gap equation in Landau gauge  QCD,''
  Phys.\ Lett.\  B {\bf 632} (2006) 282
  [arXiv:hep-ph/0510151].
  %%CITATION = PHLTA,B632,282;%%

\bibitem{Maas:2005hs}
  A.~Maas, J.~Wambach and R.~Alkofer,
  %``The high-temperature phase of Landau-gauge Yang-Mills theory,''
  Eur.\ Phys.\ J.\  C {\bf 42} (2005) 93
  [arXiv:hep-ph/0504019].
  %%CITATION = EPHJA,C42,93;%%

\bibitem{Bowman:2004jm}
  P.~O.~Bowman, U.~M.~Heller, D.~B.~Leinweber, M.~B.~Parappilly and A.~G.~Williams,
  %``Unquenched gluon propagator in Landau gauge,''
  Phys.\ Rev.\  D {\bf 70}, 034509 (2004)
  [arXiv:hep-lat/0402032].
  %%CITATION = PHRVA,D70,034509;%%

\bibitem{Maas:2004se}
  A.~Maas, J.~Wambach, B.~Gruter and R.~Alkofer,
  %``High-temperature limit of Landau-gauge Yang-Mills theory,''
  Eur.\ Phys.\ J.\  C {\bf 37}, 335 (2004)
  [arXiv:hep-ph/0408074].  
  %%CITATION = EPHJA,C37,335;%%

%\cite{Alkofer:2008tt}
\bibitem{Alkofer:2008tt}
  R.~Alkofer, C.~S.~Fischer, F.~J.~Llanes-Estrada and K.~Schwenzer,
  %``The quark-gluon vertex in Landau gauge QCD: Its role in dynamical chiral
  %symmetry breaking and quark confinement,''
  arXiv:0804.3042 [hep-ph], accepted by Ann. Phys..
  %%CITATION = ARXIV:0804.3042;%%

\bibitem{Cucchieri:1997ns}
  A.~Cucchieri,
  %``Numerical study of the fundamental modular region in the minimal Landau
  %gauge,''
  Nucl.\ Phys.\  B {\bf 521}, 365 (1998)
  [arXiv:hep-lat/9711024].
  %%CITATION = NUPHA,B521,365;%%

%\cite{Reinhardt:2008ij}
\bibitem{Reinhardt:2008ij}
  H.~Reinhardt and W.~Schleifenbaum,
  %``Hamiltonian Approach to 1+1 dimensional Yang-Mills theory in Coulomb
  %gauge,''
  arXiv:0809.1764 [hep-th].
  %%CITATION = ARXIV:0809.1764;%%

%\cite{Cucchieri:2007zm}
\bibitem{Cucchieri:2007zm}
  A.~Cucchieri, T.~Mendes, O.~Oliveira and P.~J.~Silva,
  %``Just how different are SU(2) and SU(3) Landau-gauge propagators in the IR
  %regime?,''
  Phys.\ Rev.\  D {\bf 76} (2007) 114507
  [arXiv:0705.3367 [hep-lat]].
  %%CITATION = PHRVA,D76,114507;%%

%\cite{Sternbeck:2007ug}
\bibitem{Sternbeck:2007ug}
  A.~Sternbeck, L.~von Smekal, D.~B.~Leinweber and A.~G.~Williams,
  %``Comparing SU(2) to SU(3) gluodynamics on large lattices,''
  PoS {\bf LAT2007} (2007) 340
  [arXiv:0710.1982 [hep-lat]].
  %%CITATION = POSCI,LAT2007,340;%%

\bibitem{Maas:2005ym}
  A.~Maas,
  %``Gluons at finite temperature in Landau gauge Yang-Mills theory,''
  Mod.\ Phys.\ Lett.\  A {\bf 20}, 1797 (2005)
  [arXiv:hep-ph/0506066].
  %%CITATION = MPLAE,A20,1797;%%

\bibitem{Maas:2007af}
  A.~Maas and {\v{S}}.~Olejn\'ik,
  %``A first look at Landau-gauge propagators in G2 Yang-Mills theory,''
  JHEP {\bf 0802}, 070 (2008)
  [arXiv:0711.1451 [hep-lat]].
  %%CITATION = JHEPA,0802,070;%%

\bibitem{Alkofer:2003jj}
  R.~Alkofer, W.~Detmold, C.~S.~Fischer and P.~Maris,
  %``Analytic properties of the Landau gauge gluon and quark propagators,''
  Phys.\ Rev.\  D {\bf 70}, 014014 (2004)
  [arXiv:hep-ph/0309077].
  %%CITATION = PHRVA,D70,014014;%%

\bibitem{Bowman:2007du}
  P.~O.~Bowman {\it et al.},
  %``Scaling behavior and positivity violation of the gluon propagator in full
  %QCD,''
  Phys.\ Rev.\  D {\bf 76}, 094505 (2007)
  [arXiv:hep-lat/0703022].
  %%CITATION = PHRVA,D76,094505;%%

\bibitem{Cucchieri:2004mf}
  A.~Cucchieri, T.~Mendes and A.~R.~Taurines,
  %``Positivity violation for the lattice Landau gluon propagator,''
  Phys.\ Rev.\  D {\bf 71}, 051902 (2005)
  [arXiv:hep-lat/0406020].
  %%CITATION = PHRVA,D71,051902;%%

\bibitem{Nielsen:1975fs}
  N.~K.~Nielsen,
  %``On The Gauge Dependence Of Spontaneous Symmetry Breaking In Gauge
  %Theories,''
  Nucl.\ Phys.\  B {\bf 101}, 173 (1975).
  %%CITATION = NUPHA,B101,173;%%




\end{thebibliography}
\end{document}